\documentclass[aps,physrev,preprint,groupedaddress,nofootinbib]{revtex4-2}

\usepackage[T1]{fontenc}
\usepackage{diagbox}
\usepackage{adjustbox}
\usepackage{multirow}
\usepackage{pifont}
\usepackage{array}
\usepackage{mathtools}
\usepackage{booktabs}
\usepackage[dvipsnames]{xcolor}
\usepackage{tcolorbox}
\usepackage{float}
\usepackage{placeins}
\usepackage{braket}
\usepackage{slashed}
\usepackage{amssymb}

\newcommand{\vev}{v_\text{EW}}

\usepackage[normalem]{ulem}
\newcommand{\stkout}[1]{\ifmmode\text{\sout{\ensuremath{#1}}}\else\sout{#1}\fi}

\newcommand{\s}{\sigma}

\newcommand{\mO}{\mathcal{O}}

\newcommand{\beq}{\begin{equation}}
\newcommand{\eeq}{\end{equation}}
\newcommand{\bea}{\begin{eqnarray}}
\newcommand{\eea}{\end{eqnarray}}

\begin{document}

\title{Establishing the Primary HEFT as a Precision Benchmark for UV-HEFT Matching}

\author{Zizhou Ge}
\thanks{The authors are listed in alphabetical order by last name.}
\email{gezizhou@snnu.edu.cn}
\affiliation{School of physics and Information Technology, Shaanxi Normal University,
Xi'an 710119, China}

\author{Huayang Song}
\email{huayangs1990@ibs.re.kr}
\affiliation{Particle Theory and Cosmology Group, Center for Theoretical Physics of the Universe, Institute for Basic Science (IBS), Daejeon, 34126, Korea}

\author{Xia Wan}
\email{wanxia@snnu.edu.cn}
\affiliation{School of physics and Information Technology, Shaanxi Normal University,
Xi'an 710119, China}

\date{\today}

\begin{abstract}

We match the real Higgs triplet model (RHTM) onto HEFT under different parameter choices and power-counting schemes, thereby obtaining several representative HEFT formulations and clarifying their relations.
We establish the primary HEFT (pHEFT) as a benchmark framework, demonstrating that alternative HEFT constructions can be systematically derived from it. A key advantage of the pHEFT construction is its parameter choice, which maintains linear relations between the UV Lagrangian parameters and squared heavy masses. By strictly employing the inverse squared heavy masses as the expansion parameters without imposing additional constraints, pHEFT preserves maximal ultraviolet (UV) information and ensures higher perturbative accuracy by avoiding the additional truncations inherent in more complex, non-linear formulations or extra constraints. Through the analysis of the $Z_2$-symmetric real singlet model and the 2HDM, we illustrate the criteria for identifying viable primary HEFT constructions in UV models with scalar extensions. Furthermore, for the first time, we derive the HEFT operators of the RHTM involving fermions.
\end{abstract}

%\keywords{}

\maketitle

\tableofcontents
\section{Introduction\label{sec:introduction}}

Since the Higgs boson is discovered at the Large Hadron Collider (LHC)~\cite{CMS:2012qbp,ATLAS:2012yve}, looking for new physics beyond the Standard Model (BSM) at the LHC becomes the predominant goal of high energy physics. Although experimental measurements have become increasingly precise, there is no clear evidence for the existence of BSM particles~\cite{CMS:2024yiy,ATLAS:2024zkx}. This suggests that new physics, if it exists, is likely to reside at a higher energy scale, such as the TeV scale. Effective field theories (EFTs) formulated at the electroweak scale, extending the SM, provide a systematic framework for parameterizing new-physics effects~\cite{Weinberg:1980wa, Georgi:1993mps}.

At the electroweak energy scale, two main types of Effective Field Theories (EFTs) are used: Standard Model EFT (SMEFT)~\cite{Weinberg:1979sa,Buchmuller:1985jz,Leung:1984ni,Brivio:2017vri} and Higgs EFT (HEFT)~\cite{Appelquist:1980vg, Longhitano:1980iz, Longhitano:1980tm, Feruglio:1992wf, Herrero:1993nc, Herrero:1994iu, Grinstein:2007iv, Buchalla:2012qq, Alonso:2012px, Buchalla:2013rka, Brivio:2013pma, Buchalla:2013eza, Gavela:2014vra, Pich:2015kwa, Alonso:2015fsp, Brivio:2016fzo, Alonso:2016oah, Pich:2016lew, Merlo:2016prs, Pich:2018ltt, Krause:2018cwe, Sun:2022ssa, Sun:2022snw}. Both SMEFT and HEFT include the same particle content as the SM. Their difference fundamentally stems from the assumption about the symmetry realization of the Higgs boson. In SMEFT, the Higgs and the three Goldstones together form a $SU(2)_L$ doublet as in the SM.
HEFT treats the Higgs as an $SU(2)_L$ singlet with non-linearly transforming Goldstones, and encompasses the SMEFT doublet description as a special limiting case~\cite{Alonso:2015fsp,Alonso:2016oah,Falkowski:2019tft}. Over the past decade, the SMEFT has gained popularity and developed rapidly~\cite{Brivio:2020onw,Ethier:2021bye,Grzadkowski:2010es,Murphy:2020rsh,Ren:2022tvi,Henning:2017fpj,Henning:2014wua,Isidori:2024pca,Gargalionis:2024jaw}, partly due to its linear structure, which allows for a concise formulation.
Nevertheless, the SMEFT framework is not always adequate. When new physics introduces additional sources of electroweak symmetry breaking or involves non-decoupling heavy degrees of freedom~\cite{Cohen:2020xca,Banta:2021dek}, the effects of new physics cannot be consistently captured within the SMEFT framework and require a HEFT description. There has been a notable increase in research on HEFT in recent years~\cite{Remmen:2024hry,Sun:2022ssa,Sun:2022snw,Graf:2022rco,Gomez-Ambrosio:2022why,Gomez-Ambrosio:2022qsi,Alonso:2021rac,Asiain:2021lch,Cohen:2021ucp,Herrero:2021iqt,Herrero:2022krh,Alasfar:2023xpc,Li:2025gbx,Ding:2026qto,Brivio:2025yrr,Brivio:2025sib,Alonso:2025jvv,Alonso:2025ksv}.

EFTs can be useful purely from the bottom up: one specifies the physical degrees of freedom along with their transformations under a certain set of symmetries and identifies a power counting scheme to organize the operator expansion. Though in this sense EFTs are ``model independent'' and as such they provide an approach for classifying observables which deviate from the SM predictions, the top down approach is more useful to help understanding the (more) fundamental ultraviolet (UV) description of nature. In this scenario, a certain BSM model is assumed, and a process, known as ``matching'', is carried out between the EFTs and the UV model by ``integrating out'' the ``heavy'' degrees of freedom. However, it should be noted that the expansion parameter used during matching is generally method-dependent and can differ from the power-counting parameter used in the bottom-up approach.

Within BSM physics, commonly used matching methods are performing the heavy mass expansion via either diagrammatic calculation or functional prescription~\cite{Henning:2014wua,Ellis:2023zim,Corbett:2021eux}. This exercise has been done especially for the SMEFT but also the HEFT, considering several different UV models~\cite{Robens:2015gla, Robens:2016xkb,Gunion:1989we, Branco:2011iw}. Following this direction, several automatic matching tools are developed~\cite{Criado:2017khh,DasBakshi:2018vni,Fuentes-Martin:2022jrf,Carmona:2021xtq}. However, even for the SMEFT, such expansion parameters (the inverse powers of the BSM state masses) are not fully consistent with the SMEFT metric (canonical dimension): 1) due to computational limitations, we must perform a truncated loop expansion implicitly in the above description; 2) the extra dimensional parameters $\mu_i$ in the UV theory can give potential corrections, which are generally also incalculable in full, at the order of $\mu_i^n/M_j^{d+n-4}$ to the $d$-dim operators. This discrepancy between the power counting and the matching expansion parameter becomes more evident in HEFT, since its operator expansion is organized by chiral dimension. Without other known mathematical tools, we must rely on this approach, expanding the propagators of the heavy BSM states order by order~\footnote{Though physicists usually say that there are two matching methods, diagrammatic and functional methods, the physics behind them are same, that propagators of the BSM states are introduced and shrunk to some contacted interactions.}~\footnote{One could imagine that a lattice simulation matching, like what is done in the matching between QCD and chiral perturbation theory, can in principle be performed to obtain the HEFT's Wilson coefficients without introducing a different power counting parameter besides the chiral dimension. However, due to the technical obstacles to simulating chiral fermions on the lattice and also the resource consumption, a systematic study of the lattice matching between generic UV theories and HEFT is, to our knowledge, still lacking.}. Therefore, top-down matching essentially fixes the parametric scaling of the Wilson coefficients~\footnote{We will employ the term ``power-counting scheme'' in a flexible manner to encompass both related concepts, relying on the sentence context to clarify the specific meaning.}. After electroweak symmetry breaking, the lifting of degeneracies in the BSM spectrum leads to a richer structure in the UV-HEFT matching. Further the inclusion of experimental constraints renders the power-counting scheme non-unique~\cite{Dawson:2023oce} (see Sec.~\ref{sec:HEFTs} for concrete examples). It is therefore worthwhile to study the relations among different HEFTs from the same UV completion, with the aim of providing practical guidance for phenomenological applications. A central challenge in this endeavor is whether a consistent and practically useful power-counting scheme can be formulated for UV–HEFT matching, especially one that is amenable to automation, since any systematic use of HEFT ultimately relies on automated tools.

In this work, we find that in new physics models with scalar extensions, there exists such a HEFT, named the primary HEFT, that can play this role. Starting from a UV parameter set consisting of heavy masses, mixing angles, and VEVs, we adopt a power-counting scheme that expands in inverse powers of heavy masses and organizes operators accordingly, without imposing additional restrictions. This matching procedure preserves maximal information within the specified expansion order. Other HEFTs can thus be mapped from this primary HEFT by imposing further restrictions or performing parameter transformations.

\newpage
The accuracy of an EFT is fundamentally determined by the convergence of its power-counting expansion. In this context, the primary HEFT achieves superior accuracy at a specific order (e.g., $1/M^n$) because it encapsulates the complete UV dynamics without the loss of information incurred by additional parametric simplifications. While restricted HEFTs may be easier to implement for specific phenomenological studies, they often neglect sub-leading mixing effects, or assume specific hierarchies between VEVs to justify further truncations, thereby leading to a loss of fidelity relative to the UV theory. By utilizing the primary HEFT as a benchmark, the truncation error introduced by these further approximations can be quantified, ensuring a controlled and systematic mapping from the primary HEFT to other HEFTs.

Using the real Higgs triplet model (RHTM) as an explicit example, we construct its primary HEFT and several alternative HEFT descriptions. Compared with the primary HEFT, these alternative HEFTs differ from the primary HEFT either by adopting different scaling behaviors for the same parameter set, or by employing different parameter sets that may or may not explicitly include heavy mass parameters, leading to intrinsically different power-counting schemes. By studying the mappings from the primary HEFT to these alternative HEFTs, we demonstrate how accuracy is progressively lost under different power-counting schemes and their associated truncations. We further present numerical comparisons for the $hh \to hh$ scattering process to illustrate these effects quantitatively.

In addition, by mapping the primary HEFT onto a specific parameter space that mimics the SMEFT setting, we obtain a formulation that recovers the SMEFT results at the amplitude level. The exact agreement found in our calculation suggests that, under this mapping, the HEFT becomes phenomenologically equivalent to the SMEFT (after spontaneous symmetry breaking) across a general class of processes.

The paper is organized as follows: In Sec.~\ref{sec:HEFTandRHTM} we introduce the HEFT formulation and the RHTM. Sec.~\ref{sec:HEFTs} include five parts.
In Sec.~\ref{sec:pHEFTanddHEFT} we first obtain the primary HEFT(pHEFT) and decoupling HEFT(dHEFT) of RHTM, then get their mapping relation and discuss their correspondence. In Sec.~\ref{sec:Z2HEFT}, we introduce the $Z_2$-HEFT and point out that a viable primary HEFT requires a parameterization in which the UV Lagrangian parameters is polynomial to the squared heavy masses. In Sec.~\ref{sec:xiHEFT}, We establish a hierarchical relation among the pHEFT, dHEFT and $\xi$-HEFT.
In Sec.~\ref{sec:Y2HEFT}, based on the SMEFT parameter set and power-counting scheme, we map the pHEFT to the $Y_2$-HEFT and reproduce the SMEFT predictions.
In Sec.~\ref{sec:otherUVs} we introduce the good primary HEFT for Z2RSM and 2HDM.
Sec.~\ref{sec:conclusion} gives a discuss and conclusion.

\section{HEFT and the model\label{sec:HEFTandRHTM}}
\subsection{HEFT\label{sec:HEFT}}

The Higgs sector in the SM has an approximate global symmetry $SU(2)_L\times SU(2)_R$ which is spontaneously broken down to the custodial symmetry $SU(2)_C$ by the Higgs vacuum expectation value (VEV). Using the CCWZ formalism~\cite{Coleman:1969sm}, one can build an EFT, known as the HEFT, describing the interactions of the non-linearly realized Goldstone bosons among themselves and
with other SM fields, where a real scalar denotes the Higgs field. The Goldstones $\pi_i$ are usually treated as a single object, a unitary matrix $U\equiv\exp\left(\frac{i \pi_i \sigma_i}{\vev}\right)$, where $\sigma_i$ are the three Pauli matrices. At the lowest order, the HEFT Lagrangian reads~\footnote{Only terms relevant for our purposes are shown. The kinetic terms for gauge bosons and fermions are omitted.}
\beq
\begin{aligned}[b]
\mathcal{L}_{\text{HEFT}}^{\text{LO}} \supset{}& \frac{1}{2}\partial_\mu h \partial^\mu h - \mathcal{V}( h ) + \frac{\vev^2}{4} \mathcal{F}( h ) \braket{D_\mu U^\dagger D^\mu U} \\
& + \frac{\vev^2}{4}\mathcal{G}( h ) \braket{U^\dagger D_\mu U\sigma_3} \braket{U^\dagger D^\mu U\sigma_3} \\
& - \frac{\vev}{\sqrt{2}}\Bigl(\bar{Q}_L U\mathcal{Y}_Q( h ) Q_R + \bar{L}_L U\mathcal{Y}_L( h ) L_R + \text{h.c.}\Bigr),
\end{aligned}
\label{eq:HEFT}
\eeq
where $\vev=246\,\mathrm{GeV}$ represents the electroweak VEV, $\braket{\cdots}$ denotes the trace, $D_\mu$ is the covariant derivative, and $\mathcal{V}( h )$, $\mathcal{F}( h )$, $\mathcal{G}( h )$, and $\mathcal{Y}_{Q/L}( h )$ are polynomial functions of $ h $ with the following forms,
\begin{gather}
    \mathcal{V}( h )=\frac{1}{2}m_h^2 h^2\left[1+(1+\Delta\kappa_3)\frac{h}{\vev}+\frac{1}{4}(1+\Delta\kappa_4)\frac{h^2}{\vev^2}+\cdots\right], \label{kappa34def} \\
    \mathcal{F}( h )=1+2(1+\Delta a)\frac{h}{\vev}+(1+\Delta b)\frac{h^2}{\vev^2}+\cdots, \\
    \mathcal{G}( h )=\Delta\alpha+\Delta a^\slashed{C}\frac{h}{\vev}+\Delta b^\slashed{C}\frac{h^2}{\vev^2}+\cdots, \\
    \mathcal{Y}_Q( h )=\mathrm{diag}\biggl(\sum_n Y_U^{(n)}\frac{h^n}{\vev^n},\,\sum_n Y_D^{(n)}\frac{h^n}{\vev^n}\biggr),\ \mathcal{Y}_L( h )=\mathrm{diag}\biggl(0,\,\sum_n Y_\ell^{(n)}\frac{h^n}{\vev^n}\biggr).
\end{gather}
The first term is the Higgs potential. The second and third terms describe the Goldstones dynamic term which contains the gauge boson mass terms and interactions between Higgs and massive gauge bosons; and note that the third one breaks the custodial symmetry. The four term generates fermion masses and Higgs-fermion interactions.

\subsection{The Real Higgs Triplet Model (RHTM)\label{sec:RHTM}}
\subsubsection{The model in a non-linear representation}
In the RHTM besides the $SU(2)_L$ doublet there exists a triplet with hyper-charge $Y=0$.
The Lagrangian of RHTM can be expressed as~\cite{Song:2024kos,Song:2022jns,Corbett:2021eux,Ellis:2023zim} (for phenomenological studies, cf. e.g. Refs.~\cite{FileviezPerez:2008bj, Patel:2012pi, Niemi:2018asa, Niemi:2020hto})
\beq
\begin{aligned}[b]
\mathcal{L} &\supset \left(D_{\mu} H \right)^{\dagger}\left(D^\mu H \right)+\braket{D_{\mu}\Sigma^{\dagger} D^{\mu}\Sigma} -V_S-V_Q,
% \label{LHSigma}
\\
V_S &=
Y_1^2 H^\dagger H +Z_1 ( H^\dagger H)^2
+ Y_2^2 \braket{\Sigma^\dagger \Sigma}
+ Z_2 \braket{\Sigma^\dagger \Sigma}^2
+ Z_3 H^\dagger H \braket{\Sigma^\dagger\Sigma}
+2 Y_3 H^\dagger \Sigma H, \\
V_Q &= y_u \bar Q_L \tilde{H} u_R + y_d \bar Q_L H d_R + \text{h.c.},
\end{aligned}
\label{Hpotential}
\eeq
where $\braket{\cdots}$ denotes the trace,
$Y_is,i=1,2,3$ are dimensional parameters while $Z_is,i=1,2,3$ are dimensionless. $ H $ is
the SM Higgs doublet, $\Sigma$ is the real triplet. $\tilde{H}= i \sigma_2 H^\ast$. $y_u$ and $y_d$ are diagonal Yukawa coupling matrices in flavor space for simplicity. Due to the gauge structure, there are no interactions between fermion fields and the triplet.

After spontaneously symmetry breaking (SSB), we can write the doublet and triplet in a non-linear representation as~\cite{Song:2024kos}
\begin{align}
H &= U
\begin{pmatrix}
 \chi^+ \\
 \frac{1}{\sqrt{2}}(v_H+h^0+i \chi^0) \\
\end{pmatrix},
\quad U\equiv \exp\left(\frac{i \pi_i \sigma_i}{\vev}\right)
\label{eq:H} \\
\Sigma &= U \Phi
U^{\dagger}, \quad \Phi = \phi_i \sigma_i/2=\frac{1}{2}\begin{pmatrix}
 v_\Sigma +\phi^0 & \phi_1 -i \phi_2 \\
 \phi_1 + i \phi_2 & -v_\Sigma -\phi^0
\end{pmatrix},
\label{eq:SigmaPhi}
\end{align}
with
\beq
\chi^\pm = 2 \frac{v_H}{v_\Sigma} \phi^{\pm}, \quad \chi^0 = 0,
\eeq
where $\sigma_i,i= 1,2,3$ are Pauli matrices, $v_H$ and $v_\Sigma$ are vacuum expectation values (VEV) of the doublet and the triplet, $h^0$ and $\phi^0$ are the radial modes associated to the VEV directions, and the unitary matrix $U$ encapsulate three Goldstones $\pi_i$s, $\vev \equiv\sqrt{v_H^2+4 v_\Sigma^2}\simeq 246\,\mathrm{GeV}$ are the electroweak VEV, $\phi^{\pm}=\frac{1}{\sqrt{2}}(\phi_1\mp i \phi_2)$ is a pair of charged Higgs bosons, $\phi^0 = -v_\Sigma + \phi_3$. In this non-linear representation the physical states are factored out from the Goldstone matrix $U$, so that ``integrating out'' the heavy states is straightforward.

\subsubsection{Linear Relations Between Squared Masses and UV Lagrangian Parameters\label{sec:linear}}

After minimizing the potential in
Eq.~\eqref{Hpotential},
we have the following relations
\beq
Y_1^2 = - Z_1 v_H^2 - \frac{Z_3 v_\Sigma^2}{2} + Y_3 v_\Sigma, \quad
Y_2^2 = - Z_2 v_\Sigma^2 - \frac{Z_3 v_H^2}{2} + \frac{Y_3 v_H^2}{2 v_\Sigma},
\label{Y1Y2}
\eeq
which allows us to express the couplings $Y_{1, 2}$ in terms of the vacuum expectation values $v_{H, \Sigma}$. With nonzero values of $v_H$ and $v_\Sigma$, the mass terms after symmetry breaking are given by
\beq
\mathcal{L}_\text{mass}=\frac{1}{2}\begin{pmatrix}
    h^0 & \phi^0
\end{pmatrix}
\begin{pmatrix}
    2Z_1 v_H^2 &~v_H\left(Z_3 v_\Sigma-Y_3\right) \\
    v_H\left(Z_3 v_\Sigma-Y_3\right) &~2Z_2 v_\Sigma^2+\frac{Y_3 v_H^2}{2v_\Sigma}
\end{pmatrix}
\begin{pmatrix}
    h^0 \\ \phi^0
\end{pmatrix}+\left(v_H^2+4v_\Sigma^2\right)^2\frac{Y_3}{2v_\Sigma v_H^2}\phi^+\phi^-
\label{eq:massmatrix}
\eeq
where $h^0$ and $\phi^0$ are neutral Higgs bosons and $\phi^\pm$ are charged Higgs bosons.
The neutral fields $h^0$ and $\phi^0$ can then be rotated into the mass eigenstates $h$ and $K$ via a mixing angle $\gamma$~\footnote{Throughout this work, we use the shorthand notation $s_x \equiv \sin(x)$, $c_x \equiv \cos(x)$ for any angle $x$.},
\beq
\begin{pmatrix}
 h \\
 K \\
\end{pmatrix}
=
\begin{pmatrix}
  c_\gamma & - s_\gamma \\
  s_\gamma & c_\gamma
\end{pmatrix}
\begin{pmatrix}
 h^0 \\
 \phi^0 \\
\end{pmatrix},\quad \tan(2\gamma)= \frac{v_H (Z_3 v_\Sigma-Y_3)}{Z_2 v_\Sigma^2-Z_1 v_H^2+\frac{Y_3 v_H^2}{4 {v_\Sigma}}},
\label{eq:NSDiagMatrix}
\eeq
where $h$ represents the discovered 125 GeV Higgs, and $K$ represents a heavy neutral scalar.
%It is easily seen that $\tan(2\gamma)$ behaves approximately as $-4v_\Sigma/v_H$ in the limit of tiny $v_\Sigma$ with $Z_{1, 2, 3}, Y_3/v_H\sim\mathcal{O}(1)$. Nevertheless, in order to disentangle the non-decoupling effects induced by neutral Higgs mixing, we treat the angle $\gamma$ as a free parameter later.

The masses of $h$ and $K $ are given by
\begin{align}
m_{h, K}^2=Z_1 v_H^2+Z_2 v_\Sigma^2+\frac{Y_3 v_H^2}{4 v_\Sigma}\mp\sqrt{\left(Z_1 v_H^2-Z_2 v_\Sigma^2-\frac{Y_3 v_H^2}{4 v_\Sigma}\right)^2+v_H^2 (Z_3 v_\Sigma -Y_3)^2}.
\label{eq:NSmass2}
\end{align}
%for convention, we write them as $H^\pm=\phi^\pm$.
%after normalizing their kinetic term, $(1+ 4(v_\Sigma/v_H)^2) D_\mu \phi^+ D^\mu \phi^-$,
%from the second term in Eq.~\eqref{eq:massmatrix} we get their masses as
Meanwhile, the mass of the charged Higgs bosons is given by
\beq
m_{\phi^\pm}^2=(v_H^2+4 v_\Sigma^2)\frac{Y_3}{2 v_\Sigma}=\vev^2 \frac{Y_3}{2 v_\Sigma}.
\label{eq:mHpm}
\eeq

It is useful to rewrite the Lagrangian parameters in terms of physical masses, mixing angles, and VEVs. Apart from $Y_1$ and $Y_2$, the other four parameters are given by
\beq
\begin{aligned}[b]
    Y_3&=\frac{2v_\Sigma}{v_H^2+4 v_\Sigma^2}m_{\phi^\pm}^2 \\
    Z_1&=\frac{1}{2v_H^2}\left(c_\gamma^2 m_h^2+s_\gamma^2 m_K^2\right) \\
    Z_2&=\frac{1}{2v_\Sigma^2}\left(s_\gamma^2 m_h^2+c_\gamma^2 m_K^2-\frac{v_H^2}{v_H^2+4v_\Sigma^2}m_{\phi^\pm}^2\right) \\
    Z_3&=\frac{1}{v_H v_\Sigma}\left[s_\gamma c_\gamma\left(m_K^2-m_h^2\right)+\frac{2v_H v_\Sigma}{v_H^2+4v_\Sigma^2}m_{\phi^\pm}^2\right].
\end{aligned}
\label{Zis}
\eeq
From these expressions, we see that the four parameters are linearly proportional to the squares of the scalar masses, such as $m_h^2$, $m_K^2$ and $m_{\phi^\pm}^2$. Moreover, combining these results with Eq.~\eqref{Y1Y2}, $Y_1^2$ and $Y_2^2$ are also linearly proportional to the squared masses. This behavior arises because the physical masses of neutral states are the eigenvalues of the mass matrix (see Eq.~\eqref{eq:massmatrix}), which
depends linearly on the Lagrangian parameters and is diagonalized by
an orthogonal rotation matrix characterized by the mixing angle $\gamma$, while the physical mass of charged states is also linearly proportional to a Lagrangian parameter. Parameterized by the VEVs and the mixing angle $\gamma$, this linear correspondence is a generic feature of Standard Model extensions with additional scalars~\cite{Dawson:2023oce, Buchalla:2016bse, Du:2018eaw}.

\section{HEFTs of the RHTM\label{sec:HEFTs}}

The tree-level matching procedure follows the methodology established in Refs.~\cite{Buchalla:2016bse, Buchalla:2023hqk, Dawson:2023ebe, Dawson:2023oce, Song:2024kos, Song:2025kjp}, wherein heavy states are integrated out via the functional method. Specifically, starting from the UV Lagrangian, we adopt a systematic power-counting scheme and integrate out the heavy degrees of freedom by solving their classical EoMs. The resulting operators are subsequently projected onto a complete basis~\cite{Sun:2022ssa}. Throughout this procedure, there is considerable freedom in selecting both the initial parameter set and the corresponding power-counting scheme. These choices result in several distinct HEFT formulations~\cite{Dawson:2023oce}.
To clarify their interrelations, we first examine HEFTs derived from a common parameter set under different power-counting rules. Subsequently, we investigate the variations arising from different parameter sets, each associated with its own respective power-counting scheme.

\subsection{Primary HEFT and decoupling HEFT\label{sec:pHEFTanddHEFT}}
\subsubsection{Parameter Set}

In Ref.~\cite{Song:2025kjp}, we adopted the parameter set $\{Z_1, Z_2, Z_3, Y_3, v_H, \xi\}$ and defined the power counting via an expansion in $\xi\equiv v_\Sigma/v_H$. This framework describes a decoupling limit: as the heavy degrees of freedom are integrated out, the theory reduces to the SM at the lowest order of the expansion. However, this choice is not optimal for revealing possible non-decoupling effects. To address this point, it is advantageous to adopt a parameter set in which the heavy scalar masses and the mixing angle $ s_\gamma$ (see Eq.~\eqref{eq:NSDiagMatrix}) are explicit. Unlike the previous set, where these quantities were implicit, this choice directly exposes non-decoupling effects; in particular, $ s_\gamma$ can remain finite even after the heavy states are integrated out, signaling a departure from the standard decoupling limit. Accordingly, we choose the following set of independent parameters for our analysis:
%which is approximated to zero in SMEFT since the Higgs triplet is integrated out as a whole~\cite{Ellis:2023zim,Corbett:2021eux}\Huayang{The reason is incorrect. I'm unclear about what you intend to convey.}, but in HEFT matching it could be a free parameter. Finally \Huayang{why do we use "finally"?} we choose a new parameter set
\beq
\{ m_{\phi^\pm}, m_K, m_h, s_\gamma, \vev, \xi\},
\label{setPhy}
\eeq
where $m_{\phi^\pm}$ and $m_K$ represent the masses of the heavy charged and neutral scalars, respectively.

%where the former $Z_i$s and $Y_3$ are replaced by the masses of scalars and the mixing angle (see Eq.~\eqref{Zis}). \Huayang{can be shorten and made more clear.}

\subsubsection{Power Countings}

In the UV--SMEFT matching procedure~\cite{Ellis:2023zim, Corbett:2021eux, Song:2025kjp}, the power counting is performed by expanding in the ratio $(\vev/Y_2)^2$, where $Y_2$ denotes the mass parameter of the real triplet. Upon electroweak symmetry breaking, the mass degeneracy within the triplet is lifted, resulting in two distinct physical mass eigenstates: the charged Higgs with mass $m_{\phi^\pm}$ and the heavy neutral Higgs with mass $m_K$. Unlike the SMEFT approach, the UV--HEFT matching is performed by integrating out these physical fields $\phi^\pm$ and $K$. Given their large masses, it is therefore natural to adopt their inverse mass squares, ($1/m^2_{\phi^\pm}$ and $1/m^2_K$), as the expansion parameters. This leads to a two-parameter expansion.

%...Given their large masses, it is natural to adopt their inverse mass squares ($1/m^2_{\phi^\pm, K}$) as the expansion parameters.

When multiple expansion parameters are involved, their relative importance must be regulated by specifying their scaling behaviors within a consistent power-counting scheme. %This framework allows us to assign a formal scaling weight not only to the heavy mass scales but also to other relevant parameters.
Following the approach in Refs.~\cite{Dittmaier:2021fpq, Dawson:2023ebe}, we introduce an auxiliary parameter $t$ to track these weights,
\beq
m^2_{\phi^\pm}\sim m^2_K\sim\mathcal{O}(t^{-1}), \quad m_h\sim \vev \sim s_\gamma\sim \xi\sim\mathcal{O}(t^0),
\label{scale1}
\eeq
where the large masses of the heavy fields serve as expansion parameters, while parameters such as the mixing angle $ s_\gamma$ and the VEV ratio $\xi$ remain unsuppressed, capturing the non-decoupling effects characteristic of the HEFT framework.
Moreover, as a result of the linear relations between the Lagrangian parameters and the heavy mass squares (see Eq.~\eqref{Zis}), except for the propagator, there is no additional inverse-mass expansion when integrating out the heavy states, this scheme preserves the maximal amount of UV information. Consequently, we refer to the resulting effective theory as the primary HEFT (pHEFT).

Another possible scaling is:
\beq
m^2_{\phi^\pm}\sim m^2_K\sim\mathcal{O}(t^{-1}), \quad m_h\sim \vev\sim \mathcal{O}(t^0), \quad s_\gamma\sim \xi\sim \mathcal{O}(t^1),
\label{scale2}
\eeq
where the mixing angle $ s_\gamma$ and the VEV ratio $\xi$ are suppressed by the heavy scales. In the limit $m^2_{\phi^\pm, K} \to \infty$ (i.e., $t \to 0$), these parameters vanish, thereby satisfying the requirements for a decoupling scenario. We refer to the theory under this power-counting as the decoupling HEFT (dHEFT).

\subsubsection{The matching results}

Using the parameter set in Eq.~\eqref{setPhy} the Lagrangian is written as
\beq
\begin{aligned}[b]
\mathcal{L}\supset \mathcal{L}_\text{kin}(\vev, \xi; \phi_1, \phi_2, K, h, U) &- V_S(m^2_{\phi^\pm}, m^2_K, m_h, \vev, s_\gamma, \xi; \phi_1, \phi_2, K, h, U) \\
&- V_Q(y_u, y_d; \phi_1, \phi_2, K, h, U),
\end{aligned}
\eeq
where $\phi_1,\phi_2,K$ are heavy Higgses that will be integrated out. We use two real fields $\phi_{1,2}$ instead of $\phi^\pm$ in calculation for simplicity. Their EoMs are
\beq
\partial_\mu\left[ \frac{\partial \mathcal{L}}{\partial (\partial_\mu H^a)} \right] -\frac{\partial \mathcal{L}}{\partial H^a}=0, \quad H^a=(K,\phi_1,\phi_2).
\label{eq:EoMs}
\eeq
We solve the EoMs through a $t$-dependent series expansion. The heavy states are expanded as
\beq
\begin{aligned}[b]
K&=K_{0}+ K_{1}+K_{2}+\cdots, \\
\phi_1&=\phi_{10}+\phi_{11}+\phi_{12}+\cdots, \\
\phi_2&=\phi_{20}+\phi_{21}+\phi_{22}+\cdots,
\end{aligned}
\label{field1}
\eeq
where $K_{0}\sim \mathcal{O}(t^{0})$, $K_{1}\sim \mathcal{O}(t^{1})$, $K_{2}\sim \mathcal{O}(t^{2})$, and so on.
Similar scaling behavior is assumed for $\phi_{1i}$ and $\phi_{2i}$.
As the explicit solutions for $K_i$ and $\phi_{1i,2i}$ are lengthy, we present them in Appendix~\ref{SolofpHEFT}.
Substituting the solutions for the heavy fields into the original Lagrangian, we obtain the effective HEFT Lagrangian.
In the following, we directly present the results for the pHEFT and dHEFT and analyze their relations.

With the definition $V_\mu\equiv U^\dagger D_\mu U$, the Lagrangian of the primary HEFT up to $\mO(t^0)$ takes the form:
\beq
\hspace{-3em}
\begin{aligned}[b]
\mathcal{L}^{p}_\text{HEFT}(t^{-1})={}&\frac{v_H^2\bigl[(4 \xi^2 + 1)m_K^2(s_\gamma + \xi c_\gamma)^2 - \xi^2 m_{\phi^\pm}^2\bigr]}{8(4 \xi^2 + 1)} - \frac{h^3m_{\phi^\pm}^2 s_\gamma^2(2 \xi c_\gamma + s_\gamma)}{2\xi(4 \xi^2 + 1)v_H} \\
& - \frac{h^4}{8\xi^2(4 \xi^2 + 1)^2 v_H^2 m_K^2}\Bigl\{m_{\phi^\pm}^2 s_\gamma^2\Bigl[(4 \xi^2 + 1)m_K^2\Bigl(6(2 \xi^2 - 1)c_\gamma^4 + 7 c_\gamma^2 + 18 \xi c_\gamma^3 s_\gamma - 4 \xi c_\gamma s_\gamma - 1\Bigr) \\
&\qquad + m_{\phi^\pm}^2\Bigl(9(1 - 4 \xi^2)c_\gamma^4 + 3(8 \xi^2 - 3)c_\gamma^2 - 36 \xi c_\gamma^3 s_\gamma + 12 \xi c_\gamma s_\gamma - 4 \xi^2\Bigr)\Bigr]\Bigr\} + \mathcal{O}(h^5),
\end{aligned}
\label{pHEFTLm}
\eeq
\beq
\hspace{-3em}
\begin{aligned}[b]
\mathcal{L}^{p}_\text{HEFT}(t^0)={}&\frac{1}{2}\braket{V_\mu\sigma_3}\braket{V^\mu\sigma_3}\biggl\{\xi^2 v_H^2 - 2 h \xi v_H s_\gamma + h^2 s_\gamma\biggl[s_\gamma^3 - \xi c_\gamma^3 + \frac{c_\gamma m_{\phi^\pm}^2(3 c_\gamma s_\gamma + 4 \xi - 6 \xi s_\gamma^2)}{(4 \xi^2 + 1)m_K^2}\biggr] + \mathcal{O}(h^3)\biggr\} \\
& + \frac{1}{4}\braket{V_\mu V^\mu}\biggl\{- (4 \xi^2 + 1)v_H^2 - 2 h v_H(c_\gamma - 4 \xi s_\gamma) + \frac{h^2}{\xi}\biggl[c_\gamma s_\gamma(4 \xi^2 c_\gamma^2 + s_\gamma^2) \\
&\qquad + \xi(- 5 s_\gamma^4 + 2 s_\gamma^2 - 1) + \frac{m_{\phi^\pm}^2 s_\gamma\bigl(3(8 \xi^2 - 1)c_\gamma s_\gamma^2 - 16 \xi^2 c_\gamma + 2 \xi(9 s_\gamma^2 - 8)s_\gamma\bigr)}{(4 \xi^2 + 1)m_K^2}\biggr] + \mathcal{O}(h^3)\biggr\} \\
& + \frac{1}{2}D_\mu h D^\mu h + \frac{1}{8}m_h^2 v_H^2\bigl(c_\gamma - \xi s_\gamma)^2 - \frac{1}{2}h^2 m_h^2 + \frac{h^3 m_h^2(s_\gamma^3-\xi c_\gamma^3)}{2 \xi v_H} \\
& + \frac{h^4 m_h^2}{24 \xi^2(4 \xi^2 + 1)^2 v_H^2 m_K^4}\Bigl\{4 m_{\phi^\pm}^4 s_\gamma^2\bigl(3 c_\gamma s_\gamma + 4 \xi - 6 \xi s_\gamma^2\bigr)^2 \\
&\qquad + (4 \xi^2 + 1)^2 m_K^4\Bigl[s_\gamma^2\Bigl(38 \xi c_\gamma^3 s_\gamma + 25 \xi^2 + 19(\xi^2 - 1)s_\gamma^4 + (16 - 41 \xi^2)s_\gamma^2\Bigr) - 3 \xi^2\Bigr] \\
&\qquad - 20(4 \xi^2 + 1)m_K^2 m_{\phi^\pm}^2 s_\gamma^2\Bigl[\xi c_\gamma s_\gamma(7 - 9 s_\gamma^2) - c_\gamma^2\Bigl((6 \xi^2 - 3)s_\gamma^2 - 4 \xi^2\Bigr)\Bigr]\Bigr\} + \mathcal{O}(h^5) \\
& + \bar Q_L U \begin{pmatrix}y_u&0\\0&y_d\end{pmatrix} Q_R \times \frac{1}{\sqrt{2}}\biggl\{- v_H - h c_\gamma \\
&\qquad + \frac{h^2 s_\gamma^2\bigl[(4 \xi^2 + 1)c_\gamma m_K^2(\xi c_\gamma + s_\gamma) - m_{\phi^\pm}^2(3 c_\gamma s_\gamma + 4 \xi - 6 \xi s_\gamma^2)\bigr]}{2\xi(4 \xi^2 + 1)v_H m_K^2} + \mathcal{O}(h^3)\biggr\} + \text{h.c.},
\end{aligned}
\label{pHEFTL0}
\eeq
where $v_H \equiv \vev/\sqrt{1+4 \xi^2}$
is used to avoid overly lengthy expressions,
the covariant derivative $D_\mu$ acting on the Higgs boson $h$ reduces to the ordinary partial derivative, $\partial_\mu$, since it is a gauge singlet, and the kinetic term for the Higgs field $h$ has already been brought into its canonical form via a field redefinition. The primary HEFT starts at $\mathcal{O}(t^{-1})$, with terms proportional to $m_K^2$, $m_{\phi^\pm}^2$, or their combinations.
At the order $\mathcal{O}(t^{0})$, in addition to the terms present in the Standard Model,
\beq
\begin{aligned}[b]
\mathcal{L}^\text{SM}={}&\frac{1}{2}D_\mu hD^\mu h - \frac{1}{4}(\vev + h)^2\braket{V_\mu V^\mu} + \frac{m_h^2(\vev + h)^2(\vev^2 - 2 h \vev - h^2)}{8 \vev^2} \\
& - \frac{\vev+h}{\sqrt{2}}\bar Q_L U \begin{pmatrix}y_u&0\\0&y_d\end{pmatrix} Q_R + \text{h.c.}, \\
\end{aligned}
\label{LSM}
\eeq
the remaining terms proportional to $s_\gamma$, $c_\gamma$, and $\xi$ encode the non-decoupling effects of the heavy fields.

The decoupling HEFT Lagrangian to the first few orders in the power counting is given by
\beq
\begin{aligned}[b]
\mathcal{L}^\text{d}_\text{HEFT}(t^0)={}&\frac{1}{2}D_\mu hD^\mu h - \frac{1}{4}(\vev + h)^2\braket{V_\mu V^\mu} + \frac{m_h^2(\vev + h)^2(\vev^2 - 2 h \vev - h^2)}{8 \vev^2} \\
& - \frac{\vev+h}{\sqrt{2}}\bar Q_L U \begin{pmatrix}y_u&0\\0&y_d\end{pmatrix} Q_R + \text{h.c.}, \\
\end{aligned}
\label{dHEFTL0}
\eeq
\beq
\begin{aligned}[b]
\mathcal{L}^\text{d}_\text{HEFT}(t^1)={}&\frac{1}{8}\vev^2\bigl[m_K^2(\xi + s_\gamma)^2 - \xi^2 m_{\phi^\pm}^2\bigr] - \frac{h^3 m_{\phi^\pm}^2 s_\gamma^2(2 \xi + s_\gamma)}{2 \xi \vev} \\
& + \frac{h^4 m_{\phi^\pm}^2 s_\gamma^2\bigl[m_{\phi^\pm}^2(4 \xi + 3 s_\gamma)^2 - m_K^2(12 \xi^2 + 14 \xi s_\gamma + 5 s_\gamma^2)\bigr]}{8 \xi^2 \vev^2 m_K^2}+\mathcal{O}(h^5),
\end{aligned}
\label{dHEFTL1}
\eeq
\beq
\begin{aligned}[b]
\mathcal{L}^\text{d}_\text{HEFT}(t^2)={}&\frac{1}{2}\braket{V_\mu\sigma_3}\braket{V^\mu\sigma_3}\biggl\{\xi^2 \vev^2 - 2 h \xi \vev s_\gamma + \frac{h^2 s_\gamma\bigl[m_{\phi^\pm}^2(4 \xi + 3 s_\gamma) - \xi m_K^2\bigr]}{m_K^2} + \mathcal{O}(h^3)\biggr\} \\
& + \frac{1}{4}\braket{V_\mu V^\mu}\biggl\{h \vev(4 \xi^2 + 8 \xi s_\gamma + s_\gamma^2) \\
&\qquad + \frac{h^2 s_\gamma\bigl[m_K^2(4 \xi^2 + 2 \xi s_\gamma + s_\gamma^2) - m_{\phi^\pm}^2(4 \xi + s_\gamma)(4 \xi + 3 s_\gamma)\bigr]}{\xi m_K^2} + \mathcal{O}(h^3)\biggr\} \\
& - \frac{1}{8}\vev^2 m_h^2(4 \xi^2 + 2 \xi s_\gamma + s_\gamma^2) + \frac{h^3 m_h^2( - 4 \xi^3 + 3 \xi s_\gamma^2 + 2 s_\gamma^3)}{4 \xi \vev} \\
& + \frac{h^4 m_h^2}{24 \xi^2 \vev^2 m_K^4}\Bigl[m_K^4( - 12 \xi^4 + 25 \xi^2 s_\gamma^2 + 38 \xi s_\gamma^3 + 16 s_\gamma^4) \\
&\qquad - 20 m_K^2 m_{\phi^\pm}^2 s_\gamma^2(\xi + s_\gamma)(4 \xi + 3 s_\gamma) + 4 m_{\phi^\pm}^4 s_\gamma^2(4 \xi + 3 s_\gamma)^2\Bigr] + \mathcal{O}(h^5) \\
& + \bar Q_L U \begin{pmatrix}y_u&0\\0&y_d\end{pmatrix} Q_R \times \sqrt{2}\,\biggl\{\xi^2 \vev + \frac{h s_\gamma^2}{4} \\
&\qquad + \frac{h^2 s_\gamma^2\bigl[m_K^2(\xi + s_\gamma) - m_{\phi^\pm}^2(4 \xi + 3 s_\gamma)\bigr]}{4 \xi \vev m_K^2} + \mathcal{O}(h^3)\biggr\} + \text{h.c.},
\end{aligned}
\label{dHEFTL2}
\eeq
where $\mathcal{L}^\text{d}_\text{HEFT}(t^{0})$ coincides exactly with the SM Lagrangian given in Eq.~\eqref{LSM}.
That is,
in the limit $t \to 0$, the dHEFT reduces to the SM.
Physically, this corresponds to taking the masses of the heavy states to infinity, in which case the dHEFT reproduces the SM, thereby manifesting its decoupling behavior. Custodial-symmetry violation first enters at $\mathcal{O}(t^{2})$ through the term $\braket{V_\mu \sigma_3}\braket{V^\mu \sigma_3}$ in $\mathcal{L}^\text{d}_\text{HEFT}(t^{2})$.

\subsubsection{The mapping from pHEFT to dHEFT}
Starting from the pHEFT results in Eqs.~\eqref{pHEFTLm}--\eqref{pHEFTL0}, we can re-derive an EFT Lagrangian by promoting $s_\gamma$ and $\xi$ to $\mathcal{O}(t)$. Once the terms are rearranged following the revised power‑counting, the results match the dHEFT expressions in Eqs.~\eqref{dHEFTL0}--\eqref{dHEFTL2} exactly. Specifically, since $s_\gamma$ and $\xi$ are $\mathcal{O}(t)$ in the dHEFT power-counting, the factors $c_\gamma = \sqrt{1 - s_\gamma^2}$ and $(1 + 4\xi^2)^{-1}$ must be Taylor-expanded as:
\begin{align}
c_\gamma &= 1 - \frac{s_\gamma^2}{2} - \frac{s_\gamma^4}{8} + \cdots,
\label{cgamma} \\
\frac{1}{1+4 \xi^2} &= 1 - 4 \xi^2 + 16 \xi^4 + \cdots,
\label{xi}
\end{align}
which only yield even powers of $s_\gamma$ and $\xi$. Meanwhile, in the terms in $\mathcal{L}^\text{p}_\text{HEFT}(t^{-1})$ and the shifted Lagrangian $\Delta \mathcal{L}^\text{p}_\text{HEFT}(t^{0}) \equiv \mathcal{L}^\text{p}_\text{HEFT}(t^{0}) - \mathcal{L}^\text{SM}$, the total power of $s_\gamma$ and $\xi$ is restricted to even integers starting from two.

Consequently, we obtain the following mapping:
\begin{align}
\mathcal{L}^\text{p}_\text{HEFT}(t^{-1}) &\to \mathcal{L}^\text{d}_\text{HEFT}(t^{1}) + \mathcal S_1 \mathcal{L}^\text{d}_\text{HEFT}(t^{3}) + \cdots,
\label{pHEFTtodHEFTm} \\
\Delta \mathcal{L}^\text{p}_\text{HEFT}(t^{0}) &\to \mathcal{L}^\text{d}_\text{HEFT}(t^{2}) + \mathcal S_1 \mathcal{L}^\text{d}_\text{HEFT}(t^{4}) + \cdots,
\label{pHEFTtodHEFT1}
\end{align}
where $\mathcal S_i \mathcal{L}^\text{d}_\text{HEFT}(t^{n})$ denotes a distinct term of the dHEFT Lagrangian at order $\mathcal{O}(t^n)$. Here, the index $i$ serves merely as a label to distinguish different disjoint subsets and does not imply any particular sequence. Notably, the power $t^n$ within each HEFT is governed by its own power-counting rule. %This mapping highlights that pHEFT effectively resums contributions that appear at multiple different orders in the dHEFT expansion.

A concrete example is provided by the $\braket{V_\mu \sigma_3} \braket{V^\mu \sigma_3} h^2$ operator. In pHEFT, its coefficient at $\mathcal{O}(t^{0})$ is:
\beq
\frac{s_\gamma}{2}\biggl[s_\gamma^3 - \xi c_\gamma^3 + \frac{c_\gamma m_{\phi^\pm}^2(3 c_\gamma s_\gamma + 4 \xi - 6 \xi s_\gamma^2)}{(4 \xi^2 + 1)m_K^2}\biggr].
\label{pHEFTc}
\eeq
By substituting $c_\gamma = \sqrt{1-s_\gamma^2}$ and applying the dHEFT scaling from Eq.~\eqref{scale2}, the factors $c_\gamma$ and $(1+4\xi^2)^{-1}$ are expanded according to Eqs.~\eqref{cgamma} and \eqref{xi}. The leading-order result becomes
\beq
\frac{s_\gamma}{2 m_K^2}\bigl[m_{\phi^\pm}^2(4 \xi + 3 s_\gamma) - \xi m_K^2\bigr],
\label{dHEFTc}
\eeq
which is exactly the $\mathcal{O}(t^2)$ term in dHEFT. Any remaining terms from the expansion of Eq.~\eqref{pHEFTc} consist of higher powers of $s_\gamma$ and $\xi$, contributing to the dHEFT Lagrangian starting from $\mathcal{O}(t^4)$.

The mapping at the next order follows a slightly different structure:
\beq
\mathcal{L}^\text{p}_\text{HEFT}(t^{1}) \to \mathcal S_2 \mathcal{L}^\text{d}_\text{HEFT}(t^{3}) + \mathcal S_2 \mathcal{L}^\text{d}_\text{HEFT}(t^{5}) + \cdots,
\label{pHEFTtodHEFT2}
\eeq
where $\mathcal S_2 \mathcal{L}^\text{d}_\text{HEFT}(t^{3})$ denotes the second disjoint subset of the dHEFT Lagrangian at $\mathcal{O}(t^{3})$. This partitioning accounts for the fact $\mathcal{L}^\text{p}_\text{HEFT}(t^{-1})$ has already populated the first subset, $\mathcal S_1 \mathcal{L}^\text{d}_\text{HEFT}(t^{3})$, as shown in Eq.~\eqref{pHEFTtodHEFTm}. These subsets are mutually exclusive; notably, the derivative operators at $\mathcal{O}(t^{3})$ in dHEFT belong exclusively to $\mathcal S_2$, since $\mathcal{L}^\text{p}_\text{HEFT}(t^{-1})$ consists solely of $h$ polynomials and does not generate derivative structures in its mapping.

We list the $p^2$ and $p^4$ operators of $\mathcal{L}^\text{p}_\text{HEFT}(t^{1})$ separately in Tables~\ref{tab:p2} and \ref{tab:p4}, providing the corresponding terms in $\mathcal{L}^\text{d}_\text{HEFT}(t^{3})$ for comparison. In the Wilson coefficients, we retain only the lowest-order $h$-polynomials with higher-order terms provided in a supplementary document. Given that $\mathcal{S}_2 \mathcal{L}^\text{d}_\text{HEFT}(t^{3})$ captures only the leading-order contribution of the mapping, the transformation between the second and third columns of these tables is simplified via the substitutions $c_\gamma \to 1$ and $(1 + 4\xi^2) \to 1$ as established in Eqs.~\eqref{cgamma} and \eqref{xi}. Notice that $v_H \to \vev$ as $(1 + 4\xi^2) \to 1$.

\begin{table}[htbp]
  \centering
  \begin{tabular}{c|c|c}
    \hline
    Operators &WCs ($\mathcal{L}^\text{p}_\text{HEFT}(t^{1})$) &WCs ($\mathcal{L}^\text{d}_\text{HEFT}(t^{3})$) \\
    \hline
    $\braket{V_\mu V^\mu}$
  & $\begin{aligned}
&\tfrac{h^2 m_h^2 s_\gamma(4 \xi c_\gamma + s_\gamma)}{2\xi(4 \xi^2 + 1)m_K^4}\bigl[2(4 \xi^2 + 1)c_\gamma m_K^2(\xi c_\gamma + s_\gamma) \\
&\qquad - m_{\phi^\pm}^2(3 c_\gamma s_\gamma + 4 \xi - 6 \xi s_\gamma^2)\bigr]
    \end{aligned}$
  & $\begin{aligned}
&\tfrac{h^2 m_h^2 s_\gamma(4 \xi + s_\gamma)}{2 \xi m_K^4}\bigl[2 m_K^2(\xi + s_\gamma) \\
&\qquad - m_{\phi^\pm}^2(4 \xi + 3 s_\gamma)\bigr]
    \end{aligned}$ \\
    \hline
    $\braket{V_\mu \s_3} \braket{V^\mu \s_3}$
  & $\begin{aligned}
&\tfrac{h^2 c_\gamma m_h^2 s_\gamma}{(4 \xi^2 + 1)m_K^4}\bigl[m_{\phi^\pm}^2(3 c_\gamma s_\gamma + 4 \xi - 6 \xi s_\gamma^2) \\
&\qquad - 2(4 \xi^2 + 1)c_\gamma m_K^2(\xi c_\gamma + s_\gamma)\bigr]
    \end{aligned}$
  & $\begin{aligned}
&\tfrac{h^2 m_h^2 s_\gamma}{m_K^4}\bigl[m_{\phi^\pm}^2(4 \xi + 3 s_\gamma) \\
&\qquad - 2 m_K^2(\xi + s_\gamma)\bigr]
    \end{aligned}$ \\
    \hline
  \end{tabular}
  \caption{$p^2$ operators in $\mathcal{L}^\text{p}_\text{HEFT}(t^{1})$ and their corresponding Wilson coefficients (WCs). The first column lists the operators, while the second and third columns show the WCs in pHEFT and their leading-order mapping to dHEFT at $\mathcal{O}(t^{3})$, respectively. The transformation between these columns is simplified by $c_\gamma \to 1$ and $(1 + 4\xi^2) \to 1$. For brevity, higher-order $h$-polynomials (e.g., $\mathcal{O}(h^3)$) are omitted and provided in the supplementary files.}
\label{tab:p2}
\end{table}

\begin{table}[htbp]
  \begin{tabular}{c|c|c}
    \hline
 Operators &WCs ($\mathcal{L}^\text{p}_\text{HEFT}(t^{1})$) &WCs ($\mathcal{L}^\text{d}_\text{HEFT}(t^{3})$) \\
    \hline
    $\braket{V_\mu V^\mu} \braket{V_\nu V^\nu}$
  & $v_H^2(4 \xi c_\gamma + s_\gamma)^2 / 8 m_K^2$
  & $\vev^2(4 \xi + s_\gamma)^2 / 8 m_K^2$ \\
    \hline
    $\braket{V_\mu \s_3} \braket{V^\mu \s_3} \braket{V_\nu V^\nu}$
  & $- \xi c_\gamma v_H^2(4 \xi c_\gamma + s_\gamma) / 2 m_K^2$
  & $- \xi \vev^2(4 \xi + s_\gamma) / 2 m_K^2$ \\
    \hline
    $\braket{V_\mu \s_3} \braket{V_\nu \s_3} \braket{V^\mu V^\nu}$
  & $\xi^2 v_H^2 / (4 \xi^2 + 1)m_{\phi^\pm}^2$
  & $\xi^2 \vev^2 / m_{\phi^\pm}^2$ \\
    \hline
    $\braket{V_\mu \s_3} \braket{V^\mu \s_3} \braket{V_\nu \s_3} \braket{V^\nu \s_3}$
  & $\frac{1}{2}\xi^2 v_H^2\Bigl[\frac{c_\gamma^2}{m_K^2} - \frac{1}{(4 \xi^2 + 1)m_{\phi^\pm}^2}\Bigr]$
  & $\tfrac{\xi^2 \vev^2}{2 m_K^2 m_{\phi^\pm}^2}(m_{\phi^\pm}^2 - m_K^2)$ \\
    \hline
    $\braket{V_\mu V_\nu \s_3} \braket{V^\mu \s_3} D^\nu h$
  & $4 \xi v_H(\xi c_\gamma + s_\gamma) / (4 \xi^2 + 1)\,m_{\phi^\pm}^2$
  & $4 \xi \vev(\xi + s_\gamma) / m_{\phi^\pm}^2$ \\
    \hline
    $\braket{V_\mu V_\nu} D^\mu h D^\nu h$
  & $- 4(\xi c_\gamma + s_\gamma)^2 / (4 \xi^2 + 1)\,m_{\phi^\pm}^2$
  & $- 4(\xi + s_\gamma)^2 / m_{\phi^\pm}^2$ \\
    \hline
    $\braket{V_\mu \s_3} \braket{V_\nu \s_3} D^\mu h D^\nu h$
  & $2(\xi c_\gamma + s_\gamma)^2 / (4 \xi^2 + 1)\,m_{\phi^\pm}^2$
  & $2(\xi + s_\gamma)^2 / m_{\phi^\pm}^2$ \\
    \hline
    $\braket{V_\mu V^\mu} D_\nu h D^\nu h$
  & $\begin{aligned}
&\tfrac{s_\gamma(4 \xi c_\gamma + s_\gamma)}{2\xi(4 \xi^2 + 1)m_K^4}\bigl[m_{\phi^\pm}^2(3 c_\gamma s_\gamma + 4 \xi - 6 \xi s_\gamma^2) \\
&\qquad - (4 \xi^2 + 1)c_\gamma m_K^2(\xi c_\gamma + s_\gamma)\bigr]
    \end{aligned}$
  & $\begin{aligned}
&\tfrac{s_\gamma(4 \xi + s_\gamma)}{2\xi m_K^4}\bigl[m_{\phi^\pm}^2(4 \xi + 3 s_\gamma) \\
&\qquad - m_K^2(\xi + s_\gamma)\bigr]
    \end{aligned}$ \\
    \hline
    $\braket{V_\mu \s_3} \braket{V^\mu \s_3} D_\nu h D^\nu h$
  & $\begin{aligned}
&\tfrac{c_\gamma s_\gamma}{(4 \xi^2 + 1)m_K^4}\bigl[(4 \xi^2 + 1)c_\gamma m_K^2(\xi c_\gamma + s_\gamma) \\
&\qquad - m_{\phi^\pm}^2(3 c_\gamma s_\gamma + 4 \xi - 6 \xi s_\gamma^2)\bigr]
    \end{aligned}$
  & $\begin{aligned}
&\tfrac{s_\gamma}{m_K^4}\bigl[m_K^2(\xi + s_\gamma) \\
&\qquad - m_{\phi^\pm}^2(4 \xi + 3 s_\gamma)\bigr]
    \end{aligned}$ \\
    \hline
    $D_\mu h D^\mu h D_\nu h D^\nu h$
  & $\begin{aligned}
&\tfrac{s_\gamma^2}{2 \xi^2(4 \xi^2 + 1)^2 v_H^2 m_K^6}\bigl[(4 \xi^2 + 1)c_\gamma m_K^2(\xi c_\gamma + s_\gamma) \\
&\qquad - m_{\phi^\pm}^2(3 c_\gamma s_\gamma + 4 \xi - 6 \xi s_\gamma^2)\bigr]^2
    \end{aligned}$
  & $\begin{aligned}
&\tfrac{s_\gamma^2}{2 \xi^2 \vev^2 m_K^6}\bigl[m_K^2(\xi + s_\gamma) \\
&\qquad - m_{\phi^\pm}^2(4 \xi + 3 s_\gamma)\bigr]^2
    \end{aligned}$ \\
    \hline
  \end{tabular}
  \caption{$p^4$ operators in $\mathcal{L}^\text{p}_\text{HEFT}(t^{1})$ and their corresponding Wilson coefficients. Following the same structure as Table~\ref{tab:p2}, the second column presents the pHEFT results, which reduce to the $\mathcal{O}(t^3)$ dHEFT terms in the third column under the approximations $c_\gamma \to 1$ and $(1 + 4\xi^2) \to 1$. Higher-order $h$-polynomials are relegated to the supplementary files.}
\label{tab:p4}
\end{table}

%To sum up, the Lagrangian in the decoupling HEFT could be got from the primary HEFT, just by requiring the parameters scaling as the same in the decoupling HEFT and reorganize the terms. We have checked them up to $t^3$ order in the decoupling HEFT, however, they should be consistent at all orders.

%The HEFTs that could be transmuted from the primary HEFT have to fulfil the minimal assumption of the mass separation and use the field expansion as in Eq.~\eqref{field1}, which are generally acceptable.
%Thus the matching procedure could be split into two steps, in the first step we ``integrate out'' the heavy states to get the primary HEFT, in the second step we reorganize the terms according to the scaling requirements.

%There are ten $p^4$ terms appear in $\mathcal{L}_\text{HEFT}(t^1)$, which are

%We leave their coefficients in the appendix~\ref{app:p4}.
%Together in $\mathcal{L}_\text{HEFT}(t^1)$ there are operators involving fermions, such as fermion current operators and four fermion operators, which are listed in Table~\ref{table:Q}, their detailed coefficients are left in the appendix~\ref{app:Q}.

The mapping established in Eqs.~\eqref{pHEFTtodHEFT1} and \eqref{pHEFTtodHEFT2} also applies to the quark sector. In $\mathcal{L}^\text{p}_\text{HEFT}(t^{0})$, the interaction involving quark currents is given by:
\beq
\mathcal{L}^\text{p}_\text{HEFT}(t^{0}) \supset -\frac{\vev(1+4\xi^2)^{-1/2} + h \, c_\gamma}{\sqrt{2}}\, \bar{Q}_L U \begin{pmatrix} y_u & 0 \\ 0 & y_d \end{pmatrix} Q_R + \text{h.c.},
\eeq
where higher-order polynomials in $h$ are omitted. Following the dHEFT power counting, this expression is mapped onto:
\begin{align}
\mathcal{L}^\text{d}_\text{HEFT}(t^{0}) &\supset -\frac{\vev + h}{\sqrt{2}}\, \bar{Q}_L U \begin{pmatrix} y_u & 0 \\ 0 & y_d \end{pmatrix} Q_R + \text{h.c.},
\label{dHEFTquark0} \\
\mathcal{L}^\text{d}_\text{HEFT}(t^{2}) &\supset \biggl(\sqrt{2}\, \xi^2 \vev + \frac{h s_\gamma^2}{2\sqrt{2}}\biggr)\, \bar{Q}_L U \begin{pmatrix} y_u & 0 \\ 0 & y_d \end{pmatrix} Q_R + \text{h.c.}.
\label{dHEFTquark2}
\end{align}
Specifically, Eq.~\eqref{dHEFTquark0} represents the Standard Model Yukawa sector, while the $\mathcal{O}(t^2)$ term in Eq.~\eqref{dHEFTquark2} provides the leading new physics correction in the decoupling limit.

%We list the operators of $\mathcal{L}^\text{p}_\text{HEFT}(t^{1})$ involving fermion currents and their Wilson coefficients in Table~\ref{tab:quark}. The fermionic operator structures in $\mathcal{L}^\text{d}_\text{HEFT}(t^{3})$ are identical, and their corresponding Wilson coefficients can be obtained by setting $c_\gamma \to 1$ and $(1 + 4\xi^2) \to 1$. This simple mapping arises because $\mathcal{L}^\text{p}_\text{HEFT}(t^{-1})$ contains no fermion currents, ensuring that no additional contributions from the lower-order pHEFT Lagrangian overlap with this sector.

We list the operators of $\mathcal{L}^\text{p}_\text{HEFT}(t^{1})$ involving fermion currents and their Wilson coefficients in Table~\ref{tab:quark}. Notably, the fermionic operator structures in $\mathcal{L}^\text{d}_\text{HEFT}(t^{3})$ are identical to those in the pHEFT framework. Since $\mathcal{L}^\text{p}_\text{HEFT}(t^{-1})$ contains no fermion currents, there are no contribution from lower-order pHEFT terms. Consequently, the Wilson coefficients for the dHEFT at $\mathcal{O}(t^{3})$ can be directly obtained from the second column of Table~\ref{tab:quark} by applying the leading-order substitutions $c_\gamma \to 1$ and $(1 + 4\xi^2) \to 1$.

\begin{table}[htbp]
  \centering
  \begin{tabular}{l l}
    \hline
    \hline
    Operator & WCs ($\mathcal{L}^\text{p}_\text{HEFT}(t^{1})$) \\
    \hline
    \hline
    $(\bar Q_L U Q_R) \braket{V_\mu V^\mu} + \text{h.c.}$
  & $\frac12(y_u+y_d)\frac{v_H s_\gamma(4 \xi c_\gamma + s_\gamma)}{2 \sqrt{2}m_K^2}$ \\
    $(\bar Q_L U \s_3 Q_R) \braket{V_\mu V^\mu} + \text{h.c.}$
  & $\frac12(y_u-y_d)\frac{v_H s_\gamma(4 \xi c_\gamma + s_\gamma)}{2 \sqrt{2}m_K^2}$ \\
    \hline
    $(\bar Q_L U Q_R) \braket{V_\mu \s_3} \braket{V^\mu \s_3} + \text{h.c.}$
  & $\frac12(y_u+y_d)\bigl(- \frac{\xi c_\gamma v_H s_\gamma}{\sqrt{2}m_K^2}\bigr)$ \\
    $(\bar Q_L U \s_3 Q_R) \braket{V_\mu \s_3} \braket{V^\mu \s_3} + \text{h.c.}$
  & $\frac12(y_u-y_d)\bigl(- \frac{\xi c_\gamma v_H s_\gamma}{\sqrt{2}m_K^2}\bigr)-\frac12(y_u-y_d)\frac{\sqrt{2}\xi^2 v_H}{(4 \xi^2 + 1)m_{\phi^\pm}^2}$ \\
    $(\bar Q_L U [V_\mu,\s_3] Q_R)\braket{V^\mu \s_3} + \text{h.c.}$
  & $\frac12(y_u+y_d)\frac{\sqrt{2}\xi^2 v_H}{(4 \xi^2 + 1)m_{\phi^\pm}^2}$ \\
    $(\bar Q_L U V_\mu Q_R)\braket{V^\mu \s_3} + \text{h.c.}$
  & $(y_u-y_d)\frac{\sqrt{2}\xi^2 v_H}{(4 \xi^2 + 1)m_{\phi^\pm}^2}$ \\
    \hline
    $(\bar Q_L U Q_R)\, D_\mu h D^\mu h + \text{h.c.}$
  & $\frac12(y_u+y_d)\frac{s_\gamma^2\bigl\{m_{\phi^\pm}^2\bigl[3 s_\gamma c_\gamma + 2\xi(2 - 3 s_\gamma^2)\bigr] - (4 \xi^2 + 1)c_\gamma m_K^2(\xi c_\gamma + s_\gamma)\bigr\}}{\sqrt{2}\xi(4 \xi^2 + 1)v_H m_K^4}$ \\
    $(\bar Q_L U \s_3 Q_R)\, D_\mu h D^\mu h + \text{h.c.}$
  & $\frac12(y_u-y_d)\frac{s_\gamma^2\bigl\{m_{\phi^\pm}^2\bigl[3 s_\gamma c_\gamma + 2\xi(2 - 3 s_\gamma^2)\bigr] - (4 \xi^2 + 1)c_\gamma m_K^2(\xi c_\gamma + s_\gamma)\bigr\}}{\sqrt{2}\xi(4 \xi^2 + 1)v_H m_K^4}$ \\
    \hline
    $(\bar Q_L U [V_\mu,\s_3] Q_R)\, D^\mu h + \text{h.c.}$
  & $\frac12(y_u-y_d)\frac{2 \sqrt{2}\xi(\xi c_\gamma + s_\gamma)}{(4 \xi^2 + 1)m_{\phi^\pm}^2}$ \\
    $(\bar Q_L U V_\mu Q_R)\, D^\mu h + \text{h.c.}$
  & $(y_u+y_d)\frac{2 \sqrt{2}\xi(\xi c_\gamma + s_\gamma)}{(4 \xi^2 + 1)m_{\phi^\pm}^2}$ \\
    $(\bar Q_L U \s_3 Q_R)\braket{V_\mu \s_3} D^\mu h + \text{h.c.}$
  & $-\frac12(y_u+y_d)\frac{2 \sqrt{2}\xi(\xi c_\gamma + s_\gamma)}{(4 \xi^2 + 1)m_{\phi^\pm}^2}$ \\
    \hline
    $(\bar Q_L U Q_R) + \text{h.c.}$
  & $\frac12(y_u+y_d)\frac{h^2 m_h^2 s_\gamma^2\bigl\{2(4 \xi^2 + 1)c_\gamma m_K^2(\xi c_\gamma + s_\gamma) - m_{\phi^\pm}^2\bigl[3 s_\gamma c_\gamma + 2\xi(2 - 3 s_\gamma^2)\bigr]\bigr\}}{\sqrt{2}\xi(4 \xi^2 + 1)v_H m_K^4}$ \\
    $(\bar Q_L U \s_3 Q_R) + \text{h.c.}$
  & $\frac12(y_u-y_d)\frac{h^2 m_h^2 s_\gamma^2\bigl\{2(4 \xi^2 + 1)c_\gamma m_K^2(\xi c_\gamma + s_\gamma) - m_{\phi^\pm}^2\bigl[3 s_\gamma c_\gamma + 2\xi(2 - 3 s_\gamma^2)\bigr]\bigr\}}{\sqrt{2}\xi(4 \xi^2 + 1)v_H m_K^4}$ \\
    \hline
    $(\bar Q_L U Q_R)(\bar Q_L U Q_R) + \text{h.c.}$
  & $\frac14(y_u+y_d)^2\frac{s_\gamma^2}{4 m_K^2}$ \\
    $(\bar Q_L U Q_R)(\bar Q_L U\sigma_3 Q_R) + \text{h.c.}$
  & $\frac12(y_u^2-y_d^2)\frac{s_\gamma^2}{4 m_K^2}$ \\
    $(\bar Q_L U\sigma_3 Q_R)(\bar Q_L U\sigma_3 Q_R) + \text{h.c.}$
  & $\frac14(y_u-y_d)^2\frac{s_\gamma^2}{4 m_K^2}+\frac14y_uy_d\frac{4 \xi^2}{(4 \xi^2 + 1)m_{\phi^\pm}^2}$ \\
    $(\bar Q_L \sigma^I U Q_R)(\bar Q_L \sigma^I U Q_R) + \text{h.c.}$
  & $-\frac14y_uy_d\frac{4 \xi^2}{(4 \xi^2 + 1)m_{\phi^\pm}^2}$ \\
    \hline
    $(\bar Q_L\gamma_\mu Q_L)(\bar Q_R\gamma^\mu Q_R)$
  & $\frac14(y_u^2+y_d^2)\frac{s_\gamma^2}{4 m_K^2}+\frac18(y_u^2+y_d^2)\frac{4 \xi^2}{(4 \xi^2 + 1)m_{\phi^\pm}^2}$ \\
    $(\bar Q_L\gamma_\mu\sigma^I Q_L)(\bar Q_R\gamma^\mu U^\dagger\sigma^I U Q_R)$
  & $\frac12y_u y_d \frac{s_\gamma^2}{4 m_K^2}$ \\
    $(\bar Q_L\gamma_\mu U\sigma_3U^\dagger Q_L)(\bar Q_R\gamma^\mu Q_R)$
  & $\frac14(y_u^2-y_d^2)\frac{s_\gamma^2}{4 m_K^2}-\frac18(y_u^2-y_d^2)\frac{4 \xi^2}{(4 \xi^2 + 1)m_{\phi^\pm}^2}$ \\
    $(\bar Q_L\gamma_\mu Q_L)(\bar Q_R\gamma^\mu\sigma_3 Q_R)$
  & $\frac14(y_u^2-y_d^2)\frac{s_\gamma^2}{4 m_K^2} + \frac18(y_u^2-y_d^2)\frac{4 \xi^2}{(4 \xi^2 + 1)m_{\phi^\pm}^2}$ \\
    $(\bar Q_L\gamma_\mu U\sigma_3U^\dagger Q_L)(\bar Q_R\gamma^\mu\sigma_3 Q_R)$
  & $\frac14(y_u-y_d)^2\frac{s_\gamma^2}{4 m_K^2}-\frac18(y_u^2+y_d^2)\frac{4 \xi^2}{(4 \xi^2 + 1)m_{\phi^\pm}^2}$ \\
    \hline
  \end{tabular}
  \caption{Fermionic operators in $\mathcal{L}^\text{p}_\text{HEFT}(t^{1})$ and their corresponding Wilson coefficients (second column). The dHEFT results at $\mathcal{O}(t^3)$ share the same operator structures, and their coefficients can be retrieved by setting $c_\gamma \to 1$ and $(1 + 4\xi^2) \to 1$. As in previous tables, higher-order $h$-polynomials are omitted and provided in the supplementary files.}
\label{tab:quark}
\end{table}

\subsubsection{Discussion on Framework Correspondence}

The pHEFT and dHEFT Lagrangians are derived via a common functional matching approach: the heavy states $\phi^\pm$ and $K$ are integrated out at tree level by solving their EoMs perturbatively and substituting back, which yields a series of local HEFT operators. These two HEFTs share an identical parameter set:
\beq
\mathcal{P} = \{m^2_{\phi^\pm}, m^2_K, m_h, \vev, s_\gamma, \xi \},
\label{pHEFTset}
\eeq
while following different power-counting schemes:
\begin{align*}
&\text{pHEFT:} \quad m^2_{\phi^\pm}\sim m^2_K\sim\mathcal{O}(t^{-1}), \quad m_h\sim \vev\sim s_\gamma\sim \xi\sim\mathcal{O}(t^0), \\
&\text{dHEFT:} \quad m^2_{\phi^\pm}\sim m^2_K\sim\mathcal{O}(t^{-1}), \quad m_h\sim \vev\sim \mathcal{O}(t^0),\quad s_\gamma\sim \xi\sim\mathcal{O}(t^1).
\end{align*}

A comparison of the matching results reveals that the dHEFT Lagrangian can be exactly reproduced from the pHEFT results by simply applying the dHEFT power-counting scheme to the pHEFT coefficients. Specifically, by performing the Taylor expansions of $c_\gamma$ and $(1+4\xi^2)^{-1}$ as dictated by the $s_\gamma \sim \xi \sim \mathcal{O}(t^1)$ scaling, the pHEFT expressions map exactly onto their dHEFT counterparts. Therefore, performing the pHEFT matching is sufficient: the dHEFT limit follows directly from parameter re-scaling, making a separate functional matching procedure for the decoupling case redundant.

This means that part of the matching procedure—namely, “integrating out” the heavy states and expanding in the inverse of their masses—can be separated from the full matching procedure. The pHEFT completes this core step while retaining the maximal information from the UV model. Further scalings of $\sin \gamma$ and $\xi$ then act as additional restrictions. In other words, a HEFT with an arbitrary scaling of the parameter set $\mathcal{P} = \{m^2_{\phi^\pm}, m^2_K, m_h, \vev, s_\gamma, \xi \}$ can be obtained by applying such a scaling directly to the pHEFT, without the need to solve the EoM or perform operator reduction.
By circumventing case-by-case procedures for solving the EoMs and performing operator reductions, this approach provides a unified and efficient framework for deriving the effective limits of the underlying UV theory across different scaling schemes.

\begin{figure}[h!]
    \centering
    \includegraphics[width=0.85\textwidth]{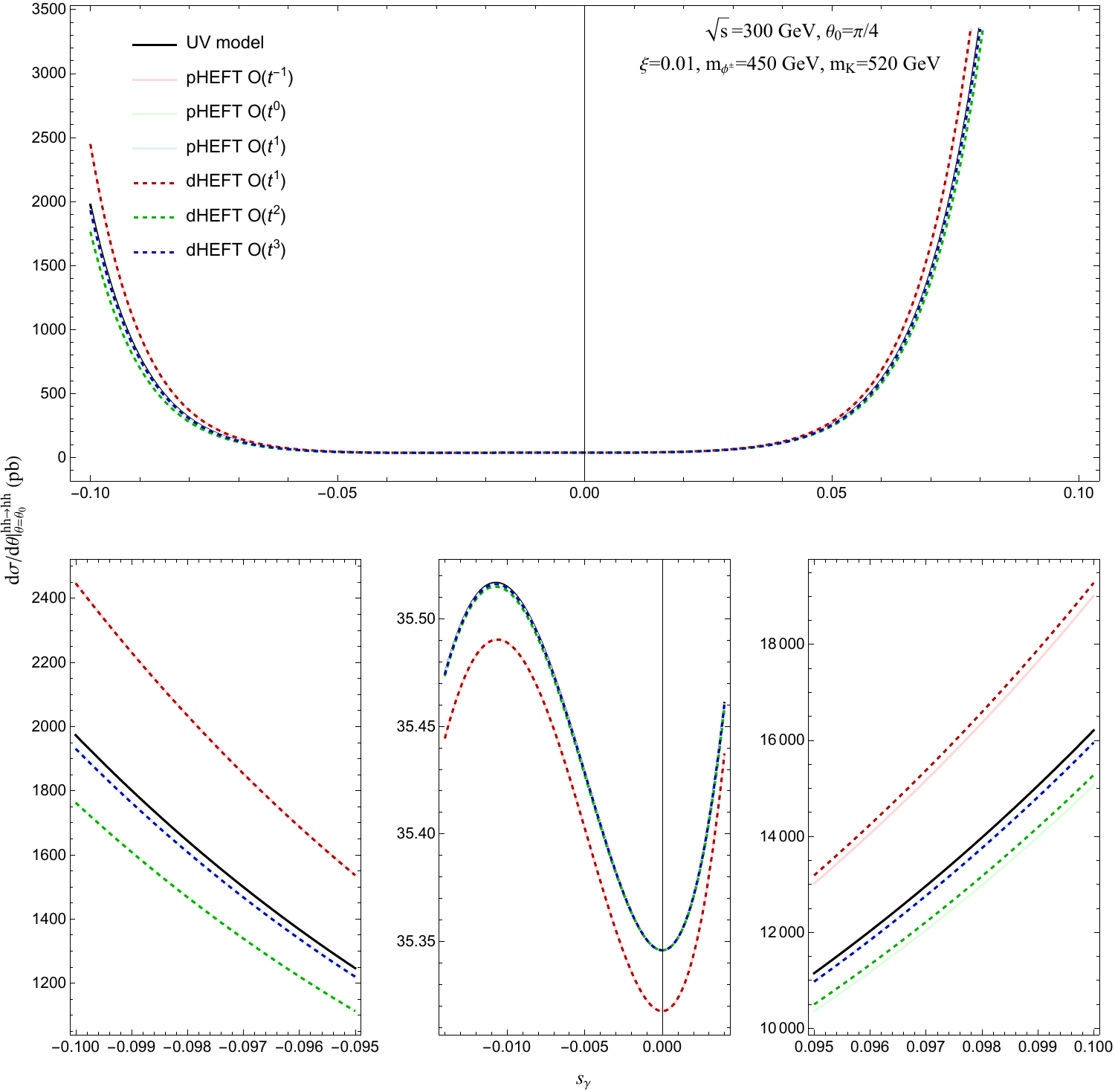}
    \caption[]{Comparison between the UV model and the pHEFT and dHEFT in the differential cross-section of $hh\rightarrow hh$, for a center-of-mass energy $\sqrt{s}=300\,\mathrm{GeV}$ and a scattering angle $\theta_0=\pi/4$. We set $\xi=0.01$, $m_{\phi^\pm}=450\,\mathrm{GeV}$, and $m_K=520\,\mathrm{GeV}$. The Higgs mass and electroweak vacuum expectation value are fixed to their experimentally measured values: $m_h=125\,\mathrm{GeV}$ and $\vev=246\,\mathrm{GeV}$. The top panel presents the cross section over $s_\gamma\in[-0.1, 0.1]$, while the three lower panels provide detailed views of three selected subintervals.}
    \label{fig:xs_hh_pHEFT_mK520}
\end{figure}

\begin{figure}[h!]
    \centering
    \includegraphics[width=0.7\textwidth]{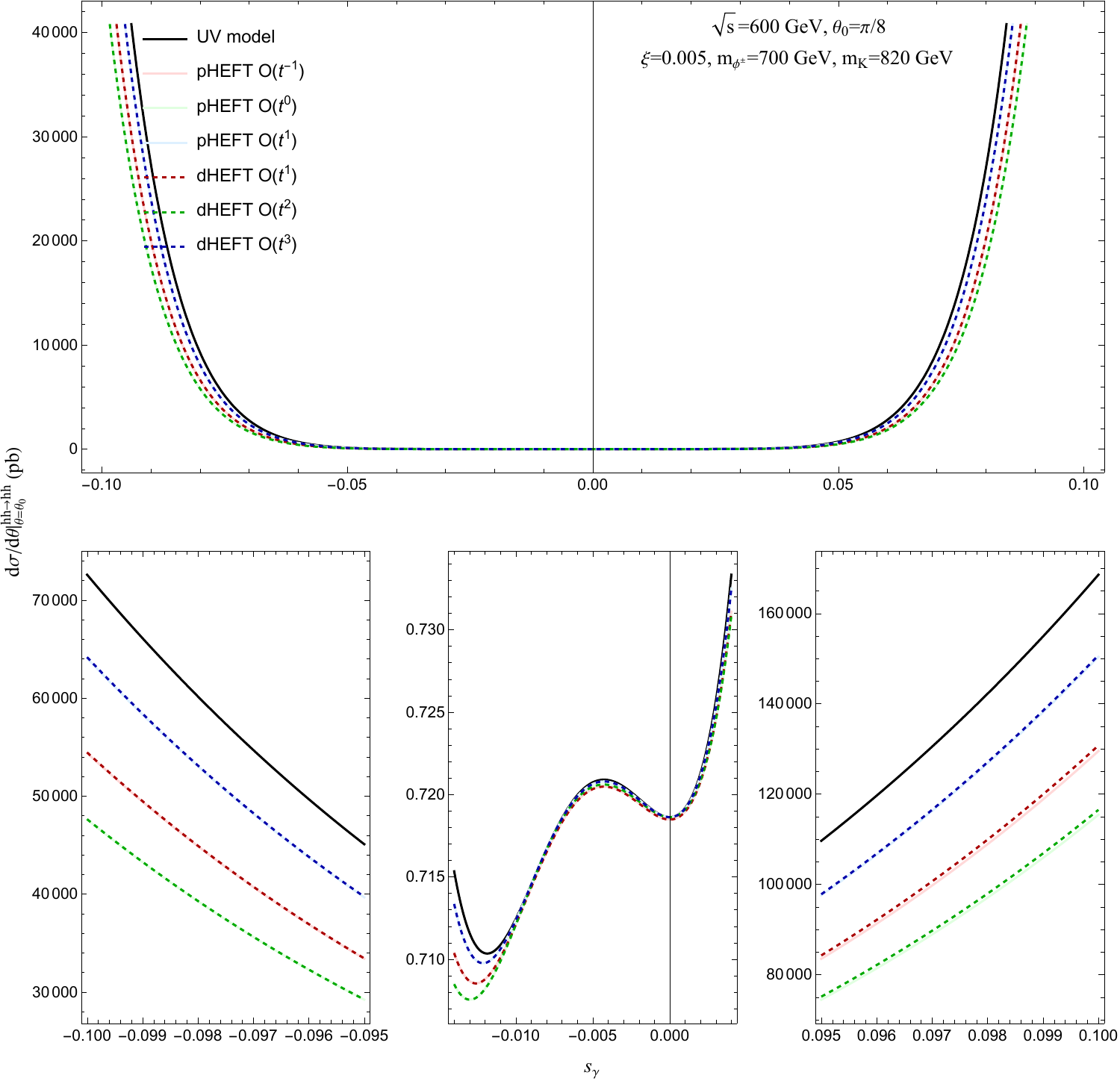}
    \caption[]{Similar to Fig.~\ref{fig:xs_hh_pHEFT_mK520}, now shown for $\sqrt{s}=600\,\mathrm{GeV}$, $\theta_0=\pi/8$, $\xi=0.005$, $m_{\phi^\pm}=700\,\mathrm{GeV}$, and $m_K=820\,\mathrm{GeV}$. When $s_\gamma>0.084$, the model becomes unstable because the boundedness-from-below conditions ($Z_1, Z_2\geq 0$ and $|Z_3|\leq 2\sqrt{Z_1 Z_2}$~\cite{Song:2025kjp}) are violated.}
    \label{fig:xs_hh_pHEFT_mK820}
\end{figure}

\begin{figure}[h!]
    \centering
    \includegraphics[width=0.7\textwidth]{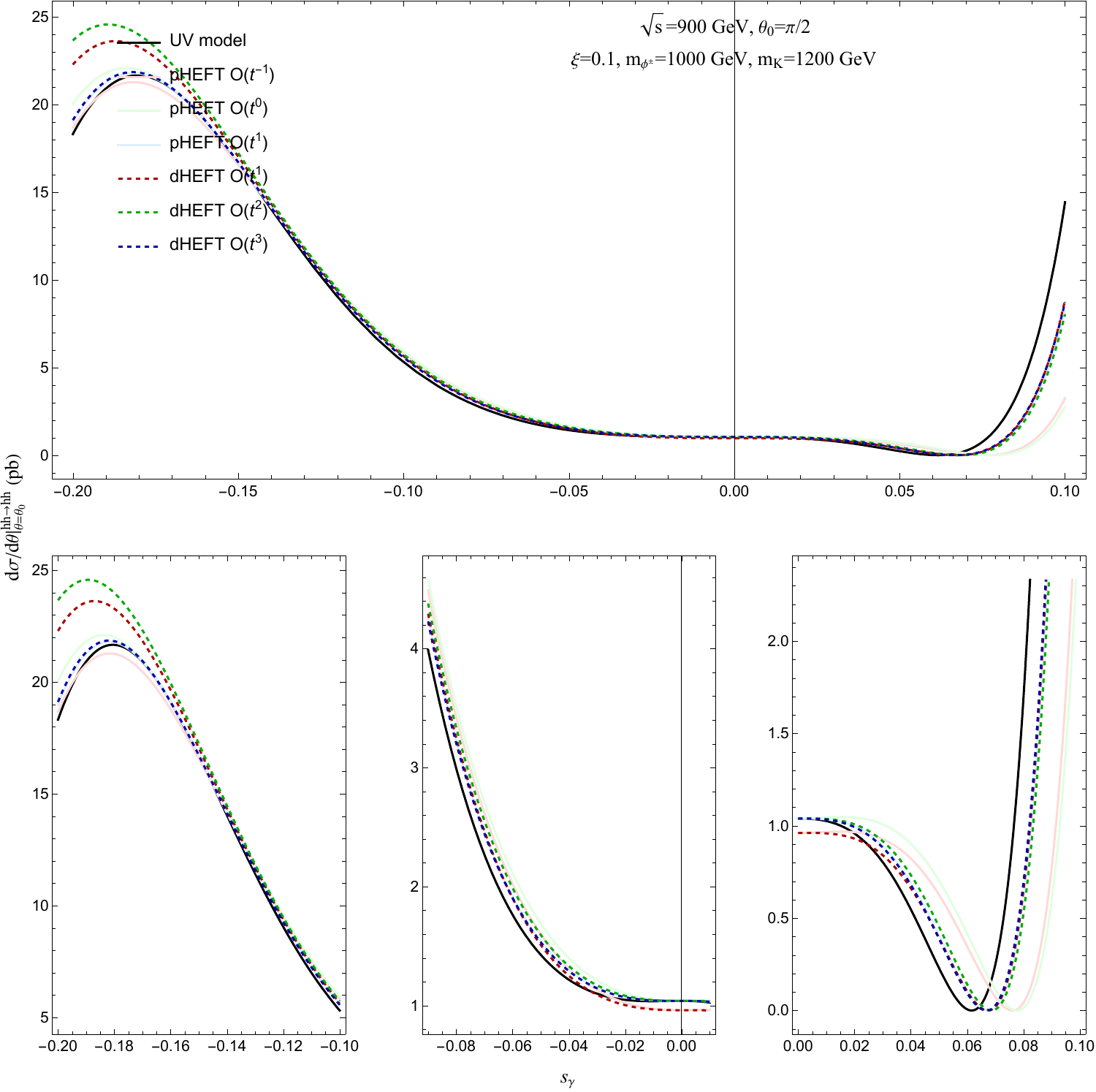}
    \caption[]{Similar to Fig.~\ref{fig:xs_hh_pHEFT_mK520}, now shown for $\sqrt{s}=900\,\mathrm{GeV}$, $\theta_0=\pi/2$, $\xi=0.1$, $m_{\phi^\pm}=1000\,\mathrm{GeV}$, and $m_K=1200\,\mathrm{GeV}$. The value $\xi = 0.1$ is chosen for illustration only and is not intended to be phenomenologically viable, having been excluded by existing measurements. When $s_\gamma>-0.064$, the model becomes unstable because of the boundedness-from-below conditions ($Z_1, Z_2\geq 0$ and $|Z_3|\leq 2\sqrt{Z_1 Z_2}$) are violated.}
    \label{fig:xs_hh_pHEFT_xi0.1}
\end{figure}

The mapping relations established in Eqs.~\eqref{pHEFTtodHEFTm},~\eqref{pHEFTtodHEFT1}, and~\eqref{pHEFTtodHEFT2} imply that one framework is systematically embedded within the other. Specifically, we observe the following inclusion relations:
\beq
\begin{aligned}[b]
\mathcal{L}^\text{p}_\text{HEFT}(t^{-1}) &\supset \mathcal{L}^\text{d}_\text{HEFT}(t^{1}), \\
\mathcal{L}^\text{p}_\text{HEFT}(t^{-1}+t^{0}) &\supset \mathcal{L}^\text{d}_\text{HEFT}(t^{1}+t^{2}), \\
\mathcal{L}^\text{p}_\text{HEFT}(t^{-1}+t^{0}+t^{1}) &\supset \mathcal{L}^\text{d}_\text{HEFT}(t^{1}+t^{2}+t^{3}), \dots
\end{aligned}
\eeq
where the ellipsis denotes that this pattern persists to all higher orders. This hierarchical structure confirms that the dHEFT is effectively contained within the pHEFT expansion. We make this correspondence visually explicit by investigating the process of Higgs self-scattering $hh\rightarrow hh$. For all subsequent plots, we fix the Higgs mass $m_h$ to 125 GeV and the electroweak VEV $\vev$ to 246 GeV. While experiments also impose strong bounds—$\xi\lesssim 0.015$ from the $\rho$ parameter~\cite{ParticleDataGroup:2024cfk} and $-0.10 \lesssim s_\gamma \lesssim 0.15$ from Higgs couplings~\cite{ParticleDataGroup:2024cfk, ATLAS:2025qxq, CMS:2025jwz}—the parameter regions displayed in our plots are not required to satisfy them fully. Usually only slight violations are allowed unless explicitly noted. Figs.~\ref{fig:xs_hh_pHEFT_mK520} and \ref{fig:xs_hh_pHEFT_mK820} demonstrate that, for typical parameter values, the $\mathcal{O}(t^n)$ result in dHEFT closely matches the $\mathcal{O}(t^{n-2})$ result in pHEFT—a consequence of the inherent power‑counting hierarchy distinguishing the two effective theories. But in extreme regions of parameter space ($\xi=0.1$, though such regions have already been ruled out by experiments, we show it for illustration), the leading-order prediction in pHEFT matches the accuracy of the second- and even third-order results in dHEFT as presented in Fig.~\ref{fig:xs_hh_pHEFT_xi0.1}. This result is expected, since a sizable non-decoupling effect in the RHTM arises from custodial-symmetry breaking and is therefore only pronounced for large values of $\xi$.

The existence of the straightforward inclusion relations described above is a direct consequence of the specific scaling assignments for $s_\gamma$ and $\xi$ in the decoupling limit. As long as these parameters are treated as $\mathcal{O}(t^n)$ with $n \geq 0$, the dHEFT can be viewed as a truncation of the pHEFT. For instance, if a model required a more suppressed scaling such as $s_\gamma \sim \mathcal{O}(t^4)$ and $\xi \sim \mathcal{O}(t^1)$, the inclusion relations would remain valid, as both parameters continue to provide additional suppression within the pHEFT expansion.
However, for more `exotic' scalings—such as $s_\gamma\sim\mO(t^{4})$ and $\xi\sim\mO(t^{-1})$—the correspondence becomes less trivial, as the enhancement from $\xi$ could potentially compete with the suppression from $s_\gamma$ or the heavy mass expansion. Fortunately, such unconventional power-counting schemes are rarely motivated by physical considerations. Thus, for the vast majority of phenomenologically relevant scenarios, the pHEFT serves as a robust and flexible framework for generating specific effective limits.

It is also of interest to examine whether the pHEFT can provide a convenient organizational starting point for HEFT formulations based on alternative parameter sets. For instance, one may consider the set
$\{m^2_{\phi^\pm}, m^2_K, m_h, \vev, Z_2, \xi \}$,
in which the physical mixing angle $s_\gamma$ is replaced by $Z_2$, a fundamental parameter appearing directly in the UV Lagrangian.
Another possibility is the set $\{Z_1, Z_2, Z_3, Y_3, \vev, \xi \}$ employed in Ref.~\cite{Song:2025kjp}, where a HEFT is constructed through a systematic $\xi$ expansion. In the following two subsections, we present a detailed analysis of these cases.

\subsection{$Z_2$-HEFT\label{sec:Z2HEFT}}

As discussed in the previous section, a Primary HEFT is obtained from UV–HEFT matching through an inverse-mass expansion without imposing additional restrictions. In principle, one may choose alternative parameter sets, including heavy-mass parameters, leading to different versions of the Primary HEFT. It is therefore important to clarify whether these formulations are equivalent and, if not, which one provides a more accurate description.

Our baseline realization, the pHEFT, utilizes the parameter set $\{m^2_{\phi^\pm}, m^2_K, m_h, \vev, s_\gamma, \xi\}$.
Among these parameters, $m^2_{\phi^\pm}$ and $m^2_K$ are kept as expansion parameters; the Higgs mass $m_h$, the electroweak VEV $\vev$ and $\xi$ are well-determined or constrained by precision experimental measurements. Thus, $s_\gamma$ emerges as the most adaptable candidate for replacement. We therefore consider an alternative primary HEFT based on the set:
\beq
\mathcal{P}_{Z_2} = \{m^2_{\phi^\pm}, m^2_K, m_h, \vev, Z_2, \xi\}
\eeq
and follow a power-counting scheme:
\beq
Z_2\text{-HEFT:} \quad m^2_{\phi^\pm}\sim m^2_K\sim\mathcal{O}(t^{-1}), \quad m_h\sim \vev\sim Z_2\sim \xi\sim\mathcal{O}(t^0).
\label{Z2PC}
\eeq
Where the physical mixing angle $s_\gamma$ is now treated as a derived quantity expressed in terms of the UV-Lagrangian parameter $Z_2$ via the relation:
\beq
s_\gamma = \pm \sqrt{\frac{
2 \xi^2 Z_2 \vev^2 - (4 \xi^2 + 1) m_K^2 + m_{\phi^{\pm}}^2
}{
(4 \xi^2 + 1)(m_h^2 - m_K^2)
}}.
\label{sgamma}
\eeq
In this expression, the $\pm$ sign corresponds to two distinct branches, each associated with a different HEFT. We refer to them as the positive $Z_2$-HEFT and the negative $Z_2$-HEFT, respectively.

The construction of the $Z_2$-HEFT then proceeds through a two-step mapping procedure. This is possible because the pHEFT already captures the full dynamics of the underlying theory, having successfully integrated out the heavy states. Consequently, the mapping requires only a parametric transformation and operator reorganization: first, we perform an substitution of $s_\gamma$ (and $c_\gamma$) into the pHEFT Lagrangian; second,
we reorganize operators according to the power counting scheme in Eq.~\eqref{Z2PC}.

In the latter step, the expression for $s_\gamma$ appearing on the right-hand side of Eq.~\eqref{sgamma} must be expanded in inverse powers of the heavy masses and consistently truncated at a given order. Such an expansion implicitly imposes several consistency conditions on the parameter space. For example, the appearance of the factor $(m_h^2 - m_K^2)$ in the denominator requires the hierarchy
\beq
m_h^2 \ll m_K^2.
\eeq
Meanwhile, regarding the numerator within the square root, a valid expansion in $m_K^2$ and $m_{\phi^{\pm}}^2$ ($m^2_{\phi^\pm}\sim m^2_K\sim\mathcal{O}(t^{-1})$) requires the following further condition:
\beq
 \left| 2 \xi^2 Z_2 \vev^2 \right| \ll \left| \left(4 \xi^2 + 1\right) m_K^2 - m_{\phi^\pm}^2\right |.
\label{Z2eqn}
\eeq

If these two hierarchies are not maintained, the heavy-mass expansion ceases to be systematically controlled, signaling that the $Z_2$-HEFT no longer provides a reliable description of the RHTM.

\begin{figure*}[h!]
    \centering
    \includegraphics[width=0.85\textwidth]{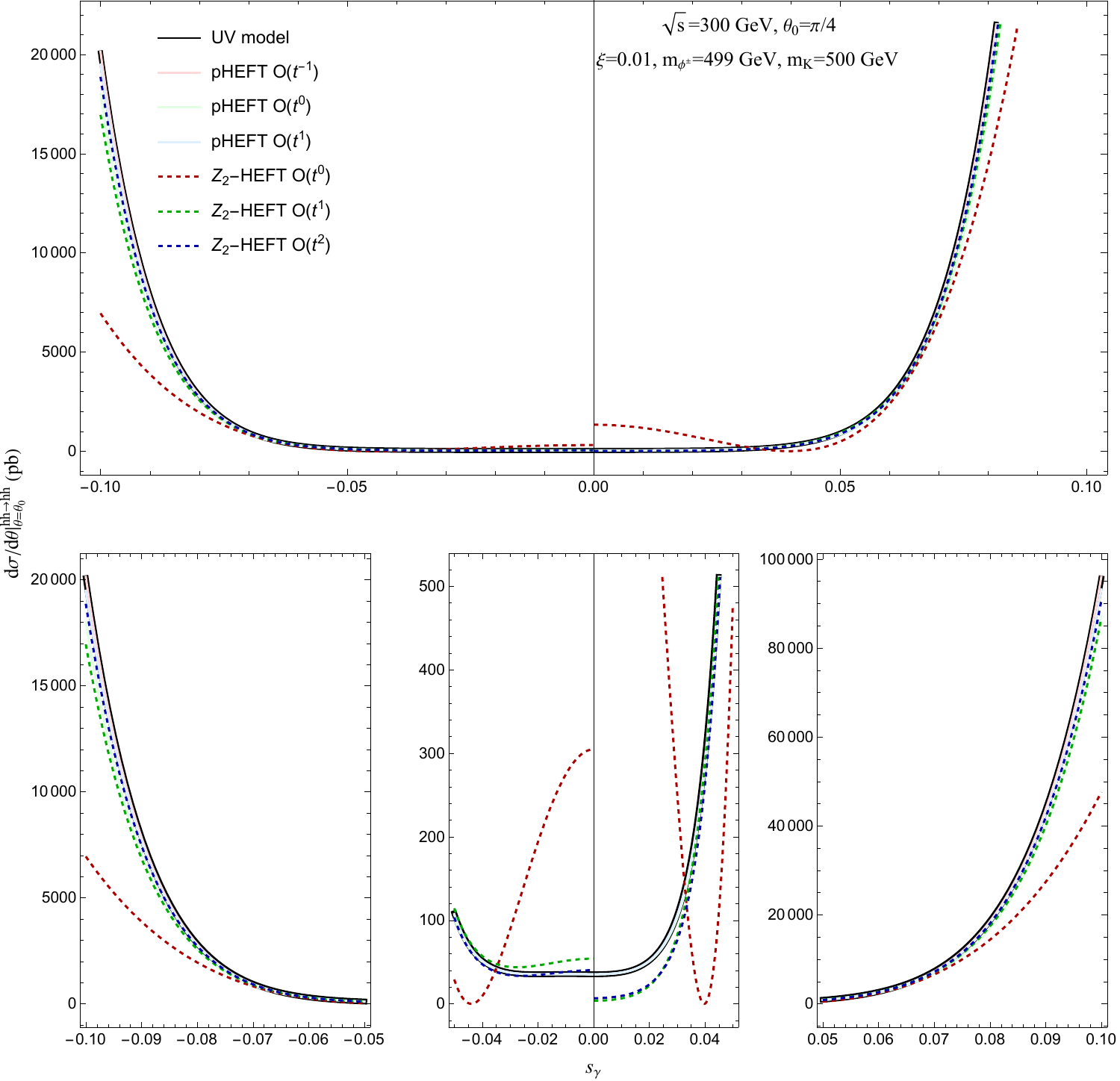}
    \caption[]{Comparison between the UV model and the pHEFT and $Z_2$-HEFT in the differential cross-section of $hh\rightarrow hh$, for a center-of-mass energy $\sqrt{s}=300\,\mathrm{GeV}$ and a scattering angle $\theta_0=\pi/4$. We set $\xi=0.01$, $m_{\phi^\pm}=499\,\mathrm{GeV}$, and $m_K=500\,\mathrm{GeV}$. The Higgs mass and electroweak vacuum expectation value are fixed to their experimentally measured values: $m_h=125\,\mathrm{GeV}$ and $\vev=246\,\mathrm{GeV}$. The top panel presents the cross section over $s_\gamma\in[-0.1, 0.1]$, while the three lower panels provide detailed views of three selected subintervals. }
    \label{fig:xs_hh_Z2HEFT_mC499}
\end{figure*}
\begin{figure*}[h!]
    \centering
    \includegraphics[width=0.7\textwidth]{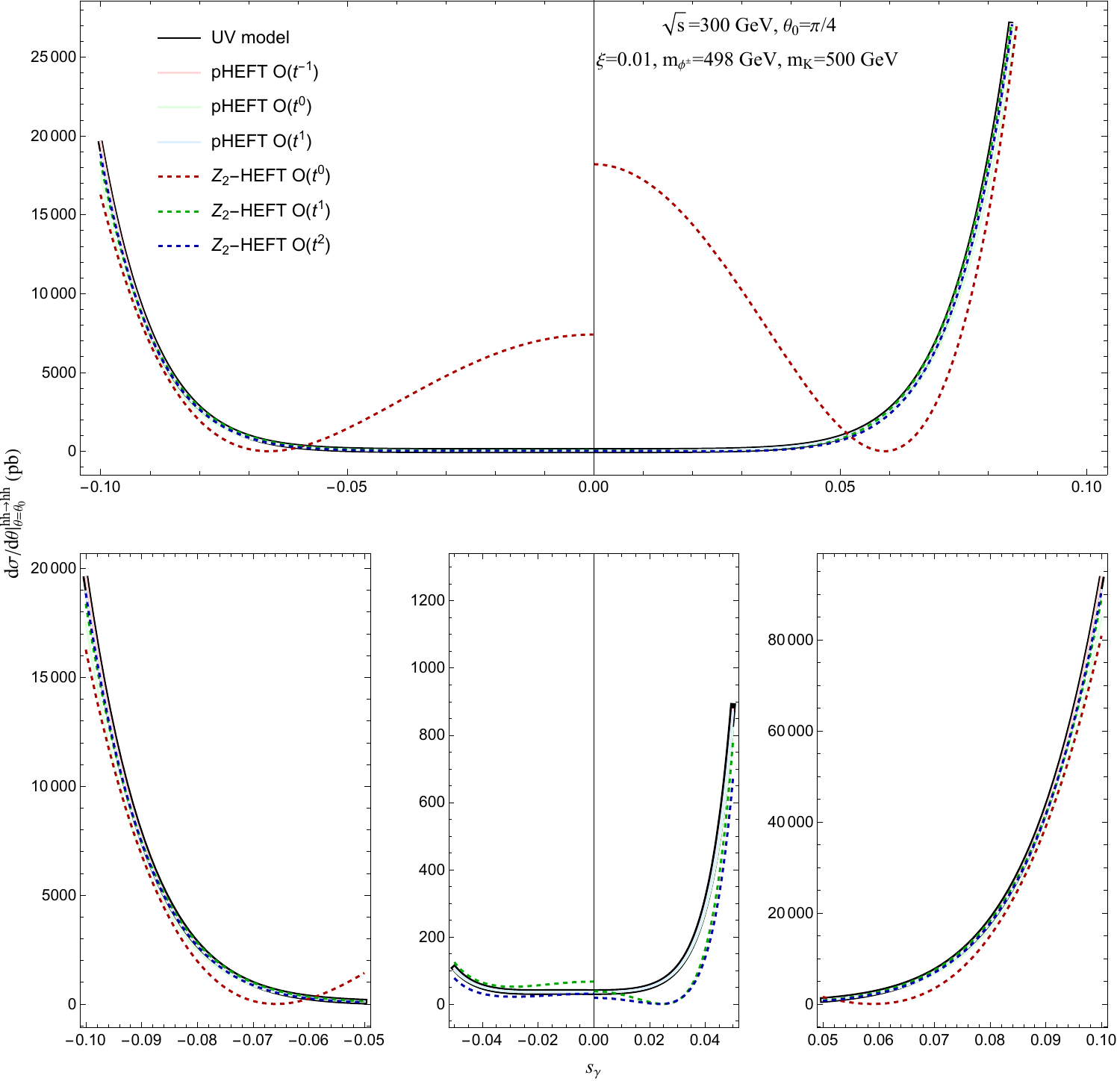}
    \caption[]{Similar to Fig.~\ref{fig:xs_hh_Z2HEFT_mC499}, now shown for $\sqrt{s}=300\,\mathrm{GeV}$, $\theta_0=\pi/4$, $\xi=0.01$, $m_{\phi^\pm}=498\,\mathrm{GeV}$, and $m_K=500\,\mathrm{GeV}$. }
    \label{fig:xs_hh_Z2HEFT_mC498}
\end{figure*}

Figs.~\ref{fig:xs_hh_Z2HEFT_mC499} and~\ref{fig:xs_hh_Z2HEFT_mC498} show the differential cross section for $hh\rightarrow hh$ obtained from $Z_2$-HEFT, compared with that from pHEFT. For $s_\gamma$ < 0, we use the negative $Z_2$-HEFT; for $s_\gamma$ > 0, we use the positive $Z_2$-HEFT. The two HEFTs do not connect smoothly at $s_\gamma = 0$, since they approach this point from opposite directions. 
In Fig.~\ref{fig:xs_hh_Z2HEFT_mC499}, the $Z_2$-HEFT reproduces the RHTM prediction well around $s_\gamma \simeq \pm 0.07$. However, the accuracy decreases as $s_\gamma$ approaches zero, with a more significant discrepancy observed for positive $s_\gamma$. This reflects the breakdown of the expansion criteria in Eq.~\ref{Z2eqn} in this parameter regime.
In Fig.~\ref{fig:xs_hh_Z2HEFT_mC498}, we increase the mass splitting between the charged and neutral Higgs bosons by setting $m_{\phi^\pm} = 498\,\mathrm{GeV}$. In this case, the $Z_2$-HEFT exhibits good agreement with the RHTM around $s_\gamma \simeq \pm 0.09$. This can be understood from the fact that a larger mixing angle enhances the mass splitting between the scalar states. It is also noticeable from both Fig.~\ref{fig:xs_hh_Z2HEFT_mC499} and Fig.~\ref{fig:xs_hh_Z2HEFT_mC498} that the boundedness-from-below condition is almost satisfied for $s_\gamma \in [-0.04,\,0.0]$, which is far from the region where the approximation works well. This further reinforces the conclusion that the $Z_2$-HEFT cannot be regarded as a viable primary HEFT.

In summary, the pHEFT framework is significantly less constrained than the $Z_2$-HEFT. The criteria for determining whether a parameter set retains the maximal information from the UV theory depends on whether the relation between the UV Lagrangian couplings ($Z_i, Y_i$) and squared heavy masses is linear or necessitates an additional inverse-mass expansion during the matching procedure. In the RHTM, the relation between the Lagrangian couplings ($Z_i, Y_i$) and the pHEFT set $\{m^2_{\phi^\pm}, m^2_K, m_h, \vev, s_\gamma, \xi\}$ is linear, as shown in Eq.~\eqref{Zis}. This linearity ensures that the UV interaction terms remain exact, avoiding the need for further expansions in inverse heavy masses. Consequently, the pHEFT parameter set preserves maximal information from the UV theory within the primary HEFT. In contrast, the $Z_2$-HEFT relies on non-linear relations; substituting Eq.~\eqref{sgamma} into Eq.~\eqref{Zis} necessitates an extra inverse-mass expansion, which introduces truncation errors and an inherent loss of precision.

It is interesting to note that in the UV models with scalar extensions, the existence of such lineary realtions between the squared heavy masses and the UV Lagrangian parameters is a common feature, as seen in the $Z_2$-symmetric singlet model and the 2HDM in our following dicussions in Sec.~\ref{sec:otherUVs}. This makes the physical basis---comprising physical masses, mixing angles, and VEVs---a naturally superior choice for defining a primary HEFT.

\subsection{$\xi$-HEFT\label{sec:xiHEFT}}
In Ref.~\cite{Song:2025kjp}, we derived the $\xi$-HEFT, which utilizes the parameter set:
\beq
\mathcal{P}_\xi=\{ Z_1, Z_2, Z_3, Y_3, v_H, \xi\}
\eeq
This framework employs a $\xi$-expansion, which is formally equivalent to the power-counting scheme:
\beq
\xi\text{-HEFT:} \quad Z_1 \sim Z_2 \sim Z_3 \sim Y_3 \sim v_H \sim \mO(t^0), \quad \xi \sim \mO(t^1).
\label{xiscale}
\eeq
In Ref.~\cite{Song:2025kjp}, the corresponding UV-to-HEFT matching results are given by
\beq
\begin{aligned}[b]
\mathcal{L}^\xi_\text{HEFT}(t^0)={}&\frac{1}{2}D_\mu h D^\mu h - \frac{1}{4} Z_1\left(-v_H^4 + 4h^2 v_H^2 + 4h^3 v_H + h^4\right) - \frac{1}{4} (v_H + h)^2\braket{V_\mu V^\mu} \\
\mathcal{L}^\xi_\text{HEFT}(t^1)={}&\frac{\xi Y_3}{4v_H}\left(-v_H^4 + 4h^2 v_H^2 + 4h^3 v_H + h^4\right) \\
\mathcal{L}^\xi_\text{HEFT}(t^2)={}&\frac{\xi^2}{4 v_H^2} \Bigl\{v_H^6 Z_3 + 8h^2 v_H^4(2Z_1-Z_3) + 8h^3 v_H^3(5Z_1-2Z_3) \\
&\qquad + \frac{14}{3}h^4 v_H^2(8Z_1-3Z_3) - 4\left(v_H^4 + 3h v_H^3 + 4h^2 v_H^2 \right)\braket{V_\mu V^\mu} \\
&\qquad + 2\left(v_H^4 + 4h v_H^3 + 6h^2 v_H^2 \right) \braket{V_\mu \sigma_3} \braket{V^\mu \sigma_3} \Bigr\},
\end{aligned}
\label{xiHEFT}
\eeq
where $h$ polynomials of higher orders have been omitted.

To facilitate a direct comparison between the pHEFT and the $\xi$-HEFT, first we start from the pHEFT and express its parameters $\{m^2_{\phi^\pm}, m^2_K, m^2_h, s_\gamma, \vev\}$ in terms of $\{ Z_1, Z_2, Z_3, Y_3, v_H\}$ through a systematic expansion in $\xi$:
\beq
\begin{aligned}[b]
m_{\phi^\pm}^2 &\to \frac{Y_3 v_H}{2 \xi}+2\xi Y_3 v_H, \\
m_K^2 &\to \frac{Y_3 v_H}{2 \xi} + 2 \xi Y_3 v_H + 2 \xi^2 v_H^2(4 Z_1 + Z_2 - 2 Z_3) + O(\xi^3), \\
m_h^2 &\to 2 Z_1 v_H^2-2 \xi Y_3 v_H-4 \xi^2 v_H^2(2 Z_1 - Z_3)+ O(\xi^3), \\
s_\gamma &\to - 2 \xi - \frac{2 (4 Z_1 - Z_3)\xi^2 v_H}{Y_3} +O(\xi^3), \\
\vev &\to v_H + 2 \xi^2 v_H + O (\xi^3).
\end{aligned}
\label{mtoZ}
\eeq
Then we substitute these expansions into the pHEFT Lagrangian (Eq.~\eqref{pHEFTL0})
and reorganize the resulting operators according to the $\xi$-HEFT power-counting scheme defined in Eq.~\eqref{xiscale}.
 Finally, we obtain a new HEFT that is identical to the $\xi$-HEFT presented in Eq.~\eqref{xiHEFT}.
This demonstrates that the $\xi$-HEFT can be systematically derived from the pHEFT via the corresponding parameter transformation and $\xi$-expansion.

It is noteworthy that the leading-order scaling in $\xi$-HEFT (Eq.~\eqref{xiscale}) is effectively equivalent to the dHEFT scaling: $m_{\phi^\pm}^2 \sim m_K^2 \sim \mO(t^{-1})$, $m_h \sim \vev \sim \mO(t^0)$, and $s_\gamma \sim \xi \sim \mO(t^1)$. This equivalence motivates a closer examination of the relationship between these two HEFTs. By applying the reparameterization in Eq.~\eqref{mtoZ} to the dHEFT, we indeed re-derive the $\xi$-HEFT results presented in Eq.~\eqref{xiHEFT}. Considering the higher-order terms discarded at each stage of the pHEFT$\,\to\,$dHEFT and dHEFT$\,\to\xi$-HEFT mappings, the hierarchical relationship among these three HEFTs can be expressed as:
\beq
\text{pHEFT} \supset \text{dHEFT} \supset \xi\text{-HEFT}.
\eeq
This nesting structure reveals that although pHEFT and dHEFT are formally reducible to $\xi$-HEFT, they encode a more fundamental matching. Specifically, the matching procedure dictates that the $\xi$ expansion must be viewed as secondary, to be performed only after the primary inverse-mass expansion. This precedence arises from the fact that the essence of UV–EFT matching lies in the treatment of heavy-state propagators. Thus, the expansion in inverse mass is fundamental, while any expansion in additional parameters is necessarily subordinate.

The pHEFT and dHEFT frameworks improve numerical agreement with the underlying UV theory by retaining—rather than truncating—subsets of higher-order terms within their physical parameter definitions, as seen through the inverse of Eq.~\eqref{mtoZ}. This approach systematically minimizes the UV–EFT mismatch at any fixed order of truncation.

\begin{figure*}[h!]
    \centering
    \includegraphics[width=0.85\textwidth]{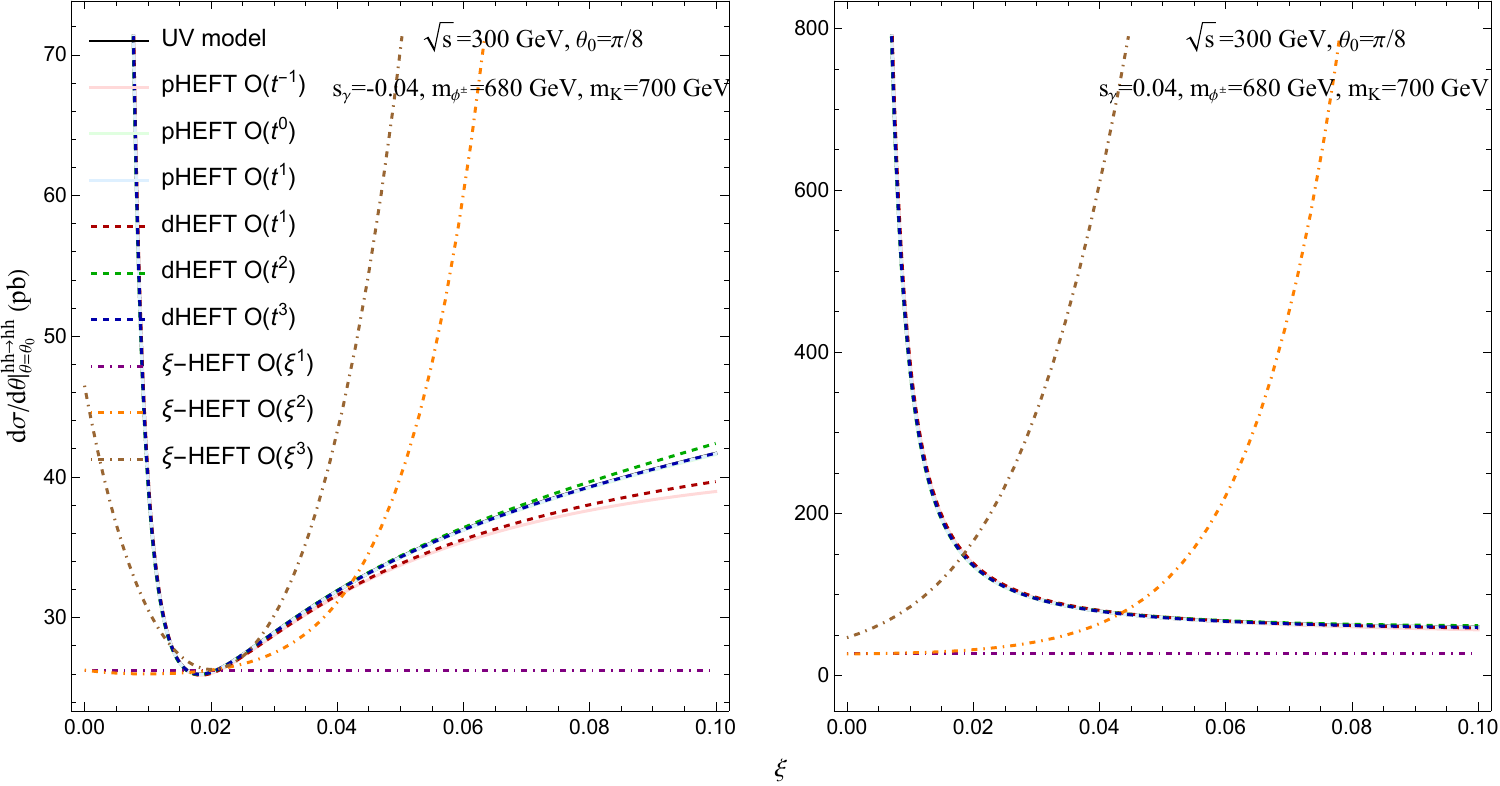}
    \caption[]{Comparison between the UV model and the pHEFT and $\xi$-HEFT in the differential cross-section of $hh\rightarrow hh$, for a center-of-mass energy $\sqrt{s}=300\,\mathrm{GeV}$ and a scattering angle $\theta_0=\pi/8$. We set $m_{\phi^\pm}=680\,\mathrm{GeV}$, and $m_K=700\,\mathrm{GeV}$. The Higgs mass and electroweak vacuum expectation value are fixed to their experimentally measured values: $m_h=125\,\mathrm{GeV}$ and $\vev=246\,\mathrm{GeV}$. The left panel presents the case where $s_\gamma=-0.04$, while the right panel shows the case where $s_\gamma=+0.04$,.}
    \label{fig:xs_hh_xi-HEFT}
\end{figure*}

In Fig.~\ref{fig:xs_hh_xi-HEFT}, we show the comparison among the pHEFT, the dHEFT and the $\xi$-HEFT~\footnote{The numerical results of the $\xi$-HEFT are obtained upon transforming $\{Z_1, v_H \}$ into $\{m_h, \vev\}$, allowing $m_h$ and $\vev$ to be fixed by experimental values.}. First, the $\xi$-HEFT is far from exactly reproducing the RHTM when $s_\gamma$ is positive (right panel). This is evident from the expansion of $s_\gamma$ in $\xi$ in Eq.~\eqref{mtoZ}, which indicates that the $\xi$ expansion is valid only for negative $s_\gamma$. Second, as shown in the left panel, $\xi$-HEFT performs comparably to other HEFTs when $\xi$ is close to $-s_\gamma/2 = 0.2$.

%Second, the $\xi$-HEFT is less precise than the dHEFT. Both the dHEFT and the $\xi$-HEFT exhibit decoupling behavior; however, the expansion in $\xi$ leads to additional truncations. Overall, expansion in inverse powers of the mass offers higher precision than expansion in other parameters.

%we could let masses implicit in the whole procedure and deal with only the Lagrangian's parameters,

%This is due to a $\xi$ expansion is a sencondary expansion after the inverse-mass expansion in the whole matching procedure. Even though we could let masses implicit in the whole procedure and deal with only the Lagrangian's parameters, e.g. $Z_i$s and $Y_i$s, the core of matching is to deal with the ``propagator'' of heavy states and thus expansion with inverse mass is unavoidable.

%The scaling method is not same as the simple expansion, but...
%the $\xi$-expansion lose precision.

\subsection{$Y_2$-HEFT vs. SMEFT\label{sec:Y2HEFT}}
\subsubsection{$Y_2$-HEFT}

In the standard UV-SMEFT matching ~\cite{Corbett:2021eux, Ellis:2023zim, Song:2025kjp}, the Wilson coefficients of SMEFT are conventionally written as functions of the UV theory's parameters, specified in the unbroken (symmetric) phase,
\beq
\mathcal{P}_{Y_2}=\{Z_1, Z_2, Z_3, Y_1, Y_2, Y_3\}.
\eeq
The power‑counting scheme proceeds as an expansion in powers of $1/Y_2^2$, with $Y_2^2$ denoting the real‑triplet mass squared before electroweak symmetry breaking. Equivalently, the parameters follow the scaling:
\beq
Y_2\text{-HEFT:} \quad Z_1 \sim Z_2 \sim Z_3 \sim Y_1 \sim Y_3 \sim \mO(t^0), \quad Y_2^2 \sim \mO(t^{-1}).
\label{SMEFTscale}
\eeq
Within this power-counting scheme, one can derive a new HEFT, which we denote as the $Y_2$-HEFT.

As discussed above, the $Y_2$-HEFT Lagrangian follows directly from the pHEFT results via the substitutions listed below:
\beq
\begin{aligned}[b]
m_{\phi^\pm}^2 &\to Y_2^2 - \frac{Y_1^2 Z_3}{2 Z_1} - \frac{Y_3^2 Y_1^2 \left(4 Z_1+Z_3\right)}{4 Y_2^2 Z_1^2} + \mO(Y_2^{-4}), \\
m_K^2 &\to Y_2^2 - \frac{Y_1^2 Z_3}{2 Z_1} - \frac{Y_3^2 Y_1^2 (4 Z_1+Z_3)}{4 Y_2^2 Z_1^2} + \mO(Y_2^{-4}), \\
m_h^2 &\to -2 Y_1^2 + \frac{Y_1^4 Y_3^2 \left(3 Z_3-8 Z_1\right)}{4 Y_2^4 Z_1^2} + \mO(Y_2^{-4}), \\
s_\gamma &\to -\frac{\sqrt{-Y_1^2} Y_3}{Y_2^2 \sqrt{Z_1}} - \frac{\sqrt{-Y_1^2} Y_3 \left(4 Y_1^2 \left(Z_3-2 Z_1\right)+Y_3^2\right)}{4 Y_2^4 Z_1^{3/2}} + \mO(Y_2^{-6}), \\
\xi &\to \frac{\sqrt{-Y_1^2}\, Y_3}{2\, Y_2^2\, \sqrt{Z_1}} + \frac{\sqrt{-Y_1^2}\, Y_3 \left( 2 Y_1^2 Z_3 + Y_3^2 \right)}{8\, Y_2^4\, Z_1^{3/2}} + \mO(Y_2^{-6}), \\
% v_H &\to \frac{\sqrt{-Y_1^2}}{\sqrt{Z_1}} + \frac{\sqrt{-Y_1^2} Y_3^2}{4 Y_2^2 Z_1^{3/2}} + \frac{3 \sqrt{-Y_1^2} \left(2 Y_1^2 Y_3^2 Z_3+Y_3^4\right)}{32 Y_2^4 Z_1^{5/2}} + \mO(Y_2^{-6}), \\
\vev &\to \frac{\sqrt{-Y_1^2}}{\sqrt{Z_1}} + \frac{\sqrt{-Y_1^2} Y_3^2}{4 Y_2^2 Z_1^{3/2}} + \frac{\sqrt{-Y_1^2} \left(2 Y_1^2 Y_3^2 \left(3 Z_3-8 Z_1\right)+3 Y_3^4\right)}{32 Y_2^4 Z_1^{5/2}} + \mO(Y_2^{-6}),
\end{aligned}
\eeq
where the physical masses, the mixing angle $s_\gamma$, and the parameter $\xi$ are expressed in terms of the Lagrangian couplings $\{Z_i, Y_i\}$. These relations are presented as a systematic expansion in $1/Y_2^2$, with only the first two leading orders explicitly shown. Note that $-Y_1^2 > 0$ is required to ensure $m_h \approx 125\,\mathrm{GeV}$ remains at the electroweak scale, while $m_{\phi^\pm}$ and $m_K$ scale primarily with the large mass parameter $Y_2^2$.

\subsubsection{SMEFT}
We present the SMEFT matching Wilson coefficients in Tab.~\ref{tab:SMEFT_couplings}~\cite{Corbett:2021eux, Ellis:2023zim, Song:2025kjp}, where terms of $\mO(Y_2^{-8})$ and higher are neglected.

\begin{table}[htbp]
\begin{center}
\begin{tabular}{c|c}
\hline
\multicolumn{2}{c}{dim-4} \\
\hline\hline
$C_{H^4}$ & $-Z_1+\frac{Y_3^2}{2Y_2^2}+\frac{2Y_1^2 Y_3^2}{Y_2^4}+\frac{6Y_1^4 Y_3^2}{Y_2^6}$ \\
\hline
\multicolumn{2}{c}{dim-6} \\
\hline\hline
$C_H$ & $\frac{Y_3^2}{2Y_2^4}\left[\left(8Z_1-Z_3\right)\left(1+\frac{4Y_1^2}{Y_2^2}\right)-\frac{4Y_3^2}{Y_2^2}\right]$ \\
$C_{HD}$ & $-\frac{2Y_3^2}{Y_2^4}\left(1+\frac{4Y_1^2}{Y_2^2}\right)$ \\
$C_{H\Box}$ & $\frac{Y_3^2}{2Y_2^4}\left(1+\frac{4Y_1^2}{Y_2^2}\right)$ \\
\hline
\multicolumn{2}{c}{dim-8} \\
\hline\hline
%$C_{H^8}$ & $\frac{Y_3^2}{4Y_2^6}\left(2\left(4Z_1-Z_3\right)^2-\frac{Y_3^2}{Y_2^2}(72Z_1+Z_2-20Z_3)+\frac{28Y_3^4}{Y_2^4}\right)$ \\
$C_{H^8}$ & $\frac{Y_3^2}{2Y_2^6}\left(4Z_1-Z_3\right)^2$ \\
$C_{H^6}^{(1)}$ & 0 \\
%& $\mO\Bigl(\frac{1}{Y_2^8}\Bigr)$ \\
$C_{H^6}^{(2)}$ & $\frac{2Y_3^2}{Y_2^6}\left(-4Z_1+Z_3\right)$ \\
$C_{H^4}^{(1)}$ & $\frac{4Y_3^2}{Y_2^6}$ \\
$C_{H^4}^{(3)}$ & $-\frac{2Y_3^2}{Y_2^6}$ \\
\hline\hline
\end{tabular}
\caption{Dimension-6 and -8 Wilson coefficients of the bosonic operators resulting from the tree-level matching of the RHTM to the SMEFT. $\mO(1/Y_2^8)$ and higher orders are omitted here. $C_{H^4}$ represents the coefficient of the Higgs doublet quartic interaction after matching.
\label{tab:SMEFT_couplings}
}
\end{center}
\end{table}

As an illustration, we compute the momentum‑independent trilinear and quartic Higgs couplings in order to compare SMEFT with $Y_2$-HEFT. After electroweak symmetry breaking, the Higgs doublet $H$ is parameterized as:
\beq
H \to \begin{pmatrix} -i G^+ \\ \frac{v_T + h + i G^0}{\sqrt{2}} \end{pmatrix},
\eeq
where $G^+, G^0$ denote the Goldstone bosons and $v_T$ is the vacuum expectation value (VEV) determined by the tadpole condition of the Higgs boson. The electroweak VEV is obtained from $m_W$ mass definition, which is given by
\beq
\vev^2=v_T^2\left[1+\frac{1}{4}v_T^4\left(C_{H^6}^{(1)}-C_{H^6}^{(2)}\right)\right].
\eeq

In the unitary gauge, the scalar potential for the Higgs field $h$ is :
\beq
V^\text{SMEFT}(h) = \frac{1}{2} Y_1^2 (v_T+h)^2 - \frac{1}{4} C_{H^4} (v_T+h)^4 - \frac{1}{8} C_{H^6} (v_T+h)^6 - \frac{1}{16} C_{H^8} (v_T+h)^8,
\eeq
while the kinetic part is:
\beq
\begin{aligned}[b]
&\mathcal{L}_\text{kin}^\text{SMEFT}(h) = \frac{1}{2} \mathcal{K}(h) \partial_\mu h \partial^\mu h, \\
&\mathcal{K}(h) = 1 + \frac{1}{2}\left(C_{HD} - 4 C_{H\Box}\right)(v_T+h)^2 + \frac{1}{4}\left(C_{H^6}^{(1)} + C_{H^6}^{(2)}\right)(v_T+h)^4.
\end{aligned}
\eeq
After performing the field redefinition $h'=\int_0^h\sqrt{\mathcal{K}(s)}\, ds$ to canonically normalize the kinetic term, we extract the triple and quartic Higgs self-couplings, $\Delta\kappa_3$ and $\Delta\kappa_4$ by the definition Eq.~\eqref{kappa34def}:
\beq
\begin{aligned}[b]
\Delta\kappa_3^\text{SMEFT} &= \frac{Y_1^2 Y_3^2 (2 Z_1-Z_3)}{2 Z_1^2 Y_2^4} + \frac{Y_1^2 Y_3^4 (Z_1-Z_3) + Y_1^4 Y_3^2 (9 Z_1 Z_3 - 12 Z_1^2 - 2 Z_3^2)}{2 Z_1^3 Y_2^6}, \\
\Delta\kappa_4^\text{SMEFT} &= \frac{Y_1^2 Y_3^2 (22 Z_1-9 Z_3)}{3 Z_1^2 Y_2^4} + \frac{Y_1^2 Y_3^4 (11 Z_1-9 Z_3) + Y_1^4 Y_3^2 (123 Z_1 Z_3 - 164 Z_1^2 - 24 Z_3^2)}{3 Z_1^3 Y_2^6},
\end{aligned}
\label{smeftk3k4}
\eeq
where higher-order terms in the $1/Y_2^2$ expansion are neglected. 

Notably, we find that up to $\mO(Y_2^{-6})$, the results for $\Delta\kappa_3^\text{SMEFT}$ and $\Delta\kappa_4^\text{SMEFT}$ derived above are analytically identical to those obtained in the $Y_2$-HEFT. Although the two approaches follow different paths---one matching in the unbroken phase (SMEFT) and the other matching in the broken phase (pHEFT) followed by reparameterization and $1/Y_2^2$ expansion ---the resulting Higgs self-interactions are perfectly consistent.
Thus, the hierarchical relation among the pHEFT, $Y_2$-HEFT, and the SMEFT can be expressed as
\beq
\text{pHEFT} \supset Y_2\text{-HEFT} \simeq \text{SMEFT},
\eeq
demonstrating that SMEFT emerges as a restricted limit within the broader HEFT framework in this example.

\section{The Primary HEFT of other UV Models\label{sec:otherUVs}}

In the previous section, through matching the RHTM to various HEFTs and exploring their connections, we have shown that the primary HEFT serves as a foundational framework that can be used to derive other specific HEFTs, while generally provides the highest level of perturbative accuracy. The idea of the primary HEFT is naturally applicable to other UV-HEFT matching procedures. In this section, we first formulate the primary HEFT within the $Z_2$-symmetric real singlet model (Z2RSM) and subsequently extend the discussion to the Two-Higgs Doublet Model (2HDM). The specific matching results for this model have been established in Ref.~\cite{Dawson:2023oce}; therefore, the present discussion is devoted exclusively to the construction of the primary HEFT and its foundational role.

\subsection{Z2RSM}
\subsubsection{Model Setup and Parameterization}

In the $Z_2$-symmetric real singlet model (Z2RSM)~\cite{Buchalla:2016bse,Dawson:2023oce, Robens:2015gla, Robens:2016xkb},
the Standard Model $SU(2)_L$ Higgs doublet is extended by a real singlet scalar field $S$, and a discrete $Z_2$ symmetry is imposed under which $S \to -S$.
The relevant part of the Lagrangian reads
\begin{align}
\mathcal{L}_\text{Z2RSM} &\supset (D^\mu H)^\dagger (D_\mu H)
+ \partial^\mu S\, \partial_\mu S
- V(H, S),
\\[6pt]
V(H, S) &=
- \frac{\mu_1^2}{2} H^\dagger H
- \frac{\mu_2^2}{2} S^2
+ \frac{\lambda_1}{4} (H^\dagger H)^2
+ \frac{\lambda_2}{4} S^4
+ \frac{\lambda_3}{2} H^\dagger H\, S^2,
\end{align}
where the parameters $\mu_i$ have mass dimension one, while the $\lambda_i$ are dimensionless couplings.

After spontaneous symmetry breaking, the scalar fields are parametrized as
\beq
H = U
\begin{pmatrix}
0 \\
\frac{1}{\sqrt{2}} (v_H + h_1)
\end{pmatrix},
\qquad
S = \frac{1}{\sqrt{2}} (v_S + h_2),
\eeq
where $v_H$ and $v_S$ denote the VEVs.
They satisfy the minimization conditions of the scalar potential,
\beq
\mu_1^2 = \frac{1}{2} \left( \lambda_1 v_H^2 + \lambda_3 v_S^2 \right),
\qquad
\mu_2^2 = \frac{1}{2} \left( \lambda_3 v_H^2 + \lambda_2 v_S^2 \right).
\eeq

The physical Higgs mass eigenstates are obtained by a rotation with mixing angle $\chi$,
\beq
\begin{pmatrix}
h \\
K
\end{pmatrix}
=
\begin{pmatrix}
 c_\chi & - s_\chi \\
 s_\chi & c_\chi
\end{pmatrix}
\begin{pmatrix}
h_1 \\
h_2
\end{pmatrix},
\eeq
where the mixing angle is given by
\beq
\tan(2\chi) =
\frac{2 \lambda_3 v_H v_S}{\lambda_2 v_S^2 - \lambda_1 v_H^2},
\eeq
and can be restricted to the range $\chi \in [-\pi/4,\, \pi/4]$ without loss of generality.

The masses of the physical Higgs bosons are
\beq
M_{h,K}^2 =
\frac{1}{4} \left[
\lambda_1 v_H^2 + \lambda_2 v_S^2
\mp \sqrt{
(\lambda_1 v_H^2 - \lambda_2 v_S^2)^2
+ 4 \lambda_3^2 v_H^2 v_S^2
}
\right],
\eeq
where $h$ denotes the observed Higgs boson with mass $m_h = 125\,\mathrm{GeV}$,
and $K$ is the heavier scalar state to be integrated out.

Expressed in terms of physical parameters, the scalar couplings can be written as
\beq
\begin{aligned}[b]
\lambda_1 &= \frac{2}{v_H^2}
\left( M_K^2 s_\chi^2 + M_h^2 c_\chi^2 \right), \\
\lambda_2 &= \frac{2}{v_S^2}
\left( M_h^2 s_\chi^2 + M_K^2 c_\chi^2 \right), \\
\lambda_3 &= \frac{2 c_\chi s_\chi}{v_H v_S}
\left( M_K^2 - M_h^2 \right),
\end{aligned}
\label{Z2RSMlambda}
\eeq
where the linear dependence of the couplings $\lambda_i$ on the heavy mass squared $M_K^2$ is explicit.

\subsubsection{Construction of the primary HEFT: parameter set and power counting}

The Lagrangian contains five independent parameters, which can be reparameterized in terms of five physical quantities.
To construct the primary HEFT, we choose the parameter set
\beq
\left( M_K,\, M_h,\, s_\chi,\, v_H,\, v_S \right),
\label{Z2RSMvSset}
\eeq
together with the power-counting assignment
\beq
M_K \sim \mathcal{O}(t^{-1}),
\qquad
M_h \sim s_\chi \sim v_H \sim v_S \sim \mathcal{O}(t^{0}).
\label{Z2RSMscale}
\eeq

From Eq.~\eqref{Z2RSMlambda}, one observes that the couplings $\lambda_i$ depend linearly on the heavy mass squared $M_K^2$.
As a result, no further expansion in inverse powers of $M_K$ is required, and the interaction terms are kept exact.
Consequently, the resulting primary HEFT preserves the maximal information from the underlying UV theory.

In contrast, if one adopts an alternative parameter set~\cite{Dawson:2023oce},
\beq
(M_K, M_h, s_\chi, v_H, \mu_2^2),
\label{Z2RSMmu22set}
\eeq
the numerical precision of the resulting EFT is compromised. This is because replacing $v_S$ with $\mu_2^2$ in the couplings $\lambda_i$ introduces an additional dependence on $M_K^2$, necessitating further expansions and truncations. From Eq.~\eqref{Z2RSMlambda}, we see that the couplings $\lambda_i$ involve the structure $1/v_S$. Using the minimization conditions, this replacement is given by:
\beq
\frac{1}{v_S} \to \frac{c_\chi^2 M_K^2 + s_\chi^2 M_h^2 - \mu_2^2}{c_\chi s_\chi v_H (M_h^2 - M_K^2)}.
\label{Z2RSMvS}
\eeq
The right-hand side contains a factor of $(M_h^2 - M_K^2)$ in the denominator. When performing an expansion in inverse powers of the heavy mass squared $M_K^2$, the term $1/(M_h^2 - M_K^2)$ must be expanded and truncated. This process inevitably discards high-order contributions that are otherwise captured exactly when $v_S$ is kept as an independent input. Consequently, the set $(M_K, M_h, s_\chi, v_H, \mu_2^2)$ is not suitable for constructing the primary HEFT.

\subsection{2HDM}
\subsubsection{General Basis vs.\ Higgs Basis}

In the 2HDM~\cite{Dawson:2023oce, Dawson:2023ebe, Arco:2023sac} (for reviews, see e.g. Refs.~\cite{Gunion:1989we, Branco:2011iw}), the Standard Model Higgs sector is enlarged by an additional $SU(2)_L$ doublet. Unlike scalar extensions involving singlets, triplets, or other representations, the two-doublet structure possesses a global $U(2)$ symmetry in the doublet space, reflecting the freedom to choose a basis for the Higgs fields. In a generic basis, the scalar Lagrangian is given by
\beq
\mathcal{L}_{\text{2HDM}} \supset \bigl( D_\mu \Phi_1 \bigr)^\dagger \bigl( D^\mu \Phi_1 \bigr) + \bigl( D_\mu \Phi_2 \bigr)^\dagger \bigl( D^\mu \Phi_2 \bigr) - V_{\text{2HDM}},
\eeq
\beq
\begin{aligned}[b]
V_{\text{2HDM}} = {} & m_{11}^2 \Phi_1^\dagger \Phi_1 + m_{22}^2 \Phi_2^\dagger \Phi_2 - m_{12}^2 \bigl(\Phi_1^\dagger \Phi_2 + \Phi_2^\dagger \Phi_1\bigr) + \frac{\lambda_1}{2} \bigl(\Phi_1^\dagger \Phi_1\bigr)^2 + \frac{\lambda_2}{2} \bigl(\Phi_2^\dagger \Phi_2\bigr)^2 \\
& + \lambda_3 \bigl(\Phi_1^\dagger \Phi_1\bigr)\bigl(\Phi_2^\dagger \Phi_2\bigr) + \lambda_4 \bigl(\Phi_1^\dagger \Phi_2\bigr)\bigl(\Phi_2^\dagger \Phi_1\bigr) + \frac{\lambda_5}{2} \left[\bigl(\Phi_1^\dagger \Phi_2\bigr)^2 + \bigl(\Phi_2^\dagger \Phi_1\bigr)^2\right],
\end{aligned}
\label{V2HDM}
\eeq
where a $\mathbb{Z}_2$ symmetry ($\Phi_1 \to \Phi_1,\, \Phi_2 \to -\Phi_2$) is imposed and softly broken by the $m_{12}^2$ term. We assume all parameters to be real, consistent with a CP-conserving scenario. The VEVs of $\Phi_1$ and $\Phi_2$ are denoted as $v_1$ and $v_2$, respectively.

Another widely used basis is the Higgs basis $\{H_1, H_2\}$, where only one doublet acquires a non-zero VEV~\cite{Lavoura:1994fv, Botella:1994cs, Branco:1999fs, Fontes:2014xva}.
The transformation from the general basis to the Higgs basis is performed via a rotation by the angle $\beta$:
\beq
\begin{pmatrix}
H_1 \\
H_2
\end{pmatrix}
=
\begin{pmatrix}
c_\beta & s_\beta \\
-s_\beta & c_\beta
\end{pmatrix}
\begin{pmatrix}
\Phi_1 \\
\Phi_2
\end{pmatrix},
\eeq
where $t_\beta = v_2/v_1$ and $v = \sqrt{v_1^2 + v_2^2} \approx 246\,\mathrm{GeV}$. In this basis, the scalar potential is parameterized as
\beq
\begin{aligned}[b]
V_\text{2HDM}^\text{H} = {} & Y_1 H_1^\dagger H_1 + Y_2 H_2^\dagger H_2 + \bigl(Y_3 H_1^\dagger H_2 + \text{h.c.}\bigr) \\
& + \frac{Z_1}{2} \bigl(H_1^\dagger H_1\bigr)^2 + \frac{Z_2}{2} \bigl(H_2^\dagger H_2\bigr)^2 + Z_3 \bigl(H_1^\dagger H_1\bigr)\bigl(H_2^\dagger H_2\bigr) + Z_4 \bigl(H_1^\dagger H_2\bigr)\bigl(H_2^\dagger H_1\bigr) \\
& + \biggl\{\frac{Z_5}{2} \bigl(H_1^\dagger H_2\bigr)^2 + Z_6 \bigl(H_1^\dagger H_1\bigr) \bigl(H_1^\dagger H_2\bigr) + Z_7 \bigl(H_2^\dagger H_2\bigr)\bigl(H_1^\dagger H_2\bigr) + \text{h.c.}\biggr\},
\end{aligned}
\label{V2HDMH}
\eeq
where all $Z_i$ and $Y_i$ parameters are real. Note that while the Higgs basis potential contains 10 potential parameters ($Y_{1,2,3}$ and $Z_{1, \dots, 7}$), the underlying $\mathbb{Z}_2$ symmetry and the minimization conditions reduce the number of independent parameters. Specifically, only five of the $Z_i$ couplings are independent due to the algebraic relations imposed by the basis transformation, leading to a total of eight independent parameters in the model.

\subsubsection{Construction of the primary HEFT: parameter set and power counting}

The construction of the primary HEFT starts with the choice of eight parameters, guided by the principle of selecting as many physical parameters as possible.
However, in the 2HDM there are only seven physical parameters, namely
$m_h, m_H, m_A, m_{H^\pm}, v, t_\beta$, and $c_{\beta-\alpha}$.
Therefore, one additional parameter must be taken from the Lagrangian.
A convenient choice is either $Y_2$ in the Higgs basis (see Eq.~\eqref{V2HDMH}) or $m_{12}$ in the general basis (see Eq.~\eqref{V2HDM}).
We denote these two choices as

\begin{align}
\text{Set 1:} \quad & (m_h, m_H, m_A, m_{H^\pm}, v, t_\beta, c_{\beta-\alpha}, Y_2),
\label{2HDMpara1} \\
\text{Set 2:} \quad & (m_h, m_H, m_A, m_{H^\pm}, v, t_\beta, c_{\beta-\alpha}, m_{12}),
\label{2HDMpara2}
\end{align}

where $h$ denotes the light Higgs boson, $H$ the heavy CP-even Higgs boson,
$A$ the heavy CP-odd Higgs boson, and $H^\pm$ the heavy charged Higgs bosons.
Here $v\equiv\sqrt{v_1^2+v_2^2}$, and $\alpha$ denotes the mixing angle between the two neutral CP-even scalar fields.

Adopting Set 1, the remaining parameters in the Higgs-basis potential are determined as:
\beq
\begin{aligned}[b]
Y_1 &= -\frac{Z_1}{2} v^2, \quad
Y_3 = -\frac{Z_6}{2} v^2, \\
Z_1 &= \frac{s_{\beta-\alpha}^2 m_h^2 + c_{\beta-\alpha}^2 m_H^2}{v^2}, \\
Z_2 &= \frac{1}{2 v^2 t_\beta^3} \Bigl[ c_{\beta-\alpha}^2 t_\beta \left( 3 t_\beta^4 - 8 t_\beta^2 + 3 \right) \left( m_h^2 - m_H^2 \right) + 2 t_\beta \left( t_\beta^2 - 1 \right)^2 \left( m_H^2 - Y_2 \right) \\
    &\qquad - m_h^2 \left( t_\beta^5 - 4 t_\beta^3 + t_\beta \right) + s_{\beta-\alpha} c_{\beta-\alpha} \left( t_\beta^6 - 7 t_\beta^4 + 7 t_\beta^2 - 1 \right) \left( m_h^2 - m_H^2 \right) \Bigr], \\
Z_3 &= \frac{2}{v^2} \left( m_{H^\pm}^2 - Y_2 \right), \\
Z_4 &= \frac{c_{\beta-\alpha}^2 \left( m_h^2 - m_H^2 \right) + m_A^2 + m_H^2 - 2 m_{H^\pm}^2}{v^2}, \\
Z_5 &= \frac{c_{\beta-\alpha}^2 \left( m_h^2 - m_H^2 \right) - m_A^2 + m_H^2}{v^2}, \\
Z_6 &= \frac{c_{\beta-\alpha} s_{\beta-\alpha} \left( m_h^2 - m_H^2 \right)}{v^2}, \\
Z_7 &= \frac{1}{2 v^2 t_\beta^2} \Bigl[ - 3 c_{\beta-\alpha}^2 t_\beta \left( t_\beta^2 - 1 \right) \left( m_h^2 - m_H^2 \right) + t_\beta \left( t_\beta^2 - 1 \right) \left( m_h^2 - 2 m_H^2 + 2 Y_2 \right) \\
    &\qquad - s_{\beta-\alpha} c_{\beta-\alpha} \left( t_\beta^4 - 4 t_\beta^2 + 1 \right) \left( m_h^2 - m_H^2 \right) \Bigr].
\end{aligned}
\eeq

From these expressions, we observe that the squared masses of the heavy states,
such as $m_H^2$, $m_A^2$, and $m_{H^\pm}^2$, appear only in the numerator.
As a result, after integrating out the heavy propagators, no additional expansion in inverse powers of the heavy masses is required.
Set 1 is well-suited for constructing the primary HEFT. Its power counting is defined as:
\beq
m_H \sim m_A \sim m_{H^\pm} \sim \mathcal{O}(t^{-1}), \quad m_h \sim v \sim t_\beta \sim c_{\beta-\alpha} \sim Y_2 \sim \mathcal{O}(t^0),
\label{2HDMscaling}
\eeq
where only the heavy masses scale as $\mathcal{O}(t^{-1})$, while the remaining parameters, including $Y_2$, are treated as $\mathcal{O}(t^0)$ and kept independent of the heavy scale.

To evaluate whether Set 2 is also suitable, we examine the relation between $Y_2$ and $m_{12}^2$. Using the minimization conditions, $Y_2$ can be expressed in terms of the parameters in Set 2 as:
\beq
Y_2 = \frac{m_{12}^2}{s_\beta c_\beta} - \frac{1}{2}\left(s_{\beta-\alpha}^2 m_h^2 + c_{\beta-\alpha}^2 m_H^2\right) - \left(m_h^2 - m_H^2\right) s_{\beta-\alpha} c_{\beta-\alpha} \cot(2\beta).
\label{Y2tom12}
\eeq
In this expression, the heavy mass squared $m_H^2$ appears exclusively in the numerator. Unlike the Z2RSM case where the replacement of $v_S$ by $\mu_2^2$ introduced inverse powers of the heavy mass (as seen in Eq.~\eqref{Z2RSMvS}), switching from $Y_2$ to $m_{12}^2$ in the 2HDM does not necessitate any additional expansion in $1/m_H^2$. Consequently, Set 2 is also suitable for constructing a primary HEFT, as it maintains the exact linear structure required for a precise matching.

\section{Conclusion and Discussion\label{sec:conclusion}}

In this work, we reanalyze the matching between the RHTM and the HEFT for various parameter sets and power counting schemes. For the first time, we obtain the complete HEFT matching results, including fermionic operators. We have established the concept of a primary Higgs Effective Field Theory (pHEFT) as a foundational and maximally precise benchmark for matching UV completions to the HEFT framework via observing the standard matching method almost based on an expansion of the effective propagators of the heavy states. We demonstrate that by choosing a parameter set consisting of physical masses, mixing angles, and vacuum expectation values—and adopting a power-counting scheme that expands solely in inverse powers of the heavy masses—we obtain an effective theory that retains the complete dynamical information from the UV completion at a given order.

The primary HEFT serves as a master framework from which other, more restricted HEFTs can be systematically derived through parameter mappings and the imposition of additional scaling rules. We explicitly mapped the pHEFT to several specific cases:
\begin{itemize}
    \item The decoupling HEFT (dHEFT), obtained by treating the mixing angle $s_\gamma$ and VEV ratio $\xi$ as suppressed quantities.
    \item The $\xi$-HEFT, recovered by applying a reparametrization and $\xi$-expansion to the pHEFT, with the decoupling limit being implicitly implemented.
    \item The $Z_2$-HEFT, a formulation based on the parameter set $\{m^2_{\phi^\pm}, m^2_K, m_h, \vev, Z_2, \xi\}$ where the mixing angle $s_\gamma$ is expressed in terms of the UV Lagrangian coupling $Z_2$. This case is particularly instructive, as it highlights that not all parameter sets are created ``equal''. Constructing the $Z_2$-HEFT requires an additional inverse-mass expansion, which inherently introduces truncation errors and makes it less precise than the pHEFT. This comparison underscores a central finding: A well-defined primary HEFT should be constructed in a parameter basis in which the UV Lagrangian parameters depend polynomially, with non-negative powers, on the heavy masses, thereby avoiding such secondary expansions.
    \item The $Y_2$-HEFT, derived from the pHEFT via a reparameterization into SMEFT variables followed by an expansion in the Lagrangian mass parameters, reproduces the SMEFT predictions. This explicitly shows that SMEFT arises as a nested limit of HEFT.
\end{itemize}
A key result is the hierarchical embedding among these EFTs: $\text{pHEFT}\supset\text{dHEFT}\supset\xi\text{-HEFT}$. This structure proves that the computationally intensive step of integrating out heavy degrees of freedom via a functional or diagrammatic matching needs to be performed only once to obtain the pHEFT. Phenomenologically motivated EFT limits can be accessed through straightforward algebraic rescalings and truncations of the pHEFT Lagrangian, thereby avoiding redundant matching calculations. We should emphasize that this concept has clear precedents, and a key takeaway for students is that expansion series are generally ``non-commutative''—a fundamental lesson for undergraduate studies~\footnote{Here, we use the term ``non-commutative'' in a practical sense: results derived under less restrictive parameter assumptions cannot be fully recovered from results obtained under stricter ones.}.

We further generalize the primary HEFT idea to other UV models with scalar extensions, namely the $Z_2$-symmetric real singlet model and the two-Higgs doublet model. The analysis, reinforced by the lessons from the $Z_2$-HEFT, led to clear criteria for selecting an optimal parameter set: a ``good'' primary HEFT is characterized by linear relations between heavy masses and the original UV Lagrangian couplings. This linearity ensures that no secondary expansion is necessary when integrating out heavy fields, preserving maximal UV information and numerical precision. Parameter sets that introduce non-linear relations (like the $Z_2$-based set) inevitably require additional inverse-mass expansions, leading to a loss of accuracy and making them unsuitable as a primary benchmark.

In summary, the primary HEFT formalism provides a unified, efficient, and precision-oriented strategy for top-down EFT matching. It clarifies the connections between various low-energy effective descriptions arising from the same UV physics and establishes a reliable benchmark for assessing the truncation errors introduced by more restrictive power-counting schemes. This work paves the way for more systematic, automated, and accurate interpretations of high-energy physics in the context of Higgs effective field theory.

\appendix
\section{Solutions for the Heavy Fields in the pHEFT}
\label{SolofpHEFT}

In this section, we present the solutions for the heavy fields that lead to the primary HEFT, which are given below.
\beq
\begin{aligned}[b]
K_{0}={}&- \frac{h^2}{2\xi(4 \xi^2 + 1)v_H m_K^2}\Bigl\{- c_\gamma^3\Bigl[(4 \xi^2 + 1)m_K^2 - 3 m_{\phi^\pm}^2\Bigr] + c_\gamma\Bigl[(4 \xi^2 + 1)m_K^2 - 3 m_{\phi^\pm}^2\Bigr] \\
&\qquad + \xi c_\gamma^2 s_\gamma\Bigl[(4 \xi^2 + 1)m_K^2 - 6 m_{\phi^\pm}^2\Bigr] + 2 \xi m_{\phi^\pm}^2 s_\gamma\Bigr\} + \mathcal{O}(h^3), \\
\phi_{10}={}&\phi_{20}=0, \\
K_{1}={}&\frac{c_\gamma\braket{V_\mu\sigma_3}\braket{V^\mu\sigma_3}}{(4 \xi^2 + 1)m_K^4}\Bigl\{(4 \xi^2 + 1)\xi v_H m_K^2 \\
&\qquad + h\Bigl[s_\gamma\Bigl((4 \xi^2 + 1)(2 c_\gamma^2 - 1)m_K^2 - 3 c_\gamma^2 m_{\phi^\pm}^2\Bigr) + 2 \xi c_\gamma\Bigl((2 - 3 c_\gamma^2)m_{\phi^\pm}^2 - (4 \xi^2 + 1)m_K^2 s_\gamma^2\Bigr)\Bigr]\Bigr\} \\
& + \frac{\braket{V_\mu V^\mu}}{2 m_K^4}\Bigl\{- v_H m_K^2(4 \xi c_\gamma + s_\gamma) + \frac{h c_\gamma}{\xi(4 \xi^2 + 1)}\Bigl[\xi s_\gamma\Bigl(2(9 c_\gamma^2 - 2)m_{\phi^\pm}^2 - 5(4 \xi^2 + 1)(2 c_\gamma^2 - 1)m_K^2\Bigr) \\
&\qquad + c_\gamma\Bigl(m_{\phi^\pm}^2\bigl(3(8 \xi^2 - 1)c_\gamma^2 - 16 \xi^2 + 3\bigr) + 2(16 \xi^4 - 1)m_K^2 s_\gamma^2\Bigr)\Bigr]\Bigr\} + \mathcal{O}(h^2) \\
& + \bar Q_L U \begin{pmatrix}y_u&0\\0&y_d\end{pmatrix} Q_R \times \frac{- s_\gamma}{\sqrt{2}\xi(4 \xi^2 + 1)v_H m_K^4}\Bigl\{(4 \xi^2 + 1)\xi v_H m_K^2 \\
&\qquad + h c_\gamma\Bigl[2(4 \xi^2 + 1)m_K^2(\xi c_\gamma^2 + c_\gamma s_\gamma - \xi) + m_{\phi^\pm}^2(- 6 \xi c_\gamma^2 - 3 c_\gamma s_\gamma + 4 \xi)\Bigr]\Bigr\} + \text{h.c.}, \\
\phi_{11}={}&\frac{\xi v_H \braket{V_\mu\sigma_1} \braket{V^\mu\sigma_3} + 2\mathrm{i}\braket{V_\mu\sigma_2}D^\mu h(\xi c_\gamma + s_\gamma)}{(4 \xi^2 + 1)m_{\phi^\pm}^2} \\
& + \frac{\mathrm{i} h(\xi c_\gamma + s_\gamma)}{(4 \xi^2 + 1)^2 m_{\phi^\pm}^2}\Bigl[(4 \xi^2 + 1)\braket{D_\mu V^\mu\sigma_2} + 2\mathrm{i} \braket{V_\mu\sigma_1} \braket{V^\mu\sigma_3}\Bigr] + \mathcal{O}(h^2) \\
& + \bar Q_L U \begin{pmatrix}0&-y_d\\y_u&0\end{pmatrix} Q_R \times \frac{\sqrt{2}\Bigl\{(4 \xi^2 + 1)\xi v_H + h\bigl[(4 \xi^2 - 1)s_\gamma - 2 \xi c_\gamma\bigr]\Bigr\}}{(4 \xi^2 + 1)^2 v_H m_{\phi^\pm}^2} + \text{h.c.}, \\
\phi_{21}={}&\frac{\xi v_H \braket{V_\mu\sigma_2} \braket{V^\mu\sigma_3} - 2\mathrm{i}\braket{V_\mu\sigma_1}D^\mu h(\xi c_\gamma + s_\gamma)}{(4 \xi^2 + 1)m_{\phi^\pm}^2} \\
& - \frac{\mathrm{i} h(\xi c_\gamma + s_\gamma)}{(4 \xi^2 + 1)^2 m_{\phi^\pm}^2}\Bigl[(4 \xi^2 + 1)\braket{D_\mu V^\mu\sigma_1} - 2\mathrm{i} \braket{V_\mu\sigma_2} \braket{V^\mu\sigma_3}\Bigr] + \mathcal{O}(h^2) \\
& + \bar Q_L U \begin{pmatrix}0&\mathrm{i} y_d\\\mathrm{i} y_u&0\end{pmatrix} Q_R \times \frac{\sqrt{2}\Bigl\{(4 \xi^2 + 1)\xi v_H + h\bigl[(4 \xi^2 - 1)s_\gamma - 2 \xi c_\gamma\bigr]\Bigr\}}{(4 \xi^2 + 1)^2 v_H m_{\phi^\pm}^2} + \text{h.c.},
\end{aligned}
\eeq

Here the solutions are further expanded around the Higgs field $h$ and $V_\mu \equiv U^\dagger D_\mu U$. The fields $\phi_1$ and $\phi_2$ vanish at lower orders and first acquire nonzero contributions at $\mathcal{O}(t^1)$. The custodial-symmetry–breaking structure $\braket{V_\mu \sigma_3} \braket{V^\mu \sigma_3}$ first appears in the solution for $K_{1}$. The fields $\phi_{11}$ and $\phi_{21}$ exhibit similar structures, as they correspond to the real and imaginary components of the charged scalar field, respectively.

%Where the solutions are further expanded around $h$, $V_\mu\equiv U^\dagger D_\mu U$. The fields $\phi_1$ and $\phi_2$ vanish at lower orders and first acquire nonzero contributions at $\mathcal{O}(t^1)$. $\phi_1$ and $\phi_2$ start from $\mO(t^1)$ order. The term of $\braket{V_\mu \sigma_3} \braket{V^\mu \sigma_3}$
%, which breaks custodial symmetry, first appear in the $K_{01}$ solution. $\phi_{11}$ and $\phi_{21}$ have similar structures since they are separately real and imaginary parts of the charged states.

\section{EoM of $U$}

When reducing operators to a complete basis, one frequently encounters terms containing $D_\mu D^\mu U$, which must be eliminated in favor of other operators using the equation of motion (EoM) of $U$. We present the corresponding derivation below.

Starting from a leading-order Lagrangian of the general form
\begin{gather*}
\mathcal{L}^\text{LO}=a(h)\braket{V_\mu\sigma_3}\braket{V^\mu\sigma_3}+b(h)\braket{V_\mu V^\mu}+\frac{1}{2}D_\mu hD^\mu h-V(h),
\end{gather*}
where $V_\mu \equiv U^\dagger D_\mu U$, $a(h)$ and $b(h)$ are polynomials in $h$, which we will henceforth denote simply as $a$ and $b$. We solve the EoM for $U$,
\beq
\frac{\partial\mathcal{L}^\text{LO}}{\partial U}=D_\mu\frac{\partial\mathcal{L}^\text{LO}}{\partial(D_\mu U)},
\eeq
which leads to

\beq
\begin{aligned}[b]
 D_\mu D^\mu U={}&-D_\mu UD^\mu U^\dagger U-\frac{D_\mu b}{b}D^\mu U-\frac{D_\mu a}{b}\braket{V^\mu\sigma_3}U\sigma_3 \\
&-\frac{a}{b}\Bigl[D_\mu\braket{V^\mu\sigma_3}U\sigma_3+\braket{V^\mu\sigma_3}D_\mu U\sigma_3-\braket{V_\mu\sigma_3}U\sigma_3V^\mu\Bigr],
\end{aligned}
\eeq

where on the right side there exists
\begin{align}
D_\mu\braket{V^\mu\sigma_3}&=\braket{D_\mu U^\dagger D^\mu U\sigma_3}+\braket{U^\dagger \boldsymbol{D_\mu D^\mu U}\sigma_3}.
\end{align}
By combining the above two equations, we finally obtain
\begin{align}
D_\mu\braket{V^\mu\sigma_3}&=-\frac{D_\mu(b+2a)}{b+2a}\braket{V^\mu\sigma_3}.
\end{align}

Similarly, for the other two components of $V^\mu$ we have
\begin{align}
D_\mu\braket{V^\mu\sigma_1}&=-\frac{D_\mu b}{b}\braket{V^\mu\sigma_1}-2\mathrm{i}\frac{a}{b}\braket{V_\mu\sigma_2}\braket{V^\mu\sigma_3} \\
D_\mu\braket{V^\mu\sigma_2}&=-\frac{D_\mu b}{b}\braket{V^\mu\sigma_1}+2\mathrm{i}\frac{a}{b}\braket{V_\mu\sigma_1}\braket{V^\mu\sigma_3}
\end{align}

\acknowledgments
 X.W. thanks Fengkun Guo and Jiannan Ding for helpful discussions. The work of H.S. is partially supported by IBS under the project code, IBS-R018-D1. X.W. is supported by the National Science Foundation of China under Grants No. 11947416.

\bibliography{reference}

@article{Dittmaier:2021fpq,
    author = "Dittmaier, Stefan and Schuhmacher, Sebastian and Stahlhofen, Maximilian",
    title = "{Integrating out heavy fields in the path integral using the background-field method: general formalism}",
    journal = "Eur. Phys. J. C",
    volume = "81",
    number = "10",
    pages = "826",
    year = "2021",
    doi = "10.1140/epjc/s10052-021-09587-7",
    eprint = "2102.12020",
    archivePrefix = "arXiv",
    primaryClass = "hep-ph"
}

@article{Weinberg:1980wa,
    author = "Weinberg, Steven",
    title = "{Effective Gauge Theories}",
    reportNumber = "HUTP-80/A001",
    doi = "10.1016/0370-2693(80)90660-7",
    journal = "Phys. Lett. B",
    volume = "91",
    pages = "51--55",
    year = "1980"
}

@article{Alasfar:2023xpc,
    author = "Alasfar, Lina and others",
    title = "{Effective Field Theory descriptions of Higgs boson pair production}",
    eprint = "2304.01968",
    archivePrefix = "arXiv",
    primaryClass = "hep-ph",
    reportNumber = "KA-TP-04-2023, P3H-23-020, OUTP-23-03P, LHCHWG-2022-004",
    doi = "10.21468/SciPostPhysCommRep.2",
    journal = "SciPost Phys. Comm. Rep.",
    volume = "2024",
    pages = "2",
    year = "2024"
}

@misc{Remmen:2024hry,
    author = "Remmen, Grant N. and Rodd, Nicholas L.",
    title = "{Positively Identifying HEFT or SMEFT}",
    eprint = "2412.07827",
    archivePrefix = "arXiv",
    primaryClass = "hep-ph",
    month = "12",
    year = "2024"
}

@article{Gargalionis:2024jaw,
    author = "Gargalionis, John and Quevillon, Jeremie and Vuong, Pham Ngoc Hoa and You, Tevong",
    title = "{Linear Standard Model extensions in the SMEFT at one loop and Tera-Z}",
    eprint = "2412.01759",
    archivePrefix = "arXiv",
    primaryClass = "hep-ph",
    reportNumber = "DESY-24-184, ADP-24-19/T1258, KCL-PH-TH-2024-72, DESY-24-184; ADP-24-19/T1258; KCL-PH-TH-2024-72;",
    doi = "10.1007/JHEP07(2025)136",
    journal = "JHEP",
    volume = "07",
    pages = "136",
    year = "2025"
}

@article{Ren:2022tvi,
    author = "Ren, Zhe and Yu, Jiang-Hao",
    title = "{A complete set of the dimension-8 Green{\textquoteright}s basis operators in the Standard Model effective field theory}",
    eprint = "2211.01420",
    archivePrefix = "arXiv",
    primaryClass = "hep-ph",
    doi = "10.1007/JHEP02(2024)134",
    journal = "JHEP",
    volume = "02",
    pages = "134",
    year = "2024"
}

@article{Isidori:2024pca,
  author = {Isidori, Gino and Wilsch, Felix and Wyler, Daniel},
  title = {The Standard Model effective field theory at work},
  journal = {Rev. Mod. Phys.},
  volume = {96},
  number = {1},
  pages = {015006},
  year = {2024},
  doi = {10.1103/RevModPhys.96.015006},
  eprint = {2303.16922},
  archivePrefix = {arXiv},
  primaryClass = {hep-ph}
}

@article{Grzadkowski:2010es,
  author = {Grzadkowski, B. and Iskrzy{\'n}ski, M. and Misiak, M. and Rosiek, J.},
  title = {Dimension-Six Terms in the Standard Model Lagrangian},
  journal = {JHEP},
  volume = {10},
  number = {085},
  year = {2010},
  doi = {10.1007/JHEP10(2010)085},
  eprint = {1008.4884},
  archivePrefix = {arXiv},
  primaryClass = {hep-ph}
}

@article{Henning:2017fpj,
  author = {Henning, Brian and Lu, Xiaochuan and Melia, Tom and Murayama, Hitoshi},
  title = {$2, 84, 30, 993, 560, \dots$: Higher dimension operators in the SM EFT},
  journal = {JHEP},
  volume = {08},
  number = {016},
  year = {2017},
  doi = {10.1007/JHEP08(2017)016},
  eprint = {1512.03433},
  archivePrefix = {arXiv},
  primaryClass = {hep-th}
}

@article{Murphy:2020rsh,
  author = {Murphy, Christopher W.},
  title = {Dimension-8 Operators in the Standard Model Effective Field Theory},
  journal = {JHEP},
  volume = {10},
  number = {174},
  year = {2020},
  doi = {10.1007/JHEP10(2020)174},
  eprint = {2005.00059},
  archivePrefix = {arXiv},
  primaryClass = {hep-ph}
}

@article{Ethier:2021bye,
    author = "Ethier, Jacob J. and Magni, Giacomo and Maltoni, Fabio and Mantani, Luca and Nocera, Emanuele R. and Rojo, Juan and Slade, Emma and Vryonidou, Eleni and Zhang, Cen",
    collaboration = "SMEFiT",
    title = "{Combined SMEFT interpretation of Higgs, diboson, and top quark data from the LHC}",
    eprint = "2105.00006",
    archivePrefix = "arXiv",
    primaryClass = "hep-ph",
    reportNumber = "OUTP-20-05P, Nikhef-2020-020, CP3-21-12, MCNET-21-07,
  MAN/HEP/2021/004",
    doi = "10.1007/JHEP11(2021)089",
    journal = "JHEP",
    volume = "11",
    pages = "089",
    year = "2021"
}

@article{Brivio:2020onw,
    author = "Brivio, Ilaria",
    title = "{SMEFTsim 3.0 {\textemdash} a practical guide}",
    eprint = "2012.11343",
    archivePrefix = "arXiv",
    primaryClass = "hep-ph",
    doi = "10.1007/JHEP04(2021)073",
    journal = "JHEP",
    volume = "04",
    pages = "073",
    year = "2021"
}

@article{Falkowski:2019tft,
    author = "Falkowski, Adam and Rattazzi, Riccardo",
    title = "{Which EFT}",
    eprint = "1902.05936",
    archivePrefix = "arXiv",
    primaryClass = "hep-ph",
    reportNumber = "LPT Orsay 19-05",
    doi = "10.1007/JHEP10(2019)255",
    journal = "JHEP",
    volume = "10",
    pages = "255",
    year = "2019"
}

@article{ATLAS:2024zkx,
    author = "Aad, Georges and others",
    collaboration = "ATLAS",
    title = "{Search for R-parity violating supersymmetric decays of the top squark to a b-jet and a lepton in s=13{\,}{\,}TeV pp collisions with the ATLAS detector}",
    eprint = "2406.18367",
    archivePrefix = "arXiv",
    primaryClass = "hep-ex",
    reportNumber = "CERN-EP-2024-136",
    doi = "10.1103/PhysRevD.110.092004",
    journal = "Phys. Rev. D",
    volume = "110",
    number = "9",
    pages = "092004",
    year = "2024"
}

@article{CMS:2024yiy,
    author = "Hayrapetyan, Aram and others",
    collaboration = "CMS",
    title = "{Search for heavy neutral Higgs bosons A and H in the tt{\textasciimacron}Z channel in proton-proton collisions at 13 TeV}",
    eprint = "2412.00570",
    archivePrefix = "arXiv",
    primaryClass = "hep-ex",
    reportNumber = "CMS-B2G-23-006, CERN-EP-2024-301",
    doi = "10.1016/j.physletb.2025.139568",
    journal = "Phys. Lett. B",
    volume = "866",
    pages = "139568",
    year = "2025"
}

@article{Banta:2021dek,
    author = "Banta, Ian and Cohen, Timothy and Craig, Nathaniel and Lu, Xiaochuan and Sutherland, Dave",
    title = "{Non-decoupling new particles}",
    eprint = "2110.02967",
    archivePrefix = "arXiv",
    primaryClass = "hep-ph",
    doi = "10.1007/JHEP02(2022)029",
    journal = "JHEP",
    volume = "02",
    pages = "029",
    year = "2022"
}

@article{Cohen:2020xca,
    author = "Cohen, Timothy and Craig, Nathaniel and Lu, Xiaochuan and Sutherland, Dave",
    title = "{Is SMEFT Enough?}",
    eprint = "2008.08597",
    archivePrefix = "arXiv",
    primaryClass = "hep-ph",
    doi = "10.1007/JHEP03(2021)237",
    journal = "JHEP",
    volume = "03",
    pages = "237",
    year = "2021"
}

@article{Sun:2022snw,
    author = "Sun, Hao and Xiao, Ming-Lei and Yu, Jiang-Hao",
    title = "{Complete NNLO operator bases in Higgs effective field theory}",
    eprint = "2210.14939",
    archivePrefix = "arXiv",
    primaryClass = "hep-ph",
    doi = "10.1007/JHEP04(2023)086",
    journal = "JHEP",
    volume = "04",
    pages = "086",
    year = "2023"
}

@article{Graf:2022rco,
    author = "Gr\'af, Luk\'a\v{s} and Henning, Brian and Lu, Xiaochuan and Melia, Tom and Murayama, Hitoshi",
    title = "{Hilbert series, the Higgs mechanism, and HEFT}",
    eprint = "2211.06275",
    archivePrefix = "arXiv",
    primaryClass = "hep-ph",
    doi = "10.1007/JHEP02(2023)064",
    journal = "JHEP",
    volume = "02",
    pages = "064",
    year = "2023"
}

@article{Gomez-Ambrosio:2022why,
    author = "G\'omez-Ambrosio, Raquel and Llanes-Estrada, Felipe J. and Salas-Bern\'ardez, Alexandre and Sanz-Cillero, Juan J.",
    title = "{SMEFT is falsifiable through multi-Higgs measurements (even in the absence of new light particles)}",
    eprint = "2207.09848",
    archivePrefix = "arXiv",
    primaryClass = "hep-ph",
    doi = "10.1088/1572-9494/ace95e",
    journal = "Commun. Theor. Phys.",
    volume = "75",
    number = "9",
    pages = "095202",
    year = "2023"
}

@article{Gomez-Ambrosio:2022qsi,
    author = "G\'omez-Ambrosio, Raquel and Llanes-Estrada, Felipe J. and Salas-Bern\'ardez, Alexandre and Sanz-Cillero, Juan J.",
    title = "{Distinguishing electroweak EFTs with WLWL\textrightarrow{}n\texttimes{}h}",
    eprint = "2204.01763",
    archivePrefix = "arXiv",
    primaryClass = "hep-ph",
    doi = "10.1103/PhysRevD.106.053004",
    journal = "Phys. Rev. D",
    volume = "106",
    number = "5",
    pages = "053004",
    year = "2022"
}

@article{Alonso:2021rac,
    author = "Alonso, Rodrigo and West, Mia",
    title = "{Roads to the Standard Model}",
    eprint = "2109.13290",
    archivePrefix = "arXiv",
    primaryClass = "hep-ph",
    doi = "10.1103/PhysRevD.105.096028",
    journal = "Phys. Rev. D",
    volume = "105",
    number = "9",
    pages = "096028",
    year = "2022"
}

@article{Asiain:2021lch,
    author = "Asi\'ain, I\~nigo and Espriu, Dom\`enec and Mescia, Federico",
    title = "{Introducing tools to test Higgs boson interactions via WW scattering: One-loop calculations and renormalization in the Higgs effective field theory}",
    eprint = "2109.02673",
    archivePrefix = "arXiv",
    primaryClass = "hep-ph",
    doi = "10.1103/PhysRevD.105.015009",
    journal = "Phys. Rev. D",
    volume = "105",
    number = "1",
    pages = "015009",
    year = "2022"
}

@article{Cohen:2021ucp,
    author = "Cohen, Timothy and Craig, Nathaniel and Lu, Xiaochuan and Sutherland, Dave",
    title = "{Unitarity violation and the geometry of Higgs EFTs}",
    eprint = "2108.03240",
    archivePrefix = "arXiv",
    primaryClass = "hep-ph",
    doi = "10.1007/JHEP12(2021)003",
    journal = "JHEP",
    volume = "12",
    pages = "003",
    year = "2021"
}

@article{Herrero:2021iqt,
    author = "Herrero, Maria J. and Morales, Roberto A.",
    title = "{One-loop renormalization of vector boson scattering with the electroweak chiral Lagrangian in covariant gauges}",
    eprint = "2107.07890",
    archivePrefix = "arXiv",
    primaryClass = "hep-ph",
    reportNumber = "IFT-UAM/CSIC-21-30",
    doi = "10.1103/PhysRevD.104.075013",
    journal = "Phys. Rev. D",
    volume = "104",
    number = "7",
    pages = "075013",
    year = "2021"
}

@article{Herrero:2022krh,
    author = "Herrero, M. J. and Morales, R. A.",
    title = "{One-loop corrections for WW to HH in Higgs EFT with the electroweak chiral Lagrangian}",
    eprint = "2208.05900",
    archivePrefix = "arXiv",
    primaryClass = "hep-ph",
    reportNumber = "IFT-UAM/CSIC-22-80",
    doi = "10.1103/PhysRevD.106.073008",
    journal = "Phys. Rev. D",
    volume = "106",
    number = "7",
    pages = "073008",
    year = "2022"
}

@article{ATLAS:2012yve,
    author = "Aad, Georges and others",
    collaboration = "ATLAS",
    title = "{Observation of a new particle in the search for the Standard Model Higgs boson with the ATLAS detector at the LHC}",
    eprint = "1207.7214",
    archivePrefix = "arXiv",
    primaryClass = "hep-ex",
    reportNumber = "CERN-PH-EP-2012-218",
    doi = "10.1016/j.physletb.2012.08.020",
    journal = "Phys. Lett. B",
    volume = "716",
    pages = "1--29",
    year = "2012"
}

@article{CMS:2012qbp,
    author = "Chatrchyan, Serguei and others",
    collaboration = "CMS",
    title = "{Observation of a New Boson at a Mass of 125 GeV with the CMS Experiment at the LHC}",
    eprint = "1207.7235",
    archivePrefix = "arXiv",
    primaryClass = "hep-ex",
    reportNumber = "CMS-HIG-12-028, CERN-PH-EP-2012-220",
    doi = "10.1016/j.physletb.2012.08.021",
    journal = "Phys. Lett. B",
    volume = "716",
    pages = "30--61",
    year = "2012"
}

@article{Buchalla:2023hqk,
    author = {Buchalla, G. and K\"onig, F. and M\"uller-Salditt, Ch. and Pandler, F.},
    title = "{Two-Higgs-doublet model matched to nonlinear effective theory}",
    eprint = "2312.13885",
    archivePrefix = "arXiv",
    primaryClass = "hep-ph",
    reportNumber = "LMU-ASC 40/23",
    doi = "10.1103/PhysRevD.110.016015",
    journal = "Phys. Rev. D",
    volume = "110",
    number = "1",
    pages = "016015",
    year = "2024"
}

@article{Coleman:1969sm,
    author = "Coleman, Sidney R. and Wess, J. and Zumino, Bruno",
    title = "{Structure of phenomenological Lagrangians. 1.}",
    doi = "10.1103/PhysRev.177.2239",
    journal = "Phys. Rev.",
    volume = "177",
    pages = "2239--2247",
    year = "1969"
}

@article{Song:2022jns,
    author = "Song, Huayang and Wan, Xia and Yu, Jiang-Hao",
    title = "{Custodial symmetry violation in scalar extensions of the standard model*}",
    eprint = "2211.01543",
    archivePrefix = "arXiv",
    primaryClass = "hep-ph",
    doi = "10.1088/1674-1137/ace5a6",
    journal = "Chin. Phys. C",
    volume = "47",
    number = "10",
    pages = "103103",
    year = "2023"
}

@article{Song:2025kjp,
    author = "Song, Huayang and Wan, Xia",
    title = "{Matching the real Higgs triplet extension of Standard Model to HEFT}",
    eprint = "2503.00707",
    archivePrefix = "arXiv",
    primaryClass = "hep-ph",
    doi = "10.1007/JHEP06(2025)249",
    journal = "JHEP",
    volume = "06",
    pages = "249",
    year = "2025"
}

@article{Song:2024kos,
    author = "Song, Huayang and Wan, Xia",
    title = "{A non-linear representation of general scalar extensions of the Standard Model for HEFT matching}",
    eprint = "2412.00355",
    archivePrefix = "arXiv",
    primaryClass = "hep-ph",
    doi = "10.1007/JHEP06(2025)021",
    journal = "JHEP",
    volume = "06",
    pages = "021",
    year = "2025"
}

@article{Sun:2022ssa,
    author = "Sun, Hao and Xiao, Ming-Lei and Yu, Jiang-Hao",
    title = "{Complete NLO operators in the Higgs effective field theory}",
    eprint = "2206.07722",
    archivePrefix = "arXiv",
    primaryClass = "hep-ph",
    doi = "10.1007/JHEP05(2023)043",
    journal = "JHEP",
    volume = "05",
    pages = "043",
    year = "2023"
}

@article{Corbett:2021eux,
    author = "Corbett, Tyler and Helset, Andreas and Martin, Adam and Trott, Michael",
    title = "{EWPD in the SMEFT to dimension eight}",
    eprint = "2102.02819",
    archivePrefix = "arXiv",
    primaryClass = "hep-ph",
    doi = "10.1007/JHEP06(2021)076",
    journal = "JHEP",
    volume = "06",
    pages = "076",
    year = "2021"
}

@article{Alonso:2015fsp,
    author = "Alonso, Rodrigo and Jenkins, Elizabeth E. and Manohar, Aneesh V.",
    title = "{A Geometric Formulation of Higgs Effective Field Theory: Measuring the Curvature of Scalar Field Space}",
    eprint = "1511.00724",
    archivePrefix = "arXiv",
    primaryClass = "hep-ph",
    reportNumber = "CERN-PH-TH-2015-257",
    doi = "10.1016/j.physletb.2016.01.041",
    journal = "Phys. Lett. B",
    volume = "754",
    pages = "335--342",
    year = "2016"
}

@article{Li:2025gbx,
    author = "Li, Hai Tao and Si, Zong-Guo and Wang, Jian and Zhang, Xiao and Zhao, Dan",
    title = "{QCD corrections to Higgs boson pair production and decay to the bb{\textasciimacron}{\ensuremath{\tau}}+{\ensuremath{\tau}}{\ensuremath{-}} final state}",
    eprint = "2503.22001",
    archivePrefix = "arXiv",
    primaryClass = "hep-ph",
    reportNumber = "CPTNP-2025-028",
    doi = "10.1016/j.physletb.2025.139776",
    journal = "Phys. Lett. B",
    volume = "868",
    pages = "139776",
    year = "2025"
}

@misc{Alonso:2025ksv,
    author = "Alonso, Rodrigo and Chattopadhyay, Susobhan and Ingoldby, James",
    title = "{The Potential of HEFT and the scale of New Physics}",
    eprint = "2512.13612",
    archivePrefix = "arXiv",
    primaryClass = "hep-ph",
    reportNumber = "IPPP/25/86, TIFR/TH/25-24",
    month = "12",
    year = "2025"
}

@misc{Alonso:2025jvv,
    author = "Alonso, Rodrigo and Englert, Christoph and Naskar, Wrishik and Rahaman, Shakeel Ur",
    title = "{Assessing (H)EFT theory errors by pitting EoM against Field Redefinitions}",
    eprint = "2511.15609",
    archivePrefix = "arXiv",
    primaryClass = "hep-ph",
    reportNumber = "DESY-25-171, IPPP/25/76",
    month = "11",
    year = "2025"
}

@misc{Brivio:2025sib,
    author = {Brivio, Ilaria and Gr{\"o}ber, Ramona and Schmid, Konstantin},
    title = "{Higgs pair production in gluon fusion to higher orders in Higgs Effective Field Theory}",
    eprint = "2511.23411",
    archivePrefix = "arXiv",
    primaryClass = "hep-ph",
    reportNumber = "COMETA-2025-52",
    month = "11",
    year = "2025"
}

@misc{Brivio:2025yrr,
    author = {Brivio, Ilaria and Gr{\"o}ber, Ramona and Schmid, Konstantin},
    title = "{The Art of Counting: a reappraisal of the HEFT expansion}",
    eprint = "2511.23410",
    archivePrefix = "arXiv",
    primaryClass = "hep-ph",
    reportNumber = "COMETA-2025-51",
    month = "11",
    year = "2025"
}

@misc{Ding:2026qto,
    author = "Ding, Jia-Le and others",
    title = "{Constraining the Higgs potential using multi-Higgs production}",
    eprint = "2601.13248",
    archivePrefix = "arXiv",
    primaryClass = "hep-ph",
    reportNumber = "LHCHWG-2025-015, MPP-2025-231",
    month = "1",
    year = "2026"
}

@article{Alonso:2016oah,
    author = "Alonso, Rodrigo and Jenkins, Elizabeth E. and Manohar, Aneesh V.",
    title = "{Geometry of the Scalar Sector}",
    eprint = "1605.03602",
    archivePrefix = "arXiv",
    primaryClass = "hep-ph",
    reportNumber = "CERN-TH-2016-116",
    doi = "10.1007/JHEP08(2016)101",
    journal = "JHEP",
    volume = "08",
    pages = "101",
    year = "2016"
}

@article{Brivio:2017vri,
    author = "Brivio, Ilaria and Trott, Michael",
    title = "{The Standard Model as an Effective Field Theory}",
    eprint = "1706.08945",
    archivePrefix = "arXiv",
    primaryClass = "hep-ph",
    doi = "10.1016/j.physrep.2018.11.002",
    journal = "Phys. Rept.",
    volume = "793",
    pages = "1--98",
    year = "2019"
}

@article{Arco:2023sac,
    author = "Arco, F. and Domenech, D. and Herrero, M. J. and Morales, R. A.",
    title = "{Nondecoupling effects from heavy Higgs bosons by matching 2HDM to HEFT amplitudes}",
    eprint = "2307.15693",
    archivePrefix = "arXiv",
    primaryClass = "hep-ph",
    reportNumber = "IFT-UAM/CSIC-23-97",
    doi = "10.1103/PhysRevD.108.095013",
    journal = "Phys. Rev. D",
    volume = "108",
    number = "9",
    pages = "095013",
    year = "2023"
}

@article{Dawson:2023oce,
    author = "Dawson, Sally and Fontes, Duarte and Quezada-Calonge, Carlos and Sanz-Cillero, Juan Jos\'e",
    title = "{Is the HEFT matching unique?}",
    eprint = "2311.16897",
    archivePrefix = "arXiv",
    primaryClass = "hep-ph",
    doi = "10.1103/PhysRevD.109.055037",
    journal = "Phys. Rev. D",
    volume = "109",
    number = "5",
    pages = "055037",
    year = "2024"
}

@article{Dawson:2023ebe,
    author = "Dawson, Sally and Fontes, Duarte and Quezada-Calonge, Carlos and Sanz-Cillero, Juan Jos\'e",
    title = "{Matching the 2HDM to the HEFT and the SMEFT: Decoupling and perturbativity}",
    eprint = "2305.07689",
    archivePrefix = "arXiv",
    primaryClass = "hep-ph",
    reportNumber = "IPARCOS-UCM-23-034",
    doi = "10.1103/PhysRevD.108.055034",
    journal = "Phys. Rev. D",
    volume = "108",
    number = "5",
    pages = "055034",
    year = "2023"
}

@article{Buchalla:2016bse,
    author = "Buchalla, G. and Cata, O. and Celis, A. and Krause, C.",
    title = "{Standard Model Extended by a Heavy Singlet: Linear vs. Nonlinear EFT}",
    eprint = "1608.03564",
    archivePrefix = "arXiv",
    primaryClass = "hep-ph",
    reportNumber = "LMU-ASC-35-16",
    doi = "10.1016/j.nuclphysb.2017.02.006",
    journal = "Nucl. Phys. B",
    volume = "917",
    pages = "209--233",
    year = "2017"
}

@article{Henning:2014wua,
    author = "Henning, Brian and Lu, Xiaochuan and Murayama, Hitoshi",
    title = "{How to use the Standard Model effective field theory}",
    eprint = "1412.1837",
    archivePrefix = "arXiv",
    primaryClass = "hep-ph",
    reportNumber = "UCB-PTH-14-40, IPMU14-0353",
    doi = "10.1007/JHEP01(2016)023",
    journal = "JHEP",
    volume = "01",
    pages = "023",
    year = "2016"
}

@article{Ellis:2023zim,
    author = "Ellis, John and Mimasu, Ken and Zampedri, Francesca",
    title = "{Dimension-8 SMEFT analysis of minimal scalar field extensions of the Standard Model}",
    eprint = "2304.06663",
    archivePrefix = "arXiv",
    primaryClass = "hep-ph",
    reportNumber = "KCL-PH-TH/2023-18, CERN-TH-2023-038",
    doi = "10.1007/JHEP10(2023)051",
    journal = "JHEP",
    volume = "10",
    pages = "051",
    year = "2023"
}

@article{Criado:2017khh,
    author = "Criado, Juan C.",
    title = "{MatchingTools: a Python library for symbolic effective field theory calculations}",
    eprint = "1710.06445",
    archivePrefix = "arXiv",
    primaryClass = "hep-ph",
    doi = "10.1016/j.cpc.2018.02.016",
    journal = "Comput. Phys. Commun.",
    volume = "227",
    pages = "42--50",
    year = "2018"
}

@article{DasBakshi:2018vni,
    author = "Das Bakshi, Supratim and Chakrabortty, Joydeep and Patra, Sunando Kumar",
    title = "{CoDEx: Wilson coefficient calculator connecting SMEFT to UV theory}",
    eprint = "1808.04403",
    archivePrefix = "arXiv",
    primaryClass = "hep-ph",
    doi = "10.1140/epjc/s10052-018-6444-2",
    journal = "Eur. Phys. J. C",
    volume = "79",
    number = "1",
    pages = "21",
    year = "2019"
}

@article{Fuentes-Martin:2022jrf,
    author = {Fuentes-Mart\'\i{}n, Javier and K\"onig, Matthias and Pag\`es, Julie and Thomsen, Anders Eller and Wilsch, Felix},
    title = "{A proof of concept for matchete: an automated tool for matching effective theories}",
    eprint = "2212.04510",
    archivePrefix = "arXiv",
    primaryClass = "hep-ph",
    reportNumber = "MITP-22-105, TUM-HEP-1443/22, ZU-TH-58/22",
    doi = "10.1140/epjc/s10052-023-11726-1",
    journal = "Eur. Phys. J. C",
    volume = "83",
    number = "7",
    pages = "662",
    year = "2023"
}

@article{Georgi:1993mps,
  author       = {Georgi, Howard},
  title        = {Effective field theory},
  journal      = {Annual Review of Nuclear and Particle Science},
  volume       = {43},
  pages        = {209--252},
  year         = {1993},
  doi          = {10.1146/annurev.ns.43.120193.001233}
}

@article{Weinberg:1979sa,
    author = "Weinberg, Steven",
    title = "{Baryon and Lepton Nonconserving Processes}",
    journal = "Phys. Rev. Lett.",
    volume = "43",
    year = "1979",
    pages = "1566--1570",
    doi = "10.1103/PhysRevLett.43.1566",
    reportNumber = "HUTP-79/A047"
}

@article{Buchmuller:1985jz,
    author = "Buchmuller, W. and Wyler, D.",
    title = "{Effective Lagrangian Analysis of New Interactions and Flavor Conservation}",
    journal = "Nucl. Phys. B",
    volume = "268",
    year = "1986",
    pages = "621--653",
    doi = "10.1016/0550-3213(86)90262-2"
}

@article{Leung:1984ni,
    author = "Leung, C. N. and Love, S. T. and Rao, S.",
    title = "{Low-Energy Manifestations of a New Interaction Scale: Operator Analysis}",
    journal = "Z. Phys. C",
    volume = "31",
    year = "1986",
    pages = "433",
    doi = "10.1007/BF01588041"
}

@article{Carmona:2021xtq,
    author = "Carmona, Adrian and Lazopoulos, Achilleas and Olgoso, Pablo and Santiago, Jose",
    title = "{Matchmakereft: automated tree-level and one-loop matching}",
    eprint = "2112.10787",
    archivePrefix = "arXiv",
    primaryClass = "hep-ph",
    doi = "10.21468/SciPostPhys.12.6.198",
    journal = "SciPost Phys.",
    volume = "12",
    number = "6",
    pages = "198",
    year = "2022"
}

@article{ParticleDataGroup:2024cfk,
    author = "Navas, S. and others",
    collaboration = "Particle Data Group",
    title = "{Review of particle physics}",
    doi = "10.1103/PhysRevD.110.030001",
    journal = "Phys. Rev. D",
    volume = "110",
    number = "3",
    pages = "030001",
    year = "2024"
}

@article{Appelquist:1980vg,
    author = "Appelquist, Thomas and Bernard, Claude W.",
    title = "{Strongly Interacting Higgs Bosons}",
    reportNumber = "YTP-80-01",
    doi = "10.1103/PhysRevD.22.200",
    journal = "Phys. Rev. D",
    volume = "22",
    pages = "200",
    year = "1980"
}

@article{Longhitano:1980iz,
    author = "Longhitano, Anthony C.",
    title = "{Heavy Higgs Bosons in the Weinberg-Salam Model}",
    reportNumber = "YTP-80-04",
    doi = "10.1103/PhysRevD.22.1166",
    journal = "Phys. Rev. D",
    volume = "22",
    pages = "1166",
    year = "1980"
}

@article{Longhitano:1980tm,
    author = "Longhitano, Anthony C.",
    title = "{Low-Energy Impact of a Heavy Higgs Boson Sector}",
    reportNumber = "YTP-80-27-REV, YTP-80-27",
    doi = "10.1016/0550-3213(81)90109-7",
    journal = "Nucl. Phys. B",
    volume = "188",
    pages = "118--154",
    year = "1981"
}

@article{Feruglio:1992wf,
    author = "Feruglio, F.",
    title = "{The Chiral approach to the electroweak interactions}",
    eprint = "hep-ph/9301281",
    archivePrefix = "arXiv",
    reportNumber = "DFPD-92-TH-50",
    doi = "10.1142/S0217751X93001946",
    journal = "Int. J. Mod. Phys. A",
    volume = "8",
    pages = "4937--4972",
    year = "1993"
}

@article{Herrero:1993nc,
    author = "Herrero, Maria J. and Ruiz Morales, Ester",
    title = "{The Electroweak chiral Lagrangian for the Standard Model with a heavy Higgs}",
    eprint = "hep-ph/9308276",
    archivePrefix = "arXiv",
    reportNumber = "FTUAM-93-24",
    doi = "10.1016/0550-3213(94)90525-8",
    journal = "Nucl. Phys. B",
    volume = "418",
    pages = "431--455",
    year = "1994"
}

@article{Herrero:1994iu,
    author = "Herrero, Maria J. and Ruiz Morales, Ester",
    title = "{Nondecoupling effects of the SM higgs boson to one loop}",
    eprint = "hep-ph/9411207",
    archivePrefix = "arXiv",
    reportNumber = "FTUAM-94-11",
    doi = "10.1016/0550-3213(94)00589-7",
    journal = "Nucl. Phys. B",
    volume = "437",
    pages = "319--355",
    year = "1995"
}

@article{Grinstein:2007iv,
    author = "Grinstein, Benjamin and Trott, Michael",
    title = "{A Higgs-Higgs bound state due to new physics at a TeV}",
    eprint = "0704.1505",
    archivePrefix = "arXiv",
    primaryClass = "hep-ph",
    reportNumber = "UCSD-PTH-07-03",
    doi = "10.1103/PhysRevD.76.073002",
    journal = "Phys. Rev. D",
    volume = "76",
    pages = "073002",
    year = "2007"
}

@article{Buchalla:2012qq,
    author = "Buchalla, Gerhard and Cata, Oscar",
    title = "{Effective Theory of a Dynamically Broken Electroweak Standard Model at NLO}",
    eprint = "1203.6510",
    archivePrefix = "arXiv",
    primaryClass = "hep-ph",
    reportNumber = "LMU-ASC-19-12, FLAVOUR-267104-ERC-9, LMU-ASC\textasciitilde{}19-12",
    doi = "10.1007/JHEP07(2012)101",
    journal = "JHEP",
    volume = "07",
    pages = "101",
    year = "2012"
}

@article{Buchalla:2013rka,
    author = "Buchalla, Gerhard and Cat\`a, Oscar and Krause, Claudius",
    title = "{Complete Electroweak Chiral Lagrangian with a Light Higgs at NLO}",
    eprint = "1307.5017",
    archivePrefix = "arXiv",
    primaryClass = "hep-ph",
    reportNumber = "LMU-ASC-42-13, LMU-ASC\textasciitilde{}42-13",
    doi = "10.1016/j.nuclphysb.2014.01.018",
    journal = "Nucl. Phys. B",
    volume = "880",
    pages = "552--573",
    year = "2014",
    note = "[Erratum: Nucl.Phys.B 913, 475--478 (2016)]"
}

@article{Buchalla:2013eza,
    author = "Buchalla, Gerhard and Cat\'a, Oscar and Krause, Claudius",
    title = "{On the Power Counting in Effective Field Theories}",
    eprint = "1312.5624",
    archivePrefix = "arXiv",
    primaryClass = "hep-ph",
    reportNumber = "LMU-ASC-81-13",
    doi = "10.1016/j.physletb.2014.02.015",
    journal = "Phys. Lett. B",
    volume = "731",
    pages = "80--86",
    year = "2014"
}

@article{Gavela:2014vra,
    author = "Gavela, M. B. and Gonzalez-Fraile, J. and Gonzalez-Garcia, M. C. and Merlo, L. and Rigolin, S. and Yepes, J.",
    title = "{CP violation with a dynamical Higgs}",
    eprint = "1406.6367",
    archivePrefix = "arXiv",
    primaryClass = "hep-ph",
    reportNumber = "FTUAM-14-22, IFT-UAM-CSIC-14-056, YITP-SB-14-17, DFPD-2014-TH-13",
    doi = "10.1007/JHEP10(2014)044",
    journal = "JHEP",
    volume = "10",
    pages = "044",
    year = "2014"
}

@article{Pich:2015kwa,
    author = "Pich, Antonio and Rosell, Ignasi and Santos, Joaquin and Sanz-Cillero, Juan Jose",
    title = "{Low-energy signals of strongly-coupled electroweak symmetry-breaking scenarios}",
    eprint = "1510.03114",
    archivePrefix = "arXiv",
    primaryClass = "hep-ph",
    reportNumber = "IFIC-15-74, FTUV-15-1012, IFT-UAM-CSIC-15-107, FTUAM-15-31",
    doi = "10.1103/PhysRevD.93.055041",
    journal = "Phys. Rev. D",
    volume = "93",
    number = "5",
    pages = "055041",
    year = "2016"
}

@article{Pich:2016lew,
    author = "Pich, Antonio and Rosell, Ignasi and Santos, Joaquin and Sanz-Cillero, Juan Jose",
    title = "{Fingerprints of heavy scales in electroweak effective Lagrangians}",
    eprint = "1609.06659",
    archivePrefix = "arXiv",
    primaryClass = "hep-ph",
    reportNumber = "IFIC-16-49, FTUV-16-0921",
    doi = "10.1007/JHEP04(2017)012",
    journal = "JHEP",
    volume = "04",
    pages = "012",
    year = "2017"
}

@article{Du:2018eaw,
    author = "Du, Yong and Dunbrack, Aaron and Ramsey-Musolf, Michael J. and Yu, Jiang-Hao",
    title = "{Type-II Seesaw Scalar Triplet Model at a 100 TeV $pp$ Collider: Discovery and Higgs Portal Coupling Determination}",
    eprint = "1810.09450",
    archivePrefix = "arXiv",
    primaryClass = "hep-ph",
    doi = "10.1007/JHEP01(2019)101",
    journal = "JHEP",
    volume = "01",
    pages = "101",
    year = "2019"
}

@article{Krause:2018cwe,
    author = "Krause, Claudius and Pich, Antonio and Rosell, Ignasi and Santos, Joaqu\'\i{}n and Sanz-Cillero, Juan Jos\'e",
    title = "{Colorful Imprints of Heavy States in the Electroweak Effective Theory}",
    eprint = "1810.10544",
    archivePrefix = "arXiv",
    primaryClass = "hep-ph",
    reportNumber = "IFIC/18-07, FTUV/18-1026, FERMILAB-PUB-18-550-T, VBSCAN-PUB-06-18",
    doi = "10.1007/JHEP05(2019)092",
    journal = "JHEP",
    volume = "05",
    pages = "092",
    year = "2019"
}

@article{Alonso:2012px,
    author = "Alonso, R. and Gavela, M. B. and Merlo, L. and Rigolin, S. and Yepes, J.",
    title = "{The Effective Chiral Lagrangian for a Light Dynamical ''Higgs Particle''}",
    eprint = "1212.3305",
    archivePrefix = "arXiv",
    primaryClass = "hep-ph",
    reportNumber = "FTUAM-12-115, IFT-UAM-CSIC-12-113, CERN-PH-TH-2012-335, DFPD-2012-TH-23",
    doi = "10.1016/j.physletb.2013.04.037",
    journal = "Phys. Lett. B",
    volume = "722",
    pages = "330--335",
    year = "2013",
    note = "[Erratum: Phys.Lett.B 726, 926 (2013)]"
}

@article{Brivio:2013pma,
    author = "Brivio, I. and Corbett, T. and \'Eboli, O. J. P. and Gavela, M. B. and Gonzalez-Fraile, J. and Gonzalez-Garcia, M. C. and Merlo, L. and Rigolin, S.",
    title = "{Disentangling a dynamical Higgs}",
    eprint = "1311.1823",
    archivePrefix = "arXiv",
    primaryClass = "hep-ph",
    reportNumber = "FTUAM-13-32, IFT-UAM-CSIC-13-119, YITP-SB-13-33, DFPD-2013-TH-20",
    doi = "10.1007/JHEP03(2014)024",
    journal = "JHEP",
    volume = "03",
    pages = "024",
    year = "2014"
}

@article{Brivio:2016fzo,
    author = "Brivio, I. and Gonzalez-Fraile, J. and Gonzalez-Garcia, M. C. and Merlo, L.",
    title = "{The complete HEFT Lagrangian after the LHC Run I}",
    eprint = "1604.06801",
    archivePrefix = "arXiv",
    primaryClass = "hep-ph",
    reportNumber = "FTUAM-16-13, IFT-UAM-CSIC-16-034, YITP-SB-16-14",
    doi = "10.1140/epjc/s10052-016-4211-9",
    journal = "Eur. Phys. J. C",
    volume = "76",
    number = "7",
    pages = "416",
    year = "2016"
}

@incollection{Pich:2018ltt,
    author = "Pich, Antonio",
    editor = "Davidson, Sacha and Gambino, Paolo and Laine, Mikko and Neubert, Matthias and Salomon, Christophe",
    title = "{Effective Field Theory with Nambu-Goldstone Modes}",
    booktitle = "{Effective Field Theory in Particle Physics and Cosmology}",
    series = "{Lecture Notes of the Les Houches Summer School}",
    volume = "108",
    pages = "3",
    eprint = "1804.05664",
    archivePrefix = "arXiv",
    primaryClass = "hep-ph",
    doi = "10.1093/oso/9780198855743.003.0003",
    publisher = "Oxford University Press",
    year = "2019"
}

@article{Merlo:2016prs,
    author = "Merlo, Luca and Saa, Sara and Sacrist\'an-Barbero, Mario",
    title = "{Baryon Non-Invariant Couplings in Higgs Effective Field Theory}",
    eprint = "1612.04832",
    archivePrefix = "arXiv",
    primaryClass = "hep-ph",
    reportNumber = "FTUAM-16-45, IFT-UAM-CSIC-16-135",
    doi = "10.1140/epjc/s10052-017-4753-5",
    journal = "Eur. Phys. J. C",
    volume = "77",
    number = "3",
    pages = "185",
    year = "2017"
}

@misc{ATLAS:2025qxq,
    collaboration = "ATLAS",
    title = "{Combined measurements of Higgs boson production and decay at $\sqrt{s} =$ 13 TeV using up to 140 fb$^{-1}$ of data collected by the ATLAS Experiment}",
    reportNumber = "ATLAS-CONF-2025-006",
    year = "2025"
}

@misc{CMS:2025jwz,
    collaboration = "CMS",
    title = "{Combined measurements and interpretations of Higgs boson production and decay at $\sqrt s$=13 TeV}",
    reportNumber = "CMS-PAS-HIG-21-018",
    year = "2025"
}

@article{FileviezPerez:2008bj,
    author = "Fileviez Perez, Pavel and Patel, Hiren H. and Ramsey-Musolf, Michael. J. and Wang, Kai",
    title = "{Triplet Scalars and Dark Matter at the LHC}",
    eprint = "0811.3957",
    archivePrefix = "arXiv",
    primaryClass = "hep-ph",
    reportNumber = "NPAC-08-20, MADPH-08-1517, IPMU-08-0091",
    doi = "10.1103/PhysRevD.79.055024",
    journal = "Phys. Rev. D",
    volume = "79",
    pages = "055024",
    year = "2009"
}

@article{Patel:2012pi,
    author = "Patel, Hiren H. and Ramsey-Musolf, Michael J.",
    title = "{Stepping Into Electroweak Symmetry Breaking: Phase Transitions and Higgs Phenomenology}",
    eprint = "1212.5652",
    archivePrefix = "arXiv",
    primaryClass = "hep-ph",
    doi = "10.1103/PhysRevD.88.035013",
    journal = "Phys. Rev. D",
    volume = "88",
    pages = "035013",
    year = "2013"
}

@article{Niemi:2018asa,
    author = "Niemi, Lauri and Patel, Hiren H. and Ramsey-Musolf, Michael J. and Tenkanen, Tuomas V. I. and Weir, David J.",
    title = "{Electroweak phase transition in the real triplet extension of the SM: Dimensional reduction}",
    eprint = "1802.10500",
    archivePrefix = "arXiv",
    primaryClass = "hep-ph",
    reportNumber = "HIP-2018-7-TH, ACFI-T18-04, HIP-2018-7/TH",
    doi = "10.1103/PhysRevD.100.035002",
    journal = "Phys. Rev. D",
    volume = "100",
    number = "3",
    pages = "035002",
    year = "2019"
}

@article{Niemi:2020hto,
    author = "Niemi, Lauri and Ramsey-Musolf, Michael J. and Tenkanen, Tuomas V. I. and Weir, David J.",
    title = "{Thermodynamics of a Two-Step Electroweak Phase Transition}",
    eprint = "2005.11332",
    archivePrefix = "arXiv",
    primaryClass = "hep-ph",
    reportNumber = "HIP-2020-11/TH, ACFI-T20-05",
    doi = "10.1103/PhysRevLett.126.171802",
    journal = "Phys. Rev. Lett.",
    volume = "126",
    number = "17",
    pages = "171802",
    year = "2021"
}

@book{Gunion:1989we,
    author = "Gunion, John F. and Haber, Howard E. and Kane, Gordon L. and Dawson, Sally",
    title = "{The Higgs Hunter's Guide}",
    reportNumber = "SCIPP-89/13, UCD-89-4, BNL-41644",
    doi = "10.1201/9780429496448",
    isbn = "978-0-429-49644-8",
    volume = "80",
    year = "2000"
}

@article{Branco:2011iw,
    author = "Branco, G. C. and Ferreira, P. M. and Lavoura, L. and Rebelo, M. N. and Sher, Marc and Silva, Joao P.",
    title = "{Theory and phenomenology of two-Higgs-doublet models}",
    eprint = "1106.0034",
    archivePrefix = "arXiv",
    primaryClass = "hep-ph",
    doi = "10.1016/j.physrep.2012.02.002",
    journal = "Phys. Rept.",
    volume = "516",
    pages = "1--102",
    year = "2012"
}

@article{Lavoura:1994fv,
    author = "Lavoura, L. and Silva, Joao P.",
    title = "{Fundamental CP violating quantities in a SU(2) x U(1) model with many Higgs doublets}",
    eprint = "hep-ph/9404276",
    archivePrefix = "arXiv",
    doi = "10.1103/PhysRevD.50.4619",
    journal = "Phys. Rev. D",
    volume = "50",
    pages = "4619--4624",
    year = "1994"
}

@article{Botella:1994cs,
    author = "Botella, F. J. and Silva, Joao P.",
    title = "{Jarlskog - like invariants for theories with scalars and fermions}",
    eprint = "hep-ph/9411288",
    archivePrefix = "arXiv",
    reportNumber = "FTUV-94-68, IFIC-94-65",
    doi = "10.1103/PhysRevD.51.3870",
    journal = "Phys. Rev. D",
    volume = "51",
    pages = "3870--3875",
    year = "1995"
}

@book{Branco:1999fs,
    author = "Branco, Gustavo C. and Lavoura, Luis and Silva, Joao P.",
    title = "{CP Violation}",
    doi = "10.1093/oso/9780198503996.001.0001",
    isbn = "978-1-383-02075-5, 978-0-19-850399-6",
    volume = "103",
    year = "1999"
}

@article{Fontes:2014xva,
    author = "Fontes, Duarte and Rom{\~a}o, J. C. and Silva, Jo{\~a}o P.",
    title = "{$h \rightarrow Z \gamma$ in the complex two Higgs doublet model}",
    eprint = "1408.2534",
    archivePrefix = "arXiv",
    primaryClass = "hep-ph",
    doi = "10.1007/JHEP12(2014)043",
    journal = "JHEP",
    volume = "12",
    pages = "043",
    year = "2014"
}

@article{Robens:2015gla,
    author = "Robens, Tania and Stefaniak, Tim",
    title = "{Status of the Higgs Singlet Extension of the Standard Model after LHC Run 1}",
    eprint = "1501.02234",
    archivePrefix = "arXiv",
    primaryClass = "hep-ph",
    reportNumber = "SCIPP-15-02",
    doi = "10.1140/epjc/s10052-015-3323-y",
    journal = "Eur. Phys. J. C",
    volume = "75",
    pages = "104",
    year = "2015"
}

@article{Robens:2016xkb,
    author = "Robens, Tania and Stefaniak, Tim",
    title = "{LHC Benchmark Scenarios for the Real Higgs Singlet Extension of the Standard Model}",
    eprint = "1601.07880",
    archivePrefix = "arXiv",
    primaryClass = "hep-ph",
    reportNumber = "SCIPP-16-03",
    doi = "10.1140/epjc/s10052-016-4115-8",
    journal = "Eur. Phys. J. C",
    volume = "76",
    number = "5",
    pages = "268",
    year = "2016"
}

\end{document}